\documentclass[11pt]{article}
\usepackage[hmargin={3.0cm,3.0cm},vmargin={3.0cm,3.0cm}]{geometry}

\usepackage[hang,small,center]{caption}
\usepackage{amsmath,amssymb}
\usepackage{amsfonts}
\usepackage{mathrsfs}
\usepackage{dsfont}

\usepackage{epic}
\usepackage{eepic}
\usepackage{graphicx}

\usepackage{pgf}
\usepackage{tikz}
\usetikzlibrary{arrows.meta}

\usepackage{amsmath}
\usepackage{mathtools}

\usepackage{footnote}

\usepackage{cite}
\usepackage{enumerate}
\usepackage[curve]{xy}

\usepackage[titletoc,title]{appendix}

\newcommand{\ds}{\displaystyle}

\renewcommand{\author}[1]{\large\rm #1\\ \bigskip}
\newcommand{\address}[1]{{\normalsize\it #1\\}\bigskip}
\renewcommand{\title}[1]{\bigskip\bigskip\Large\bf #1\bigskip\bigskip\\}

\newcommand{\Bigpsi}[3]{\phantom{\Psi}_2 \kern -.05em
\Psi_2\left(\genfrac{}{}{0pt}{}{#1}{#2}\biggl|#3\right)}

\newcommand{\ihat}{\mathbf{e}_i}
\newcommand{\jhat}{\mathbf{e}_j}
\newcommand{\khat}{\mathbf{e}_k}

\newcommand{\bea}{\begin{eqnarray}}
\newcommand{\eea}{\end{eqnarray}}

\newcommand{\beq}{\begin{equation}}
\newcommand{\eeq}{\end{equation}}

\newcommand{\n}{{\boldsymbol{n}}}
\newcommand{\x}{{\boldsymbol{x}}}

\newcommand{\oW}{\overline{W}}

\newcommand{\lag}{{\mathcal L}}
\newcommand{\ol}{\overline{\mathcal{L}}}

\newcommand{\G}{{\mathscr G}}
\renewcommand{\L}{{\mathscr L}}

\newcommand{\olam}{{\overline{\Lambda}}}

\def\re{\mathop{\hbox{\rm Re}}\nolimits}
\def\im{\mathop{\hbox{\rm Im}}\nolimits}

\newcounter{app}
\newcounter{sapp}[app]

\begin{document}

\vglue 1.5cm

\begin{center}

\title{Exactly solved models on planar graphs with vertices in $\mathbb{Z}^3$}
\author{Andrew P.~Kels}
\address{Institute of Physics, University of Tokyo,\\
 Komaba, Tokyo 153-8902, Japan}

\end{center}

\vskip 0.0cm 
\begin{abstract}

It is shown how exactly solved edge interaction models on the square lattice, may be extended onto more general planar graphs, with edges connecting a subset of next nearest neighbour vertices of $\mathbb{Z}^3$.  This is done by using local deformations of the square lattice, that arise through the use of the star-triangle relation.  Similar to Baxter's Z-invariance property, these local deformations leave the partition function invariant up to some simple factors coming from the star-triangle relation.  The deformations used here extend the usual formulation of Z-invariance, by requiring the introduction of oriented rapidity lines which form directed closed paths in the rapidity graph of the model.  The quasi-classical limit is also considered, in which case the deformations imply a classical Z-invariance property, as well as a related local closure relation, for the action functional of a system of classical discrete Laplace equations.

\end{abstract}
\vspace{-0.1cm}
\tableofcontents




\section{Introduction}

The star-triangle relation is a special form of the Yang-Baxter equation, for two-dimensional edge interaction models of statistical mechanics, where spins are located at vertices of a lattice, and interactions take place between neighbouring spins connected by lattice edges.  Well known examples of such edge interaction models, include the chiral Potts model \cite{AuY87,BPY87}, Kashiwara-Miwa model \cite{Kashiwara:1986,HY90}, Fateev-Zamolodchikov $Z_N$-model \cite{FZ82}, and the Ising model \cite{O44,Baxterbook}.  Each of these models involve interacting spins which take a discrete set of integer values.

There are also edge interaction models involving spins which take a continuous set of real values, including the Zamolodchikov ``fishing-net'' model \cite{Zam-fish}, that describes certain planar Feynman diagrams in quantum field theory, and the Faddeev-Volkov model \cite{FV95,BMS07a,BMS07b}, connected with the quantization of discrete conformal transformations, and a few others \cite{Spiridonov-statmech,K14,GS15,GahramanovKels}.  The most general known models in this class were recently obtained \cite{Bazhanov:2010kz,Bazhanov:2011mz,K15,Yamazaki2013}, which include all of the above mentioned discrete/continuous spin edge interaction models as limiting cases.

For edge interaction models, the star-triangle relation is a condition of integrability, implying that the row-to-row transfer matrices of the model commute, which, through the method of Baxter \cite{Bax72}, allows for an exact calculation of the partition function in the thermodynamic limit.  The same star-triangle relation implies a related property for integrable models known as Z-invariance \cite{Bax1,Baxter:1986prs}, whereby the partition function is invariant up to simple factors, under continuous deformations of the rapidity graph associated to the model.  Using this property, an integrable lattice model can be defined on general ``Baxter'' graphs \cite{Bax1,Baxter:1986prs,Bax2}, and the properties of the latter model can be related to the usual square lattice model.  Due to its association with integrability, the use of the Z-invariance property is quite extensive, for example it has played key roles in various studies of the Ising model \cite{AUYANG198744,REYESMARTINEZ1997203,Martı́nez1998463,Costa-Santos,Au-Yang2007,Boutillier2010,Boutillier2011}, the derivation of the order parameter of the chiral Potts model \cite{JimboMiwaNakayashiki,Bax2,Baxter:2005jt,Au-YangPerk2011}, relating integrable lattice models with isoradial embeddings of graphs and circle patterns \cite{Costa-Santos,BMS07a}, and relating continuous spin lattice models with supersymmetric quiver gauge theories \cite{Yamazaki2012}.
\\

This paper introduces a new extension of the concept of Z-invariance for edge interaction models.  This is done by first reformulating the edge interaction model on the square lattice, as a type of face interaction model, where spins are connected on a diagonal of two types of oriented elementary squares, shown below.  These squares are used as building blocks of a two-dimensional surface, which contains an edge interaction model defined on a general planar graph, with edges connecting a subset of next nearest neighbour vertices of $\mathbb{Z}^3$.  The partition function of the latter model is invariant, up to some simple factors, under local deformations of the faces, which take the form of ``cubic flips''.  Particularly, by using the deformations, this edge interaction model may be related to the usual model defined on the square lattice.

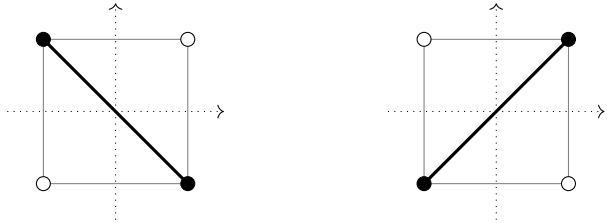
\begin{figure}[htb]

\centering
\begin{tikzpicture}[scale=2.4]

\draw[-,gray] (-0.4,1.6)--(0.4,1.6)--(0.4,2.4)--(-0.4,2.4)--(-0.4,1.6);
\draw[-,very thick] (0.4,1.6)--(-0.4,2.4);
\draw[->,dotted] (-0.6,2.0)--(0.6,2.0);
\draw[->,dotted] (0,1.4)--(0,2.6);
\filldraw[fill=black,draw=black] (0.4,1.6) circle (1.1pt);
\filldraw[fill=black,draw=black] (-0.4,2.4) circle (1.1pt);
\filldraw[fill=white,draw=black] (-0.4,1.6) circle (1.1pt);
\filldraw[fill=white,draw=black] (0.4,2.4) circle (1.1pt);

\begin{scope}[xshift=60pt]
\draw[-,gray] (-0.4,1.6)--(0.4,1.6)--(0.4,2.4)--(-0.4,2.4)--(-0.4,1.6);
\draw[-,very thick] (-0.4,1.6)--(0.4,2.4);
\draw[->,dotted] (-0.6,2.0)--(0.6,2.0);
\draw[->,dotted] (0,1.4)--(0,2.6);
\filldraw[fill=black,draw=black] (-0.4,1.6) circle (1.1pt);
\filldraw[fill=black,draw=black] (0.4,2.4) circle (1.1pt);
\filldraw[fill=white,draw=black] (0.4,1.6) circle (1.1pt);
\filldraw[fill=white,draw=black] (-0.4,2.4) circle (1.1pt);

\end{scope}

\end{tikzpicture}

\end{figure}

All of the required local deformations of the surface associated to the lattice model, are depicted in Appendix \ref{app:str}, and are used to show Z-invariance in Section \ref{sec:z-invar}.  This formulation of Z-invariance allows for somewhat surprising changes to be made to the square lattice model.  For instance, any interior edge of the lattice, can be taken an arbitrary distance out of the plane of the square lattice, at the cost of only introducing some simple factors to the partition function.  Graphically this corresponds to repeated applications of the deformations appearing in Figure \ref{1cube}.  Note also that the deformations in the latter figure, involve rapidity lines which form directed closed paths around the faces and the vertices (see also Figure \ref{newd}).  This is something that distinguishes the deformations considered here, from the usual case of Z-invariance, where such configurations of rapidity lines are not permitted.

There is however still a limitation on the types of allowed deformations, since the Yang-Baxter equation is not satisfied when rapidity lines form directed closed paths around vertices.  A consequence of this, is that the deformations of the surface must follow a particular ordering (although this is already required to an extent, since sequences of deformations are generally non-commuting), and there is also a specific condition on the allowed configuration of elementary squares on the surface.  This is explained in more detail in Section \ref{sec:sigp}.
\\

The second part of this paper considers the quasi-classical (low temperature) limit of the above models of statistical mechanics, which has been shown \cite{BMS07a,Bazhanov:2010kz,BKS2} to lead to classical discrete integrable equations that satisfy the integrability condition known as ``3D-consistency'' \cite{nijhoffwalker,BoSur,AdlerBobenkoSuris}.  In this limit, the fluctuating spins of the lattice model approach a fixed ground state configuration, which is determined as the solution of a classical discrete integrable equation.

For example, the leading order quasi-classical expansion of the star-triangle relation, is described by a classical star-triangle relation, that is satisfied on solutions of a discrete integrable equation in the so called three-leg form.  The latter three-leg equations have been identified \cite{BKS2} as classical equations in the classification given by Adler, Bobenko, and Suris (ABS) \cite{AdlerBobenkoSuris}.  Similarly, the quasi-classical asymptotics of the partition function of the model, are determined by maximising an action functional for the system of classical ``discrete Laplace equations'' \cite{MR2467378,BG11}.  Consequently, the lattice model is interpreted as a quantum counterpart of the discrete Laplace system associated to an ABS equation, while the star-triangle relation itself may be interpreted as a quantum counterpart of a single ABS quad equation.

In Section \ref{sec:qcl}, it is shown that in the quasi-classical limit, the resulting action functional for the classical discrete Laplace equations also satisfies a {\it classical} Z-invariance property, quite analogous to that described for the corresponding statistical mechanical model.  Similarly to the statistical mechanics case, the classical Z-invariance property is shown to be a consequence of a classical star-triangle relation.  The classical Z-invariance is in fact closely connected to a so-called ``closure relation'' for the action functional, that was introduced by Lobb and Nijhoff \cite{LN09} for particular ABS systems involving three-point Lagrangians.  It is shown here how the classical star-triangle relation, may be equivalently interpreted as the corresponding closure relation for the discrete Laplace system, defined in terms of two-point Lagrangians.

\section{The star-triangle relation for exactly solved models}\label{sec:modeldef}
In this section, the star-triangle relation for edge interaction models is introduced, which plays the central role in the formulation of Z-invariance of this paper.  The definition of the edge interaction model will be general in order to allow the Z-invariance property to be applicable to as wide a range of models as possible.

\subsection{Square lattice model}

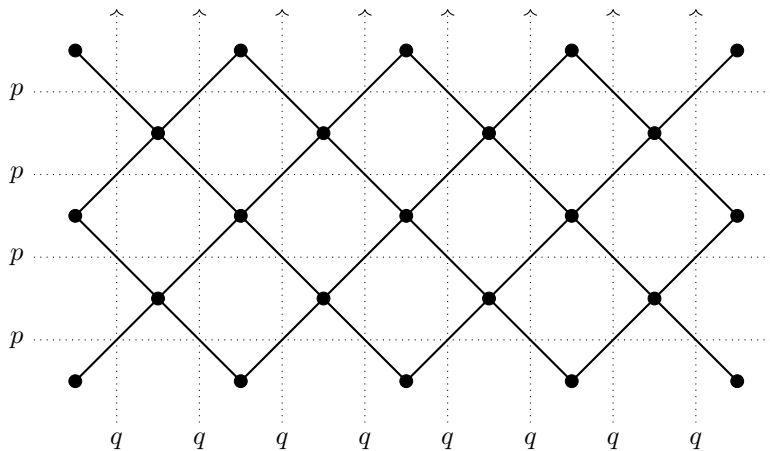
\begin{figure}[htb]
\centering
\begin{tikzpicture}[scale=1.1]

\draw[-,thick] (-0.5,-0.5)--(3.5,3.5);
\draw[-,thick] (-0.5,3.5)--(3.5,-0.5);
\draw[-,thick] (-0.5,1.5)--(1.5,3.5)--(3.5,1.5)--(1.5,-0.5)--(-0.5,1.5);
\draw[-,thick] (-4.5,-0.5)--(-0.5,3.5);
\draw[-,thick] (-4.5,3.5)--(-0.5,-0.5);
\draw[-,thick] (-4.5,1.5)--(-2.5,3.5)--(-0.5,1.5)--(-2.5,-0.5)--(-4.5,1.5);
\foreach \x in {-4,...,3}{
\draw[->,dotted] (\x,-1) -- (\x,4);
\fill[white!] (\x,-1) circle (0.1pt)
node[below=0.05pt]{\color{black}\small $q$};}
\foreach \y in {0,...,3}{
\draw[->,dotted] (-5,\y) -- (4,\y);
\fill[white!] (-5,\y) circle (0.1pt)
node[left=0.05pt]{\color{black}\small $p$};}
\foreach \y in {-0.5,1.5,3.5}{
\foreach \x in {-4.5,-2.5,...,3.5}{
\filldraw[fill=black,draw=black] (\x,\y) circle (2.2pt);}}
\foreach \y in {0.5,2.5}{
\foreach \x in {-3.5,-1.5,...,2.5}{
\filldraw[fill=black,draw=black] (\x,\y) circle (2.2pt);}}

\end{tikzpicture}
\caption{Square lattice $L$ (solid lines) and the directed rapidity graph $\L$ (dotted lines).}
\label{squarelattice}
\end{figure}

Denote the square lattice by $L$, with sets of edges, and vertices, $E(L)$, and $V(L)$, respectively.  Vertices $i\in V(L)$, are represented by solid (black) circles in Figure \ref{squarelattice}, and two nearest neighbour vertices $i,j\in V(L)$ are connected by an edge $(ij)\in E(L)$.

\begin{figure}[hbt]
\centering
\begin{tikzpicture}[scale=2.6]

\draw[-,very thick] (-0.5,2)--(0.5,2);
\draw[->,thick,dotted] (0.4,1.6)--(-0.4,2.4);
\fill[white!] (0.4,1.6) circle (0.01pt)
node[below=0.5pt]{\color{black}\small $q$};
\draw[->,thick,dotted] (-0.4,1.6)--(0.4,2.4);
\fill[white!] (-0.4,1.6) circle (0.01pt)
node[below=0.5pt]{\color{black}\small $p$};
\filldraw[fill=black,draw=black] (-0.5,2) circle (0.9pt)
node[left=3pt]{\color{black} $x_i$};
\filldraw[fill=black,draw=black] (0.5,2) circle (0.9pt)
node[right=3pt]{\color{black} $y_j$};

\fill (0,1.3) circle(0.01pt)
node[below=0.05pt]{\color{black} $ W_{pq}(x_i,x_j)$};

\begin{scope}[xshift=60pt,yshift=57pt]
\draw[-,very thick] (0,-0.5)--(0,0.5);
\draw[->,thick,dotted] (-0.4,-0.4)--(0.4,0.4);
\fill[white!] (-0.4,-0.4) circle (0.01pt)
node[below=0.5pt]{\color{black}\small $p$};
\draw[->,thick,dotted] (0.4,-0.4)--(-0.4,0.4);
\fill[white!] (0.4,-0.4) circle (0.01pt)
node[below=0.5pt]{\color{black}\small $q$};
\filldraw[fill=black,draw=black] (0,-0.5) circle (0.9pt)
node[below=3pt]{\color{black} $x_i$};
\filldraw[fill=black,draw=black] (0,0.5) circle (0.9pt)
node[above=3pt]{\color{black} $x_j$};

\fill (0,-0.7) circle(0.01pt)
node[below=0.05pt]{\color{black} $\oW_{pq}(x_i,x_j)$};
\end{scope}
\end{tikzpicture}
\caption{An edge of the first type, in $E^{(1)}(L)$ (left), an edge of the second type, in $E^{(2)}(L)$ (right), and their respective Boltzmann weights.}
\label{2boltzmannweights}
\end{figure}
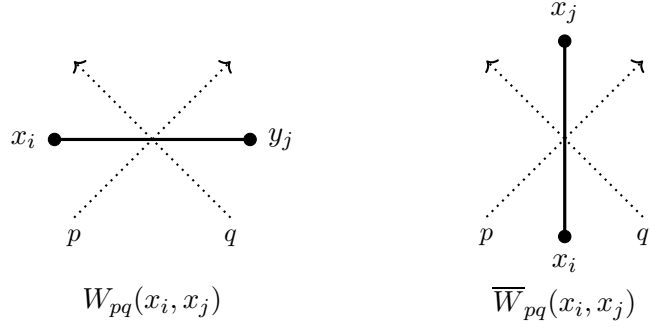

A directed rapidity graph, denoted by $\L$, is depicted in Figure \ref{squarelattice} as directed dotted lines crossing each edge of $L$ at 45 degree angles. The horizontally and vertically directed rapidity lines, are respectively labelled by rapidity variables $p$, and $q$.  There are two different types of edges of the lattice, shown in Figure \ref{2boltzmannweights}, which are distinguished by the crossing of the two rapidity lines.  The set of edges of the first type in $L$, will be denoted by $E^{(1)}(L)$, and the set of edges of the second type will be denoted by $E^{(2)}(L)$, as indicated in Figure \ref{2boltzmannweights}.

A spin variable $x_i$ is assigned to each vertex $i\in V(L)$.  These spin variables take values in some subset of the integers, or the real numbers.  Interactions take place between two spins $x_i,x_j$, at nearest neighbour vertices $i,j$.  This interaction is characterised by the two types of Boltzmann weights $ W_{pq}(x_i,x_j)$, $\oW_{pq}(x_i,x_j)$, that are associated respectively with edges $(ij)\in E^{(1)}(L)$ and $(ij)\in E^{(2)}(L)$, as indicated in Figure \ref{2boltzmannweights}.  The Boltzmann weights are functions of the two spin variables $x_i,x_j$, and the rapidity variables $p$, and $q$, associated to an edge $(ij)$.

The lattice model also depends on the vertex Boltzmann weight $ S(x_i)$, that is a function of a single spin variable $x_i$, and is independent of any rapidity variables.  Depending on the model, this weight may be unity, or otherwise has a non-trivial dependence on the spin variable $x_i$.  The spins located on the boundary of the lattice $L$ are assigned some fixed values.

The partition function of the lattice model, is defined as
\beq
\label{zdef}
Z_0=\sum_{\bf x} \prod_{(ij)\in E^{(1)}(L)} W_{pq}(x_i,x_j) \prod_{(ij)\in E^{(2)}(L)}\oW_{pq}(x_i,x_j) \prod_{i\in V_{int}(L)}  S(x_i)\,,
\eeq
where the first and second products, are products of all Boltzmann weights associated to edges $(ij)\in E^{(1)}(L)$, and $(ij)\in E^{(2)}(L)$ respectively, and the third product, is a product of Boltzmann weights $S(x_i)$, associated to spins at interior vertices $i\in V_{int}(L)$.  The sum $\sum_\x$, represents a sum over all values of interior spins $\sum_{x_1}\sum_{x_2},\ldots,\sum_{x_n}$, where $x_1,x_2,\ldots,x_n$, are spins assigned to vertices $ i_1,i_2,\ldots,i_n\in V_{int}(L)$.  For example, in the case of an N-state spin model, the $x_i$ are summed over values $x_i\in\mathbb{Z}\mbox{ mod }N$.  The expression for the partition function \eqref{zdef}, is for integer valued spin models, in the case of real valued continuous spin models, the sum should be replaced by an integration over all interior spin configurations.

\subsection{Star-triangle relation}

An integrability condition for the lattice model, is that the Boltzmann weights $ W_{pq}(x_i,x_j)$, $\oW_{pq}(x_i,x_j)$, satisfy the Yang-Baxter equation.  In this case, the Yang-Baxter equation takes the form of the following star-triangle relation
\beq
\label{STR}
\begin{array}{c}
\ds\sum_{x_0} \,  S(x_0)\,\oW_{qr}(x_1,x_0)\, W_{pr}(x_2,x_0)\,\oW_{pq}(x_0,x_3)=R_{pqr}\, W_{qr}(x_2,x_3)\,\oW_{pr}(x_1,x_3)\, W_{pq}(x_2,x_1)\,, \\[0.6cm]
\ds\sum_{x_0} \,  S(x_0)\,\oW_{qr}(x_0,x_1)\, W_{pr}(x_0,x_2)\,\oW_{pq}(x_3,x_0)=R_{pqr}\, W_{qr}(x_3,x_2)\,\oW_{pr}(x_3,x_1)\, W_{pq}(x_1,x_2)\,.
\end{array}
\eeq

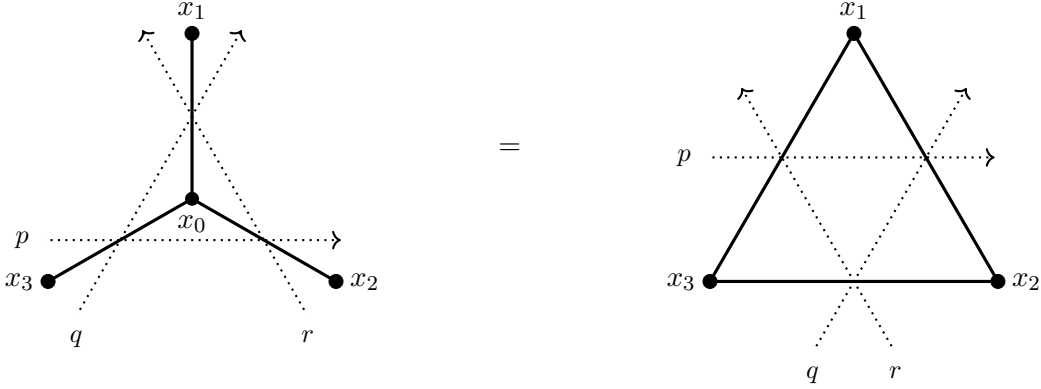
\begin{figure}[tbh]
\centering
\begin{tikzpicture}[scale=2.2]

\draw[-,very thick] (-2,0)--(-2,1);
\draw[-,very thick] (-2,0)--(-2.87,-0.5);
\draw[-,very thick] (-2,0)--(-1.13,-0.5);
\draw[->,black,thick,dotted] (-2.9,-0.25)--(-1.1,-0.25);
\fill[white!] (-2.9,-0.25) circle (0.5pt)
node[left=1.5pt]{\color{black}\small $p$};
\draw[->,black,thick,dotted] (-2.7,-0.71)--(-1.7,1.02);
\fill[white!] (-2.7,-0.71) circle (0.5pt)
node[below=1.5pt]{\color{black}\small $q$};
\draw[->,black!30!black,thick,dotted] (-1.3,-0.71)--(-2.3,1.02);
\fill[white!] (-1.3,-0.71) circle (0.5pt)
node[below=1.5pt]{\color{black}\small $r$};
\fill (-2,0) circle (1.2pt)
node[below=2.5pt]{\color{black} $x_0$};
\filldraw[fill=black,draw=black] (-2,1) circle (1.2pt)
node[above=1.5pt] {\color{black} $x_1$};
\filldraw[fill=black,draw=black] (-2.87,-0.5) circle (1.2pt)
node[left=1.5pt] {\color{black} $x_3$};
\filldraw[fill=black,draw=black] (-1.13,-0.5) circle (1.2pt)
node[right=1.5pt] {\color{black} $x_2$};

\fill[white!] (0.05,0.3) circle (0.01pt)
node[left=0.05pt] {\color{black}$=$};

\draw[-,very thick] (2,1)--(1.13,-0.5);
\draw[-,very thick] (1.13,-0.5)--(2.87,-0.5);
\draw[-,very thick] (2.87,-0.5)--(2,1);
\draw[->,black,thick,dotted] (1.1,0.25)--(2.85,0.25);
\fill[white!] (1.1,0.25) circle (0.5pt)
node[left=1.5pt]{\color{black}\small $p $};
\draw[->,black,thick,dotted] (1.75,-0.93)--(2.68,0.67);
\fill[white!] (1.75,-0.93) circle (0.5pt)
node[below=1.5pt]{\color{black}\small $q$};
\draw[->,black!30!black,thick,dotted] (2.25,-0.93)--(1.32,0.67);
\fill[white!] (2.25,-0.93) circle (0.5pt)
node[below=1.5pt]{\color{black}\small $r$};
\filldraw[fill=black,draw=black] (2,1) circle (1.2pt)
node[above=1.5pt]{\color{black} $x_1$};
\filldraw[fill=black,draw=black] (1.13,-0.5) circle (1.2pt)
node[left=1.5pt]{\color{black} $x_3$};
\filldraw[fill=black,draw=black] (2.87,-0.5) circle (1.2pt)
node[right=1.5pt]{\color{black} $x_2$};

\end{tikzpicture}
\caption{The second expression for star-triangle relation \eqref{STR}.  The first expression is equivalent to reversing the orientation of the three rapidity lines $p,q,r$.}
\label{fig3}
\end{figure}

The second expression for the star-triangle relation \eqref{STR}, is depicted graphically in Figure \ref{fig3}.  The first expression is equivalent to reversing the orientation of the three rapidity lines, that appear on both sides of Figure \ref{fig3}.  For non-chiral models, where $W_{pq}(x_i,x_j)=W_{pq}(x_j,x_i)$, and $\oW_{pq}(x_i,x_j)=\oW_{pq}(x_j,x_i)$, the two expressions in \eqref{STR} are equivalent.

For an $N$-state spin model, the sum in \eqref{STR} is taken over all values of the interior spin $x_0\in\mathbb{Z}\mbox{ mod }N$, for fixed values of the boundary spins $x_1,x_2,x_3$.  The factor $R_{pqr}$, is a spin independent factor that depends only on the value of the rapidity variables $p,q,r$.  This can be factorised in the form $R_{pqr}=f_{pq}f_{qr}/f_{pr}$, where the $f_{pq}$ depend only on the rapidity variables $p,q$ \cite{MatSmirn90,Bax02rip}.  The star-triangle relation \eqref{STR}, is a local condition on the Boltzmann weights, such that the row-to-row transfer matrices of the model commute, from which an exact expression for the partition function \eqref{zdef} may be determined in the thermodynamic limit \cite{Bax72}.

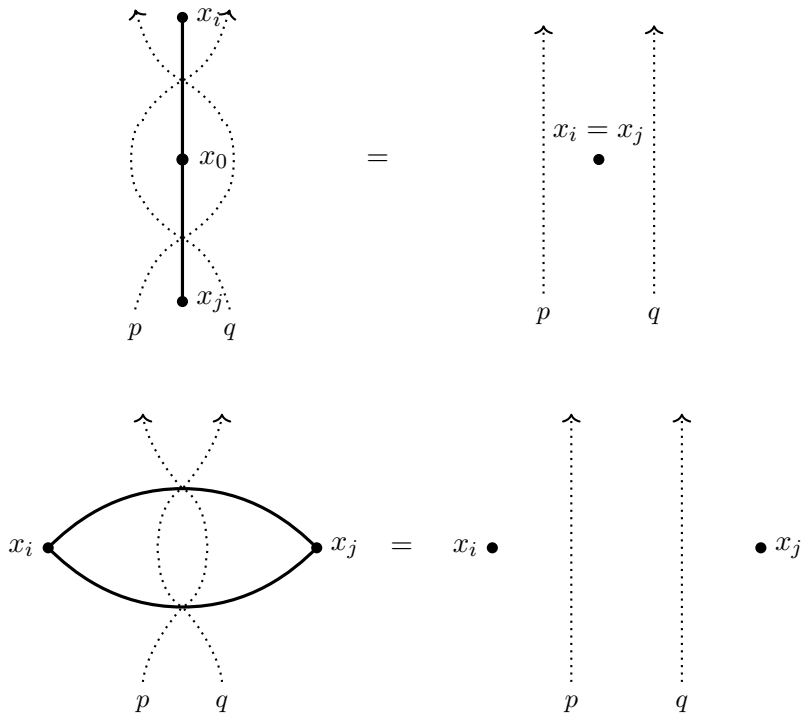
\begin{figure}[htb]
\centering
\begin{tikzpicture}[scale=1.05]

\fill[black!] (-1.7,0) circle (2.0pt)
node[left=1.5pt]{\color{black}$x_i$};
\fill[black!] (1.7,0) circle (2.0pt)
node[right=1.5pt]{\color{black}$x_j$};

\draw[-,very thick] (-1.7,0) .. controls (-1.4,0.3) and (-0.85,0.75)  .. (0,0.75) .. controls (0.85,0.75) and (1.4,0.3) .. (1.7,0);
\draw[-,very thick] (-1.7,0) .. controls (-1.4,-0.3) and (-0.85,-0.75)  .. (0,-0.75) .. controls (0.85,-0.75) and (1.4,-0.3) .. (1.7,0);
\draw[->,thick,dotted] (-0.5,-1.7) .. controls (-0.5,-1.6) and (-0.4,-1.2) .. (0,-0.75) .. controls (0.2,-0.65) and (0.3,-0.35) .. (0.3,-0.15) .. controls (0.32,-0.05) and (0.32,0.05) .. (0.3,0.15) .. controls (0.3,0.35) and (0.2,0.65) .. (0,0.75) .. controls (-0.4,1.2) and (-0.5,1.6) .. (-0.5,1.7);
\draw[white!] (-0.5,-1.7) circle (0.01pt)
node[below=1.5pt]{\color{black}\small $p$};
\draw[->,thick,dotted] (0.5,-1.7) .. controls (0.5,-1.6) and (0.4,-1.2) .. (0,-0.75) .. controls (-0.2,-0.65) and (-0.3,-0.35) .. (-0.3,-0.15) .. controls (-0.32,-0.05) and (-0.32,0.05) .. (-0.3,0.15) .. controls (-0.3,0.35) and (-0.2,0.65) .. (0,0.75) .. controls (0.4,1.2) and (0.5,1.6) ..  (0.5,1.7);
\draw[white!] (0.5,-1.7) circle (0.01pt)
node[below=1.5pt]{\color{black}\small $q$};

\draw[white!] (2.5,0) circle (0.01pt)
node[right=0.1pt]{\color{black}=};

\begin{scope}[xshift=160pt]

\fill[black!] (-1.7,0) circle (2.0pt)
node[left=1.5pt]{\color{black}$x_i$};
\fill[black!] (1.7,0) circle (2.0pt)
node[right=1.5pt]{\color{black}$x_j$};

\draw[->,thick,dotted] (-0.7,-1.7) -- (-0.7,1.7);
\draw[white!] (-0.7,-1.7) circle (0.01pt)
node[below=1.5pt]{\color{black}\small $p$};
\draw[->,thick,dotted] (0.7,-1.7) -- (0.7,1.7);
\draw[white!] (0.7,-1.7) circle (0.01pt)
node[below=1.5pt]{\color{black}\small $q$};

\end{scope}

\begin{scope}[yshift=140]
\draw[-,very thick] (0,-1.8) -- (0,1.8);

\fill[black!] (0,1.8) circle (2.0pt)
node[right=1.5pt]{\color{black}$x_i$};
\fill[black!] (0,-1.8) circle (2.0pt)
node[right=1.5pt]{\color{black}$x_j$};

\filldraw[fill=black!,draw=black!] (0,0) circle (2.0pt)
node[right=2.5pt]{\color{black}$x_0$};

\draw[->,thick,dotted] (-0.6,-1.9) .. controls (-0.5,-1.6) and (-0.4,-1.2) .. (0,-1) .. controls (0.2,-0.9) and (0.5,-0.6) .. (0.5,-0.5) .. controls (0.7,-0.4) and (0.7,0.4) .. (0.5,0.5) .. controls (0.5,0.6) and (0.2,0.9) .. (0,1) .. controls (-0.4,1.2) and (-0.5,1.6) .. (-0.6,1.9);
\draw[white!] (-0.6,-1.9) circle (0.01pt)
node[below=1.5pt]{\color{black}\small $p$};
\draw[->,thick,dotted] (0.6,-1.9) .. controls (0.5,-1.6) and (0.4,-1.2) .. (0,-1) .. controls (-0.2,-0.9) and (-0.5,-0.6) .. (-0.5,-0.5) .. controls (-0.7,-0.4) and (-0.7,0.4) .. (-0.5,0.5) .. controls (-0.5,0.6) and (-0.2,0.9) .. (0,1) .. controls (0.4,1.2) and (0.5,1.6) ..  (0.6,1.9);
\draw[white!] (0.6,-1.9) circle (0.01pt)
node[below=1.5pt]{\color{black}\small $q$};


\draw[white!] (2.2,0) circle (0.01pt)
node[right=0.1pt]{\color{black}=};

\begin{scope}[xshift=150pt]

\fill[black!] (0,0) circle (2.0pt)
node[above=1.5pt]{\color{black} $x_i=x_j$};

\draw[->,thick,dotted] (-0.7,-1.7) -- (-0.7,1.7);
\draw[white!] (-0.7,-1.7) circle (0.01pt)
node[below=1.5pt]{\color{black}\small $p$};
\draw[->,thick,dotted] (0.7,-1.7) -- (0.7,1.7);
\draw[white!] (0.7,-1.7) circle (0.01pt)
node[below=1.5pt]{\color{black}\small $q$};

\end{scope}
\end{scope}
\end{tikzpicture}

\caption{The first inversion relation (top), and second inversion relation (bottom), in (\ref{invrels}).}
\label{invfig}
\end{figure}

Along with the star-triangle relation \eqref{STR}, the Boltzmann weights satisfy the following inversion relations
\beq
\label{invrels}
\begin{array}{rcl}
\ds W_{pq}(x_i,x_j)\,W_{qp}(x_i,x_j)&\!\!\!\!=\!\!\!\!&1\,, \\[0.3cm]
\ds\sum_{x_0}\,\oW_{pq}(x_i,x_0)\, S(x_0)\,\oW_{qp}(x_0,x_j)&\!\!\!\!=\!\!\!\!& f_{pq}f_{qp}\,S(x_i)^{-1}\delta_{x_i,x_j}\,,
\end{array}
\eeq
for all values of the spins $x_i,x_j$.  The inversion relations \eqref{invrels} have the graphical interpretation given in Figure \ref{invfig}.  The star-triangle relation and inversion relations in Figures \ref{fig3}, and \ref{invfig} respectively, are also known to correspond to Reidemeister moves in knot theory \cite{AuYangPerk2016}.

For the purposes of this paper, the second inversion relation in \eqref{invrels} is actually only required for the case $x_i=x_j$ {\it i.e.}, it is only required that the Boltzmann weights satisfy
\beq
\label{invrel2}
\begin{array}{rcl}
\ds\sum_{x_0}\,\oW_{pq}(x_i,x_0)\, S(x_0)\,\oW_{qp}(x_0,x_i)&\!\!\!\!=\!\!\!\!& f_{pq}f_{qp}\,S(x_i)^{-1}\delta_{x_i,x_i}\,,
\end{array}
\eeq
rather than the full relation given in \eqref{invrels}.  For discrete spin models the delta function on the right hand side is a constant $\delta_{x_i,x_i}=1$, but for continuous spin models, this should formally be replaced with a Dirac delta function $\delta(0)$.  Using \eqref{invrel2} thus adds a divergent quantity to the partition function, which may be problematic, therefore the case where \eqref{invrel2} is not required is also discussed towards the end of the next section.

As was mentioned in the introduction, a non-trivial consequence of the star-triangle relation \eqref{STR} and inversion relations \eqref{invrels}, is the property of Z-invariance \cite{Bax1,Baxter:1986prs}, whereupon the partition function \eqref{zdef} remains invariant (up to simple factors) under continuous deformations of the rapidity graph $\L$, as long as the lines remain fixed at their boundary, and no directed closed paths are formed.  The latter restriction arises, since the star-triangle and inversion relations are not satisfied, in cases where the rapidity lines form such directed closed paths.  On the other hand, it will be seen that the introduction of certain rapidity lines, which result in directed closed paths in the rapidity graph, is essential to the formulation of Z-invariance of this paper.  These types of rapidity lines and corresponding deformations are shown in Figures \ref{newd}, and \ref{1cube}, and may be derived from the star-triangle relation \eqref{STR}, and inversion relations \eqref{invrels}.

\section{Edge interaction model with vertices in $\mathbb{Z}^3$}\label{sec:z3model}

In this section it will be shown how to generalise the model on the square lattice, to an edge interaction model on a planar graph $\G$, which has edges connecting a subset of next nearest neighbour vertices of $\mathbb{Z}^3$.  This is done by reformulating the edge interaction model, as a type of face interaction model, and describing Z-invariance in terms of ``cubic'' deformations of the face model, given in Appendix \ref{app:str}.  As is usually the case with Z-invariance, the partition function of the edge interaction model on the graph $\G$, will be shown to be equivalent to the partition function of the model on the square lattice $L$, up to some simple factors that come from the star-triangle relation \eqref{STR}, and inversion relations \eqref{invrels}.

\subsection{Two-dimensional surface associated to the square lattice model}\label{sec:sig}

\begin{figure}[htb]
\centering
\begin{tikzpicture}[scale=1.76]

\draw[white!] (0.5,0) circle (0.01pt);

\foreach \x in {0.5,1.5,...,5.5}{
\draw[->,black,dotted] (\x-0.15,-0.3) -- (\x+1.15,2.3);
\fill[white!] (\x-0.15,-0.3) circle (0.1pt)
node[below=0.05pt]{\color{black}\small $q$};}

\foreach \y in {0.25,0.75,1.25,1.75}{
\draw[->,black,dotted] (0.5*\y-0.4,\y) -- (0.5*\y+6+0.4,\y);
\fill[white!] (0.5*\y-0.4,\y) circle (0.1pt)
node[left=0.05pt]{\color{black}\small $p$};}

\draw[-,thin,gray] (0,0)--(6,0)--(7,2)--(1,2)--(0,0);
\foreach \x in {1,2,...,5}
\draw[-,thin,gray] (\x,0)--(\x+1,2);

\foreach \y in {0.5,1,1.5}
\draw[-,thin,gray] (0.5*\y,\y)--(6+0.5*\y,\y);

\foreach \x in {0,2,4}{
\filldraw[draw=black,fill=black] (\x+1,2) circle (1.5pt);
\filldraw[draw=black,fill=white] (\x+2,2) circle (1.5pt);
\foreach \y in {0,1}{
\filldraw[draw=black,fill=black] (\x+\y*0.5,\y) circle (1.5pt);
\filldraw[draw=black,fill=white] (\x+\y*0.5+1,\y) circle (1.5pt);
\filldraw[draw=black,fill=white] (\x+0.25+\y*0.5,\y+0.5) circle (1.5pt);
\filldraw[draw=black,fill=black] (\x+0.25+\y*0.5+1,\y+0.5) circle (1.5pt);
}}
\foreach \y in {0,1}{
\filldraw[draw=black,fill=black] (6+\y*0.5,\y) circle (1.5pt);
\filldraw[draw=black,fill=white] (6+\y*0.5+0.25,\y+0.5) circle (1.5pt);
}
\filldraw[draw=black,fill=black] (7,2) circle (1.5pt);

\draw[-, thick] (0.5,1)--(3,2);\draw[-, thick] (0,0)--(5,2);\draw[-, thick] (2,0)--(7,2);\draw[-, thick] (4,0)--(6.5,1);\draw[-, thick] (2,0)--(0.5,1);\draw[-, thick] (4,0)--(1,2);\draw[-, thick] (6,0)--(3,2);\draw[-, thick] (6.5,1)--(5,2);

\fill[white] (7.86,0) circle (0.01pt)
node[below=1.5pt]{\color{black} $+\ihat$};
\fill[white] (7.66,0.33) circle (0.01pt)
node[right=1.5pt]{\color{black} $+\jhat$};
\fill[white] (7.5,0.75*0.5) circle (0.01pt)
node[left=1.5pt]{\color{black} $+\khat$};
\draw[->, thick] (7.5,0)--(8.25,0);\draw[->, thick] (7.5,0)--(7.83,0.67);\draw[->, thick] (7.5,0)--(7.5,0.75);

\end{tikzpicture}
\caption{A surface $\sigma_0$, made up of faces and edges edges that connect the black and white nearest neighbour vertices of $\mathbb{Z}^2$.  The square lattice model of Figure \ref{squarelattice}, is defined on next nearest neighbour black vertices, connected by edges on the diagonals of the faces of $\sigma_0$.}
\label{Glattice}
\end{figure}
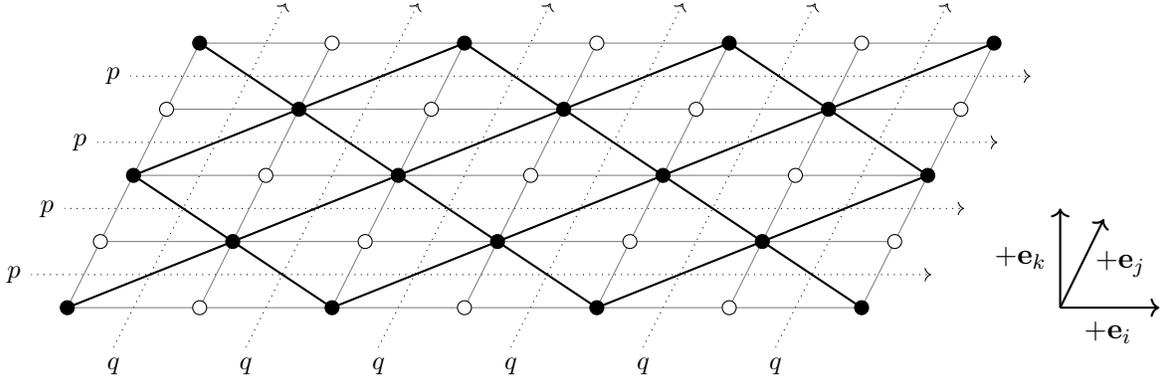

To begin with, the lattice model of the previous section will be redefined, and associated to a two-dimensional surface, denoted by $\sigma_0$, as shown in Figure \ref{Glattice}.  The main reason for associating the square lattice model with such a surface, is that the deformations of edge interaction models to be considered in this section, are more easily described in terms of ``cubic'' deformations of the faces of this surface, rather than in terms of deformations of the actual edges of the model.

The surface $\sigma_0$ consists of elementary squares lying in a plane, with edges connecting nearest neighbour vertices of $\mathbb{Z}^2$.  Let $V(\sigma_0)$, $E(\sigma_0)$, and $F(\sigma_0)$, denote respectively the sets of vertices, edges, and elementary squares (faces), of $\sigma_0$.  Vertices $i\in V(\sigma_0)$, are depicted as both the black and white vertices in Figure \ref{Glattice}.  Black vertices are connected by edges on diagonals of elementary squares, these diagonal edges do not belong to $E(\sigma_0)$, but together form a separate square lattice, $L$, equivalent to that in Figure \ref{squarelattice}.

The directed rapidity graph, $\L$, is shown in Figure \ref{Glattice}, as the continuous directed dotted lines that cross edges $(ij)\in E(\sigma_0)$ perpendicularly, and cross edges $(ij)\in E(L)$ at 45 degree angles.  Two unit vectors $\ihat$, and $\jhat$, are associated to two orthogonal lattice directions of $\sigma_0$, indicated in Figure \ref{Glattice}.  Rapidity lines that are oriented in the $+\ihat$ direction, are labelled by the variable $p$, and rapidity lines that are oriented in the $+\jhat$ direction, are labelled by the variable $q$.

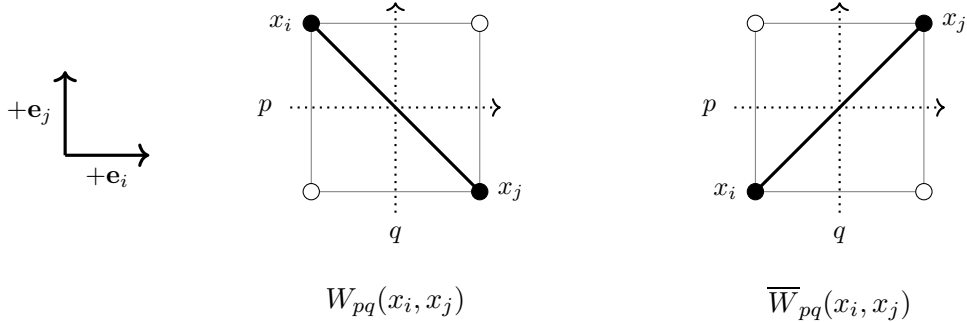
\begin{figure}[htb]
\centering
\begin{tikzpicture}[scale=2.8]
\begin{scope}[xshift=10pt,yshift=55pt,scale=0.8]
\draw[->,very thick] (-2.4,-0.2)--(-1.9,-0.2);
\fill[white!] (-2.15,-0.2) circle (0.01pt)
node[below=1pt]{\color{black}\small $+\ihat$};
\draw[->,very thick] (-2.4,-0.2)--(-2.4,0.3);
\fill[white!] (-2.4,0.05) circle (0.01pt)
node[left=1pt]{\color{black}\small $+\jhat$};
\end{scope}

\draw[-,gray] (-0.4,1.6)--(0.4,1.6)--(0.4,2.4)--(-0.4,2.4)--(-0.4,1.6);
\draw[-,very thick] (0.4,1.6)--(-0.4,2.4);
\draw[->,thick,dotted,black] (0,1.5)--(0,2.5);
\draw[->,thick,dotted,black] (-0.5,2)--(0.5,2);
\fill[white!] (0,1.5) circle (0.01pt)
node[below=1.5pt]{\color{black}\small $q$};
\fill[white!] (-0.5,2) circle (0.01pt)
node[left=3.1pt]{\color{black}\small $p$};
\filldraw[fill=black,draw=black] (0.4,1.6) circle (1.1pt)
node[right=3pt]{\color{black}\small $x_j$};
\filldraw[fill=black,draw=black] (-0.4,2.4) circle (1.1pt)
node[left=3pt]{\color{black}\small $x_i$};
\filldraw[fill=white,draw=black] (-0.4,1.6) circle (1.1pt);
\filldraw[fill=white,draw=black] (0.4,2.4) circle (1.1pt);

\fill (0,1.2) circle(0.01pt)
node[below=0.05pt]{\color{black} $ W_{pq}(x_i,x_j)$};

\begin{scope}[xshift=60pt]
\draw[white!] (1.3,1) circle (0.1pt);
\draw[-,gray] (-0.4,1.6)--(0.4,1.6)--(0.4,2.4)--(-0.4,2.4)--(-0.4,1.6);
\draw[-,very thick] (-0.4,1.6)--(0.4,2.4);
\draw[->,thick,dotted,black] (0,1.5)--(0,2.5);
\draw[->,thick,dotted,black] (-0.5,2)--(0.5,2);
\fill[white!] (0,1.5) circle (0.01pt)
node[below=1.5pt]{\color{black}\small $q$};
\fill[white!] (-0.5,2) circle (0.01pt)
node[left=3.1pt]{\color{black}\small $p$};
\filldraw[fill=black,draw=black] (-0.4,1.6) circle (1.1pt)
node[left=3pt]{\color{black}\small $x_i$};
\filldraw[fill=black,draw=black] (0.4,2.4) circle (1.1pt)
node[right=3pt]{\color{black}\small $x_j$};
\filldraw[fill=white,draw=black] (0.4,1.6) circle (1.1pt);
\filldraw[fill=white,draw=black] (-0.4,2.4) circle (1.1pt);

\fill (0,1.2) circle(0.01pt)
node[below=0.05pt]{\color{black} $\oW_{pq}(x_i,x_j)$};

\end{scope}

\end{tikzpicture}
\caption{Elementary squares, with a single edge $(ij)\in E^{(1)}(L)$, connecting the top left and bottom right vertices (left), and a single edge $(ij)\in E^{(2)}(L)$, connecting the bottom left and top right vertices (right).  These squares are the building blocks of the surface $\sigma_0$, in Figure \ref{Glattice}.}
\label{strsquare}
\end{figure}

Spin variables $x_i$, are assigned only to black vertices $i\in V(L)$, while white vertices don't play any role for the definition of the lattice model.  Spins on the boundary of $\sigma_0$ are prescribed some fixed values.  The crossing of the directed rapidity lines on $\sigma_0$, distinguishes the two types of elementary squares shown in Figure \ref{strsquare}.  These elementary squares have respectively the edges $(ij)\in E^{(1)}(L)$, and $(ij)\in E^{(2)}(L)$, on their diagonals, with associated Boltzmann weights $W_{pq}(x_i,x_j)$, $\oW_{pq}(x_i,x_j)$, according to Figure \ref{2boltzmannweights}.  The partition function of the edge interaction model on $L$, is given by \eqref{zdef}, and the Boltzmann weights are required to satisfy the star-triangle relation \eqref{STR}, and inversion relations \eqref{invrels}.  This defines the integrable square lattice model of the previous section, on the surface $\sigma_0$.

\subsection{General surface for an edge interaction model with vertices in $\mathbb{Z}^3$}\label{sec:sigp}

In order to consider more general surfaces in three-dimensional space, that are not restricted to lie in a plane, it is useful to introduce some additional notations.

Let $\n=(n_i,n_j,n_k)$, denote the coordinate of a vertex $i(\n)$, in the orthonormal basis $\{\ihat,\jhat,\khat\}$.  Oriented elementary squares, will be denoted by $\sigma_{ij}(\n)$, with the subscripts referring to the lattice basis vectors.  Specifically, in terms of vertex coordinates, an oriented elementary square is specified by the quadruple $\sigma_{ij}(\n)=(\n, \n+\ihat, \n+\ihat+\jhat, \n+\jhat)$.  For brevity, the coordinate $\n$ is not given if this doesn't cause ambiguity, and the elementary square will then be written simply as $\sigma_{ij}$.  The four other types of oriented elementary squares used here, are denoted by $\sigma_{ik},\sigma_{jk},\sigma_{ki},\sigma_{kj}$, and are always distinguished from each other by the unit vectors indicated in the subscripts.  A sixth possible type of oriented elementary square, $\sigma_{ji}$, will not be used.

\begin{figure}[htb]
\centering
\begin{tikzpicture}[scale=2.25]
\draw[->,very thick] (-0.8,0.8)--(-0.8,1.2);
\fill[white!] (-0.8,0.75) circle (0.01pt)
node[below=1.5pt]{\color{black}\small $+\khat$};
\draw[->,very thick] (-0.1,0.8)--(0.3,0.8);
\fill[white!] (-0.1,0.8) circle (0.01pt)
node[left=1.0pt]{\color{black}\small $+\ihat$};
\draw[->,very thick] (1.7,0.8)--(2.1,0.8);
\fill[white!] (1.7,0.8) circle (0.01pt)
node[left=1.0pt]{\color{black}\small $+\jhat$};
\draw[->,very thick] (3.45,0.8)--(3.85,0.8);
\fill[white!] (3.45,0.8) circle (0.01pt)
node[left=1.0pt]{\color{black}\small $-\ihat$};
\draw[->,very thick] (5.2,0.8)--(5.6,0.8);
\fill[white!] (5.2,0.8) circle (0.01pt)
node[left=1.0pt]{\color{black}\small $-\jhat$};

\draw[-,gray] (-0.4,1.6)--(0.4,1.6)--(0.4,2.4)--(-0.4,2.4)--(-0.4,1.6);
\draw[-,very thick] (0.4,1.6)--(-0.4,2.4);
\draw[->,thick,dotted,black] (0,1.5)--(0,2.5);
\draw[->,thick,dotted,black!30!black] (-0.5,2)--(0.5,2);
\fill[white!] (0,1.5) circle (0.01pt)
node[below=1.5pt]{\color{black}\small $q$};
\fill[white!] (-0.5,2) circle (0.01pt)
node[left=3.1pt]{\color{black}\small $r$};
\filldraw[fill=black,draw=black] (0.4,1.6) circle (1.1pt)
node[right=3pt]{\color{black}\small $x_j$};
\filldraw[fill=black,draw=black] (-0.4,2.4) circle (1.1pt)
node[left=3pt]{\color{black}\small $x_i$};
\filldraw[fill=white,draw=black] (-0.4,1.6) circle (1.1pt);
\filldraw[fill=white,draw=black] (0.4,2.4) circle (1.1pt);

\fill (0,1.3) circle(0.01pt)
node[below=0.05pt]{\color{black} $ W_{rq}(x_i,x_j)$};

\draw[-,gray] (-0.4,-0.4)--(0.4,-0.4)--(0.4,0.4)--(-0.4,0.4)--(-0.4,-0.4);
\draw[-,very thick] (-0.4,-0.4)--(0.4,0.4);
\draw[->,thick,dotted,black] (0,-0.5)--(0,0.5);
\draw[->,thick,dotted,black!30!black] (-0.5,0)--(0.5,0);
\fill[white!] (0,-0.5) circle (0.01pt)
node[below=1.5pt]{\color{black}\small $q$};
\fill[white!] (-0.5,0) circle (0.01pt)
node[left=3.1pt]{\color{black}\small $r$};
\filldraw[fill=black,draw=black] (-0.4,-0.4) circle (1.1pt)
node[left=3pt]{\color{black}\small $x_i$};
\filldraw[fill=black,draw=black] (0.4,0.4) circle (1.1pt)
node[right=3pt]{\color{black}\small $x_j$};
\filldraw[fill=white,draw=black] (-0.4,0.4) circle (1.1pt);
\filldraw[fill=white,draw=black] (0.4,-0.4) circle (1.1pt);

\fill (0,-0.7) circle(0.01pt)
node[below=0.05pt]{\color{black} $\oW_{rq}(x_i,x_j)$};

\begin{scope}[xshift=50pt]
\draw[-,gray] (-0.4,1.6)--(0.4,1.6)--(0.4,2.4)--(-0.4,2.4)--(-0.4,1.6);
\draw[-,very thick] (-0.4,1.6)--(0.4,2.4);
\draw[->,thick,dotted,black] (0,2.5)--(0,1.5);
\draw[->,thick,dotted,black!30!black] (-0.5,2)--(0.5,2);
\fill[white!] (0,2.5) circle (0.01pt)
node[above=1.5pt]{\color{black}\small $p$};
\fill[white!] (-0.5,2) circle (0.01pt)
node[left=3.1pt]{\color{black}\small $r$};
\filldraw[fill=black,draw=black] (-0.4,1.6) circle (1.1pt)
node[left=3pt]{\color{black}\small $x_j$};
\filldraw[fill=black,draw=black] (0.4,2.4) circle (1.1pt)
node[right=3pt]{\color{black}\small $x_i$};
\filldraw[fill=white,draw=black] (0.4,1.6) circle (1.1pt);
\filldraw[fill=white,draw=black] (-0.4,2.4) circle (1.1pt);

\fill (0,1.3) circle(0.01pt)
node[below=0.05pt]{\color{black} $ W_{pr}(x_i,x_j)$};

\draw[-,gray] (-0.4,-0.4)--(0.4,-0.4)--(0.4,0.4)--(-0.4,0.4)--(-0.4,-0.4);
\draw[-,very thick] (0.4,-0.4)--(-0.4,0.4);
\draw[->,thick,dotted,black] (0,0.5)--(0,-0.5);
\draw[->,thick,dotted,black!30!black] (-0.5,0)--(0.5,0);
\fill[white!] (0,0.5) circle (0.01pt)
node[above=1.5pt]{\color{black}\small $p$};
\fill[white!] (-0.5,0) circle (0.01pt)
node[left=3.1pt]{\color{black}\small $r$};
\filldraw[fill=black,draw=black] (0.4,-0.4) circle (1.1pt)
node[right=3pt]{\color{black}\small $x_j$};
\filldraw[fill=black,draw=black] (-0.4,0.4) circle (1.1pt)
node[left=3pt]{\color{black}\small $x_i$};
\filldraw[fill=white,draw=black] (0.4,0.4) circle (1.1pt);
\filldraw[fill=white,draw=black] (-0.4,-0.4) circle (1.1pt);

\fill (0,-0.7) circle(0.01pt)
node[below=0.05pt]{\color{black} $\oW_{pr}(x_i,x_j)$};
\end{scope}

\begin{scope}[xshift=100pt]
\draw[-,gray] (-0.4,1.6)--(0.4,1.6)--(0.4,2.4)--(-0.4,2.4)--(-0.4,1.6);
\draw[-,very thick] (-0.4,2.4)--(0.4,1.6);
\draw[->,thick,dotted,black] (0,2.5)--(0,1.5);
\draw[->,thick,dotted,black!30!black] (-0.5,2)--(0.5,2);
\fill[white!] (0,2.5) circle (0.01pt)
node[above=1.5pt]{\color{black}\small $q$};
\fill[white!] (-0.5,2) circle (0.01pt)
node[left=3.1pt]{\color{black}\small $r$};
\filldraw[fill=white,draw=black] (-0.4,1.6) circle (1.1pt);
\filldraw[fill=white,draw=black] (0.4,2.4) circle (1.1pt);
\filldraw[fill=black,draw=black] (0.4,1.6) circle (1.1pt)
node[right=3pt]{\color{black}\small $x_j$};
\filldraw[fill=black,draw=black] (-0.4,2.4) circle (1.1pt)
node[left=3pt]{\color{black}\small $x_i$};

\fill (0,1.3) circle(0.01pt)
node[below=0.05pt]{\color{black} $\oW_{qr}(x_i,x_j)$};

\draw[-,gray] (-0.4,-0.4)--(0.4,-0.4)--(0.4,0.4)--(-0.4,0.4)--(-0.4,-0.4);
\draw[-,very thick] (-0.4,-0.4)--(0.4,0.4);
\draw[->,thick,dotted,black] (0,0.5)--(0,-0.5);
\draw[->,thick,dotted,black!30!black] (-0.5,0)--(0.5,0);
\fill[white!] (0,0.5) circle (0.01pt)
node[above=1.5pt]{\color{black}\small $q$};
\fill[white!] (-0.5,0) circle (0.01pt)
node[left=3.1pt]{\color{black}\small $r$};
\filldraw[fill=white,draw=black] (0.4,-0.4) circle (1.1pt);
\filldraw[fill=white,draw=black] (-0.4,0.4) circle (1.1pt);
\filldraw[fill=black,draw=black] (0.4,0.4) circle (1.1pt)
node[right=3pt]{\color{black}\small $x_i$};
\filldraw[fill=black,draw=black] (-0.4,-0.4) circle (1.1pt)
node[left=3pt]{\color{black}\small $x_j$};

\fill (0,-0.7) circle(0.01pt)
node[below=0.05pt]{\color{black} $ W_{qr}(x_i,x_j)$};
\end{scope}

\begin{scope}[xshift=150pt]
\draw[-,gray] (-0.4,1.6)--(0.4,1.6)--(0.4,2.4)--(-0.4,2.4)--(-0.4,1.6);
\draw[-,very thick] (-0.4,1.6)--(0.4,2.4);
\draw[->,thick,dotted,black] (0,1.5)--(0,2.5);
\draw[->,thick,dotted,black!30!black] (-0.5,2)--(0.5,2);
\fill[white!] (0,1.5) circle (0.01pt)
node[below=1.5pt]{\color{black}\small $p$};
\fill[white!] (-0.5,2) circle (0.01pt)
node[left=3.1pt]{\color{black}\small $r$};
\filldraw[fill=white,draw=black] (0.4,1.6) circle (1.1pt);
\filldraw[fill=white,draw=black] (-0.4,2.4) circle (1.1pt);
\filldraw[fill=black,draw=black] (-0.4,1.6) circle (1.1pt)
node[left=3pt]{\color{black}\small $x_i$};
\filldraw[fill=black,draw=black] (0.4,2.4) circle (1.1pt)
node[right=3pt]{\color{black}\small $x_j$};

\fill (0,1.3) circle(0.01pt)
node[below=0.05pt]{\color{black} $\oW_{rp}(x_i,x_j)$};

\draw[-,gray] (-0.4,-0.4)--(0.4,-0.4)--(0.4,0.4)--(-0.4,0.4)--(-0.4,-0.4);
\draw[-,very thick] (0.4,-0.4)--(-0.4,0.4);
\draw[->,thick,dotted,black] (0,-0.5)--(0,0.5);
\draw[->,thick,dotted,black!30!black] (-0.5,0)--(0.5,0);
\fill[white!] (0,-0.5) circle (0.01pt)
node[below=1.5pt]{\color{black}\small $p$};
\fill[white!] (-0.5,0) circle (0.01pt)
node[left=3.1pt]{\color{black}\small $r$};
\filldraw[fill=white,draw=black] (-0.4,-0.4) circle (1.1pt);
\filldraw[fill=white,draw=black] (0.4,0.4) circle (1.1pt);
\filldraw[fill=black,draw=black] (-0.4,0.4) circle (1.1pt)
node[left=3pt]{\color{black}\small $x_i$};
\filldraw[fill=black,draw=black] (0.4,-0.4) circle (1.1pt)
node[right=3pt]{\color{black}\small $x_j$};

\fill (0,-0.7) circle(0.01pt)
node[below=0.05pt]{\color{black} $ W_{rp}(x_i,x_j)$};
\end{scope}

\end{tikzpicture}
\caption{From left to right, rapidity lines and Boltzmann weights, that are associated to elementary squares $\sigma_{ik}$, $\sigma_{jk}$, $\sigma_{ki}$, $\sigma_{kj}$.  Each elementary square is pictured with vertices ordered counter-clockwise, and the Boltzmann weights are read according to Figure \ref{2boltzmannweights}.  The rapidity lines are oriented to be consistently directed on neighbouring elementary squares of the surface $\sigma$.}
\label{4BW}
\end{figure}
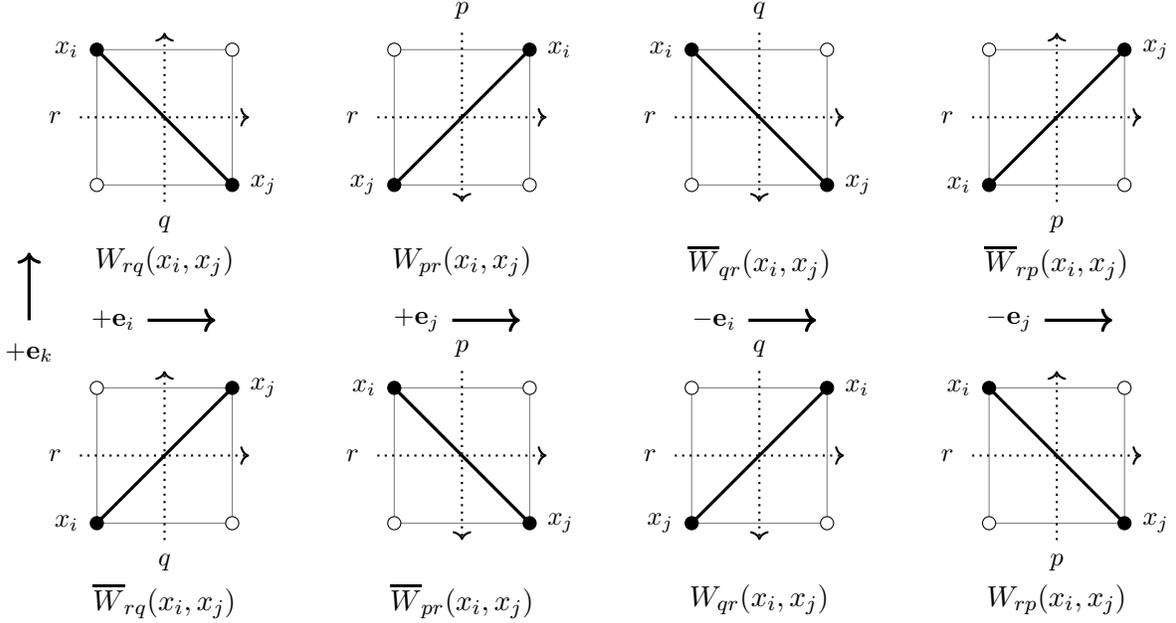

Directed rapidity lines labelled $p$, $q$, and $r$, are assigned to the different elementary squares, according to Figures \ref{strsquare}, and \ref{4BW}.  A new type of rapidity line labelled $r$ has been introduced, with its orientation chosen such that it is continuously directed along neighbouring elementary squares of the type $\sigma_{ik},\sigma_{jk},\sigma_{ki},\sigma_{kj}$, and it forms a directed closed path with itself.  Consequently, the rapidity lines $r$ always lie in a plane orthogonal to $\khat$.  The rapidity lines $r$, can be either positively or negatively oriented, depending on the arrangement of different elementary squares $\sigma_{ik},\sigma_{jk},\sigma_{ki},\sigma_{kj}$, that they are associated with.  For a fixed coordinate $n_k$, the union of different elementary squares $\sigma_{ik}(\n),\sigma_{jk}(\n),\sigma_{ki}(\n),\sigma_{kj}(\n)$ on a surface, will not generally be connected, and there may be several instances of rapidity lines labelled $r$, for the single coordinate $n_k$. The rapidity lines $r$, only intersect the $p$ and $q$ rapidity lines, and different instances of rapidity lines $r$, never intersect each other.

The elementary squares $\sigma_{ij},\sigma_{ik},\sigma_{ki},\sigma_{jk},\sigma_{kj}$, are used as building blocks of a surface $\sigma$, as follows.  First let $F(\sigma_0)=\{\sigma_{ij}(n_{i_1},n_{j_1},0)),\sigma_{ij}(n_{i_2},n_{j_2},0)),\ldots,\}$, be the elementary squares of some flat surface $\sigma_0$, as defined in the previous subsection.  Then for an admissible set of integers $n_{k_1},n_{k_2},\ldots$, $\sigma$ is the unique surface consisting of elementary squares
\beq
F(\sigma)=\{\sigma_{ij}(n_{i_1},n_{j_1},n_{k_1})),\sigma_{ij}(n_{i_2},n_{j_2},n_{k_2})),\ldots,\}\cup F_k\,,
\eeq
where $F_k$ contains elementary squares of the type $\sigma_{ik},\sigma_{jk},\sigma_{ki},\sigma_{kj}$, which are chosen such that $\sigma$ has the same boundary as $\sigma_0$, $\sigma$ is simply connected, and oriented, and the following condition is satisfied:
\beq
\label{condition}
\begin{array}{l}
\textit{\parbox{14.3cm}{For any $\n$, $\sigma$ cannot contain pairs of elementary squares $\sigma_{ki}(\n),\sigma_{kj}(\n+\ihat)$, that are associated to a positively oriented rapidity line $r$. Similarly, for any $\n$, $\sigma$ cannot contain pairs of elementary squares $\sigma_{ki}(\n+\jhat),\sigma_{kj}(\n)$, that are associated to a negatively oriented rapidity line $r$.}}
\end{array}
\eeq
The integers $n_{k_1},n_{k_2},\ldots$, should be chosen so that the above properties can be satisfied.  The condition \eqref{condition} is required, because the star-triangle relation is not satisfied where three rapidity lines form a directed closed path around the central vertex in Figure \ref{fig3}, and \eqref{condition} ensures that problematic ``corners'' with this associated rapidity configuration are avoided.

The vertices and edges of $\sigma$ form a bipartite graph, in which case $V(\sigma)$ may be split into disjoint ``black'' and ``white'' subsets, $V^{(1)}(\sigma)$, and $V^{(2)}(\sigma)$ respectively, such that an edge $(ij)\in E(\sigma)$, always connects a black vertex $i\in V^{(1)}(\sigma)$, with a white vertex $j\in V^{(2)}(\sigma)$.  Let $\G$ be the graph formed by connecting black vertices $i,j\in V^{(1)}(\sigma)$, with edges $(ij)\in E(\G)$ on a single diagonal of each elementary square of $\sigma$.  The elementary squares of $\sigma$, then coincide exactly with the different elementary squares depicted in Figures \ref{strsquare}, and \ref{4BW}. 

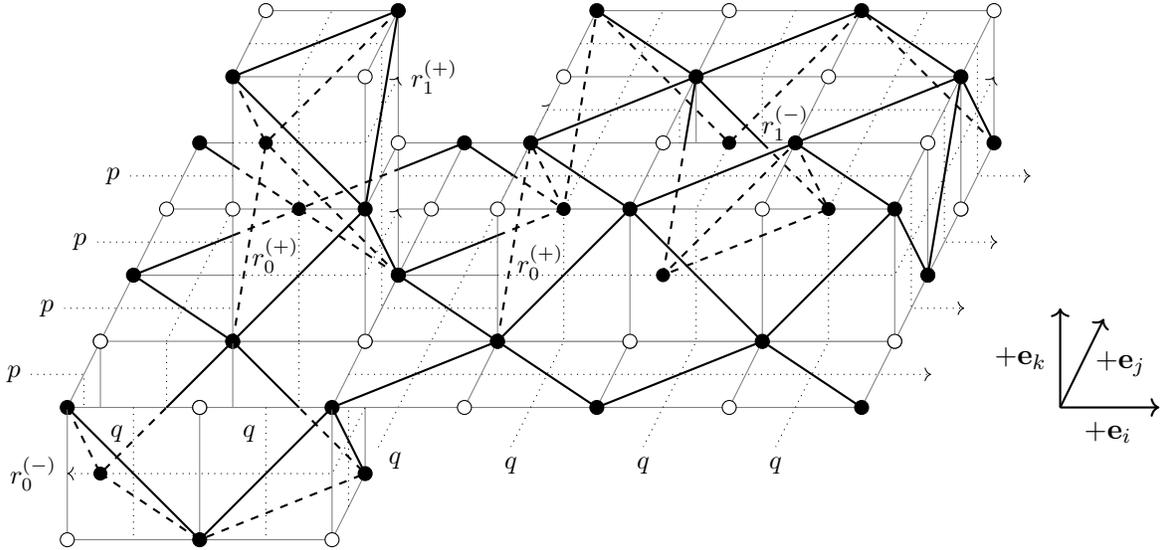
\begin{figure}[htb]
\centering
\begin{tikzpicture}[scale=1.76]

\draw[white!] (0.5,0) circle (0.01pt);


\draw[->,black,dotted] (2.5-0.15,-0.3) -- (2.5+1.15,2.3);
\fill[white!] (2.5-0.15,-0.4) circle (0.1pt)
node[right=0.05pt]{\color{black}\small $q$};
\draw[->,black,dotted] (0.75,0.5)--(0.5+1.15,2.3);

\foreach \y in {0.75,1.25,1.75}{
\draw[-,black,dotted] (0.5*\y-0.4,\y) -- (0.5*\y+3,\y);
\fill[white!] (0.5*\y-0.4,\y) circle (0.1pt)
node[left=0.05pt]{\color{black}\small $p$};}

\draw[-,thin,gray] (0,0)--(6,0)--(6.25,0.5)--(3.25,0.5);\draw[-,thin,gray] (4,2)--(1,2)--(0,0);
\draw[-,thin,gray] (2,0)--(3,2);

\foreach \y in {0.5,1,1.5}
\draw[-,thin,gray] (0.5*\y,\y)--(3+0.5*\y,\y);

\draw[-,thin,gray] (3,0)--(3.25,0.5);\draw[-,thin,gray] (4,0)--(4.25,0.5);\draw[-,thin,gray] (5,0)--(5.25,0.5);

\foreach \x in {0,2,4}{
\filldraw[draw=black,fill=black] (\x+1,2) circle (1.5pt);
\filldraw[draw=black,fill=white] (\x+2,2) circle (1.5pt);
\foreach \y in {0,1}{
\filldraw[draw=black,fill=black] (\x+\y*0.5,\y) circle (1.5pt);
\filldraw[draw=black,fill=white] (\x+\y*0.5+1,\y) circle (1.5pt);
\filldraw[draw=black,fill=white] (\x+0.25+\y*0.5,\y+0.5) circle (1.5pt);
\filldraw[draw=black,fill=black] (\x+0.25+\y*0.5+1,\y+0.5) circle (1.5pt);
}}
\foreach \y in {0,1}{
\filldraw[draw=black,fill=black] (6+\y*0.5,\y) circle (1.5pt);
\filldraw[draw=black,fill=white] (6+\y*0.5+0.25,\y+0.5) circle (1.5pt);
}

\draw[-, thick] (0.5,1)--(3,2);\draw[-, thick] (2.5,1)--(3.75,1.5);\draw[-, thick] (2,0)--(3.25,0.5);\draw[-, thick] (4,0)--(5.25,0.5);\draw[-, thick] (1.25,0.5)--(0.5,1);\draw[-, thick] (4,0)--(1,2);\draw[-, thick] (6,0)--(5.25,0.5);\draw[-, thick] (3,2)--(3.75,1.5);\draw[-, thick] (4.5,1)--(4.75,2.5)--(5.75,1.5);

\filldraw[-,thin,fill=white,draw=gray] (3.25,0.5)--(6.25,0.5)--(7,2)--(7,3)--(4,3)--(3.25,1.5)--(3.25,0.5);
\foreach \x in {4.25,5.25,6.25}
\draw[-,thin,gray] (\x,1.5)--(\x+0.75,3);
\foreach \y in {0,0.5,1}
\draw[-,thin,gray] (3.25+\y*0.5,\y+1.5)--(6.25+\y*0.5,\y+1.5)--(6.25+\y*0.5,\y+0.5);
\draw[-,thin,gray] (4.25,0.5)--(4.25,1.5);\draw[-,thin,gray] (5.25,0.5)--(5.25,1.5);
\draw[-,thin,gray] (4.75,2.5)--(4.75,2);
\foreach \x in {3.25,5.25} {
\filldraw[fill=black,draw=black] (\x,0.5) circle (1.5pt);
\filldraw[fill=white,draw=black] (\x+1,0.5) circle (1.5pt);
\foreach \y in {0,1} {
\filldraw[fill=white,draw=black] (\x+\y*0.5,\y+1.5) circle (1.5pt);
\filldraw[fill=black,draw=black] (\x+\y*0.5+1,\y+1.5) circle (1.5pt);
\filldraw[fill=black,draw=black] (\x+\y*0.5+0.25,\y+1.5+0.5) circle (1.5pt);
\filldraw[fill=white,draw=black] (\x+\y*0.5+1+0.25,\y+1.5+0.5) circle (1.5pt); }}
\filldraw[fill=black,draw=black] (6.5,1) circle (1.5pt);
\filldraw[fill=white,draw=black] (6.75,1.5) circle (1.5pt);
\filldraw[fill=black,draw=black] (7,2) circle (1.5pt);
\draw[-,black,dotted] (3.5-0.15,-0.3)--(3.75,0.5)--(3.75,1.5)--(4.5,3);
\fill[white!] (3.5-0.15,-0.3) circle (0.1pt)
node[below=0.05pt]{\color{black}\small $q$};
\draw[-,black,dotted] (4.5-0.15,-0.3)--(4.75,0.5)--(4.75,1.5)--(5,2);\draw[-,black,dotted] (5.25,2)--(5.25,2.5)--(5.5,3);
\fill[white!] (4.5-0.15,-0.3) circle (0.1pt)
node[below=0.05pt]{\color{black}\small $q$};
\draw[-,black,dotted] (5.5-0.15,-0.3)--(5.75,0.5)--(5.75,1.5)--(6.5,3);
\fill[white!] (5.5-0.15,-0.3) circle (0.1pt)
node[below=0.05pt]{\color{black}\small $q$};
\draw[->,black,dotted] (3.25,1)--(6.25,1)--(7,2.5);
\fill[white!] (5.7,2.15) circle (0.01pt)
node[left=1pt]{\color{black}\small $r_1^{(-)}$};
\fill[white!] (3.3,1.1) circle (0.01pt)
node[right=1pt]{\color{black}\small $r_0^{(+)}$};
\draw[->,black,dotted] (3.25+0.5*0.25,1.75)--(6.25+0.5*0.25,1.75)--(6.25+0.5*0.25,0.75)--(6.25+0.4+0.5*0.25,0.75);
\draw[-,black,dotted] (3.25+0.25+0.5*0.25,2.25)--(4.25+0.25+0.5*0.25,2.25)--(4.25+0.25+0.5*0.25,2);\draw[->,black,dotted] (5.25+0.25+0.5*0.25,2.25)--(6.25+0.25+0.5*0.25,2.25)--(6.25+0.25+0.5*0.25,1.25)--(6.25+0.25+0.4+0.5*0.25,1.25);
\draw[->,black,dotted] (3.25+0.5+0.5*0.25,2.75)--(6.25+0.5+0.5*0.25,2.75)--(6.25+0.5+0.5*0.25,1.75)--(6.25+0.5+0.4+0.5*0.25,1.75);
\draw[-,thick] (3.25,0.5)--(4.25,1.5)--(5.25,0.5)--(6.25,1.5)--(6.5,1)--(6.75,2.5)--(7,2);
\draw[-,thick] (4,3)--(4.75,2.5)--(3.5,2)--(4.25,1.5)--(5.5,2)--(6.25,1.5);\draw[-,thick] (5.5,2)--(6.75,2.5)--(6,3)--(4.75,2.5);\draw[-, thick] (4.5+0.25*2/3,2)--(4.75,2.5)--(5.75-0.5,2);

\draw[-,thick,dashed] (3.25,0.5)--(3.5,2)--(3.75,1.5)--(4,3)--(5,2)--(6,3)--(7,2);\draw[-,thick,dashed] (4.5+0.25*2/3,2)--(4.5,1)--(5.75,1.5)--(5.5,2)--(4.5,1);\draw[-,thick,dashed] (5.75-0.5,2)--(5.75,1.5);
\fill[black] (3.75,1.5) circle (1.5pt);\fill[black] (5,2) circle (1.5pt);\fill[black] (5.75,1.5) circle (1.5pt);\fill[black] (4.5,1) circle (1.5pt);

\filldraw[-,thin,fill=white,draw=gray] (1.25,0.5)--(2.25,0.5)--(2.5,1)--(2.5,3)--(1.5,3)--(1.25,2.5)--(1.25,0.5);
\draw[-,thin,gray] (2.25,0.5)--(2.25,2.5);\draw[-,thin,gray](1.25,2.5)--(2.25,2.5)--(2.5,3);\draw[-,thin,gray] (1.25,1.5)--(2.25,1.5)--(2.5,2);
\filldraw[fill=black,draw=black] (1.25,0.5) circle (1.5pt);
\filldraw[fill=white,draw=black] (2.25,0.5) circle (1.5pt);
\filldraw[fill=black,draw=black] (2.5,1) circle (1.5pt);
\filldraw[fill=white,draw=black] (1.25,1.5) circle (1.5pt);
\filldraw[fill=black,draw=black] (2.25,1.5) circle (1.5pt);
\filldraw[fill=white,draw=black] (2.5,2) circle (1.5pt);
\filldraw[fill=black,draw=black] (1.25,2.5) circle (1.5pt);
\filldraw[fill=white,draw=black] (2.25,2.5) circle (1.5pt);
\filldraw[fill=black,draw=black] (2.5,3) circle (1.5pt);
\filldraw[fill=white,draw=black] (1.5,3) circle (1.5pt);
\draw[-,black,dotted] (1.75,0.5)--(1.75,2.5)--(2,3);
\draw[-,black,dotted] (1.25+0.5*0.25,2.75)--(2.25+0.5*0.25,2.75)--(2.25+0.5*0.25,0.75);
\draw[->,black,dotted] (1.25,1)--(2.25,1)--(2.5,1.5);\draw[->,black,dotted] (1.25,2)--(2.25,2)--(2.5,2.5);
\fill[white!] (1.3,1.15) circle (0.01pt)
node[right=1pt]{\color{black}\small $r_0^{(+)}$};
\fill[white!] (2.5,2.5) circle (0.01pt)
node[right=1pt]{\color{black}\small $r_1^{(+)}$};
\draw[-, thick] (1.25,0.5)--(2.25,1.5)--(2.5,1);\draw[-, thick] (1.25,2.5)--(2.25,1.5)--(2.5,3)--(1.25,2.5);

\draw[-,thick,dashed] (1.25,0.5)--(1.5,2)--(2.5,1);\draw[-,thick,dashed] (1.25,2.5)--(1.5,2)--(2.5,3);
\fill[black] (1.5,2) circle (1.5pt);

\draw[-,thin,gray] (0,0)--(0,-1)--(1,-1)--(1,0);\draw[-,thin,gray] (1,-1)--(2,-1)--(2,0);\draw[-,thin,gray] (2,-1)--(2.25,-0.5)--(2.25,0);\draw[-,thin,gray] (0.25,0.5)--(0.25,0);\draw[-,thin,gray] (1.25,0.5)--(1.25,0);
\filldraw[fill=white,draw=black] (0.25,0.5) circle (1.5pt);
\filldraw[fill=white,draw=black] (1,0) circle (1.5pt);
\filldraw[fill=white,draw=black] (0,-1) circle (1.5pt);
\filldraw[fill=black,draw=black] (1,-1) circle (1.5pt);
\filldraw[fill=white,draw=black] (2,-1) circle (1.5pt);
\filldraw[fill=black,draw=black] (2.25,-0.5) circle (1.5pt);
\draw[-,black,dotted] (0.5,0)--(0.5,-1);\draw[-,black,dotted] (0.75,0)--(0.75,0.5);
\fill[white!] (0.5,-0.2) circle (0.01pt)
node[left=0.05pt]{\color{black}\small $q$};
\draw[-,black,dotted] (1.5,0)--(1.5,-1);\draw[-,black,dotted] (1.75,0)--(1.75,0.5);
\fill[white!] (1.5,-0.2) circle (0.01pt)
node[left=0.05pt]{\color{black}\small $q$};
\draw[<-,black,dotted] (0,-0.5)--(2,-0.5)--(2.25,0);
\fill[white!] (0,-0.5) circle (0.01pt)
node[left=0.05pt]{\color{black}\small $r_0^{(-)}$};
\draw[-,black,dotted] (0.5*0.25-0.4,0.25)--(0.5*0.25,0.25)--(0.5*0.25,0);\draw[-,black,dotted] (2+0.25*0.5,-0.75)--(2+0.25*0.5,0);\draw[->,black,dotted] (2+0.5*0.25,0.25)--(6+0.5*0.25+0.4,0.25);
\fill[white!] (0.5*0.25-0.4,0.25) circle (0.01pt)
node[left=0.05pt]{\color{black}\small $p$};
\draw[-, thick] (0,0)--(1,-1)--(2,0)--(2.25,-0.5);\draw[-,thick] (0.75,0)--(1.25,0.5)--(1.75,0);

\draw[-, thick,dashed] (0,0)--(0.25,-0.5)--(0.75,0);\draw[-,thick,dashed] (0.25,-0.5)--(1,-1)--(2.25,-0.5)--(1.75,0);
\fill[black] (0.25,-0.5) circle (1.5pt);

\draw[-,thick,dashed] (1.25,1+0.5*0.75/1.25)--(1.75,1.5)--(1.25,2-0.5/3);\draw[-,thick,dashed] (2.5,1)--(1.75,1.5)--(2.5,2-0.5*0.5/1.25);
\fill[black] (1.75,1.5) circle (1.5pt);
\draw[-,thick,dashed] (3.25,1+0.5*0.75/1.25)--(3.75,1.5)--(3+0.75*0.5,1.75);

\fill[white] (7.86,0) circle (0.01pt)
node[below=1.5pt]{\color{black} $+\ihat$};
\fill[white] (7.66,0.33) circle (0.01pt)
node[right=1.5pt]{\color{black} $+\jhat$};
\fill[white] (7.5,0.75*0.5) circle (0.01pt)
node[left=1.5pt]{\color{black} $+\khat$};
\draw[->, thick] (7.5,0)--(8.25,0);\draw[->, thick] (7.5,0)--(7.83,0.67);\draw[->, thick] (7.5,0)--(7.5,0.75);

\end{tikzpicture}
\caption{An example of a surface $\sigma$, made up of faces and edges connecting a subset of nearest neighbour vertices of $\mathbb{Z}^3$.  An edge interaction model is defined on the next nearest neighbour black vertices, which are connected by edges on diagonals of elementary squares of $\sigma$.  Boltzmann weights on elementary squares of $\sigma$, are given in Figures \ref{strsquare} and \ref{4BW}.  Here the rapidity lines $r$ are given some additional labelling (introduced in the next subsection), distinguishing positively (+) and negatively (-) oriented lines, and an index $i=0,1,\ldots,$ to distinguish lines $r_i$ that are on the interior of lines $r_j$, where $i>j$.  The latter index will be associated to the order of deformations of the square lattice.}
\label{Glattice2}
\end{figure}

An example of such a surface $\sigma$, associated to $\sigma_0$ in Figure \ref{Glattice}, is shown in Figure \ref{Glattice2}.  The condition that $\sigma$ is oriented, requires that for each interior edge $(ij)$, shared by two elementary squares of $\sigma$, the orientation of $(ij)$ with respect to one elementary square, is opposite with respect to the other elementary square.  Since $\sigma$ is also simply connected and has the same boundary as $\sigma_0$, elementary squares $\sigma_{ik},\sigma_{ki}$, will only appear on $\sigma$ in pairs, such that for each instance of an elementary square $\sigma_{ik}(\n)\in F(\sigma)$, there exists also a unique elementary square $\sigma_{ki}(\n+a\jhat)\in F(\sigma)$, for some $a\in\mathbb{Z}-\{0\}$.  Similarly, the elementary squares $\sigma_{jk}$, and $\sigma_{kj}$, will only appear in pairs.  The assignment of rapidity lines in Figures \ref{strsquare}, and \ref{4BW}, is chosen so that the $p$ and $q$ rapidity lines on $\sigma$ are continuous, always beginning and ending at opposite boundaries of $\sigma$ (rapidity lines always remain fixed at the boundaries under Z-invariance), and they are always consistently directed along the elementary squares of $\sigma$.

As was the case for $\sigma_0$, spin variables $x_i$ are assigned to black vertices $i\in V(\G)(=V^{(1)}(\sigma))$ only, and the white vertices play no role in the definition of the edge interaction model.  Spins on the boundary of $\sigma$ are assigned some fixed values.  The graph $\G$, defined on $\sigma$, is for an edge interaction model with Boltzmann weights associated to edges connecting two spins $x_i,x_j$, as shown in Figures \ref{strsquare}, and \ref{4BW}.

The partition function is given by \eqref{zdef}, with products over $E^{(1)}(L)$, $E^{(2)}(L)$, and $V_{int}(L)$, replaced with products over $E^{(1)}(\G)$, $E^{(2)}(\G)$, and $V_{int}(\G)$ respectively, and the sum $\sum_{\bf x}$, now taken over the values of all spins $x_i$, at vertices $i\in V_{int}(\G)$.  This may be expressed as
\beq
\label{zsig}
Z=\sum_{\bf x} \prod_{(ij)\in E^{(1)}(\G)} W_{p_{i}p_{j}}(x_i,x_j) \prod_{(ij)\in E^{(2)}(\G)}\oW_{p_{i}p_{j}}(x_i,x_j) \prod_{i\in V_{int}(\G)}  S(x_i)\,,
\eeq
where the $p_i,p_j\in\{p,q,r\}$, are rapidity variables associated to an edge $(ij)$.  The Boltzmann weights are also required to satisfy the star-triangle relation \eqref{STR}, and inversion relations \eqref{invrels}, and this defines the integrable edge interaction model on $\G$.

It should also be noted that although the model on $\G$, may be obtained from deformations of a physical square lattice model on $L$, it will typically contain some non-physical local edge interactions.   The reason for this, is that the value associated to different pairs of rapidity lines, say $rq$ and $pr$, can be freely chosen such that Boltzmann weights $W_{rq},W_{pr},\oW_{rq},\oW_{pr}$, are positive valued, but then exchanging the order of rapidity lines, means that Boltzmann weights $\oW_{qr}$, and $\oW_{rp}$, are not guaranteed to be positive (they typically assume a mixture of positive and negative real values).  This is an unfortunate side-effect of the Z-invariance property in general \cite{Bax1}, as the second inversion relation in \eqref{invrels}, involves two Boltzmann weights $\oW_{pq}$, and $\oW_{qp}$, which cannot both be chosen to be positively valued.

\subsection{Z-invariance} \label{sec:z-invar}

The main result of this section, is showing that the partition function of the edge interaction model on $\G$, is equivalent to the partition function of the edge interaction model on the square lattice $L$, up to some simple factors.  This is done by deforming the surface $\sigma_0$, until it coincides with $\sigma$, using only the deformations described in Appendix \ref{app:str}.  As is the usual case for Z-invariance, these deformations leave the partition function invariant up to the factors coming from the star-triangle relation \eqref{STR}, and inversion relations \eqref{invrels}. 

The deformations given in Appendix \ref{app:str}, exchange $m$ elementary squares of $\sigma$, with $n$ different elementary squares, where $m=1,2,3,4,5$, and $n=6-m$, and the union of the $m$ squares with the $n$ squares forms a cube.  Consequently such deformations will sometimes be referred to as ``cubic flips''.  With the convention used in Appendix \ref{app:str}, a deformation of $\sigma$, that involves the addition of a positively oriented rapidity line $r$, always translates an elementary square $\sigma_{ij}(\n)$ to $\sigma_{ij}(\n+\khat)$, while a deformation that involves the addition of a negatively oriented rapidity line $r$, always translates an elementary square $\sigma_{ij}(\n)$ to $\sigma_{ij}(\n-\khat)$.  

The deformations are generally non-commuting, and therefore there is a preferred order to perform sequences of deformations.  Moreover, as was mentioned in the previous subsection, there are limitations on the type of deformations that can be derived from the star-triangle relation \eqref{STR}, and inversion relations \eqref{invrels}.  Consequently, another purpose of the ordering, is to avoid in some cases needing to add a vertex surrounded by a combination of $p,q,r$, rapidity lines, that together form a directed closed path, which cannot be done directly by using the star-triangle relation.  Note however that these types of vertices may be added in special cases (in fact they are required), with subsequent applications of the star-triangle relation, and inversion relations.  This is done in the deformations pictured in Figure \ref{1cube} (see also Figure \ref{newd}), that correspond to the Equations \eqref{bc1eq}-\eqref{bc4eq}.
\\[-0.3cm]

\begin{figure}[htb]
\centering
\begin{tikzpicture}[scale=0.8]

\fill[white!] (6.1,-1) circle (0.01pt)
node[right=0pt]{\color{black}\small $\sigma$};
\draw[-,very thin] (-2.9,-4.4)--(6.1,-4.4)--(6.1,2.4)--(-2.9,2.4)--cycle;

\fill[white!] (-1.4,0) circle (0.01pt)
node[left=4pt]{\color{black}\small $r_0^+$};
\fill[white!] (-1.0,0) circle (0.01pt)
node[right=4pt]{\color{black}\small $r_1^+$};
\draw[thick] (-1.55,-0.00)--(-1.4,-0.15)--(-1.25,-0.00);
\draw[thick] (-1.15,-0.00)--(-1.0,-0.15)--(-0.85,-0.00);
\draw[very thick,dotted,rounded corners=10] (-1,-1) -- (1,-1) -- (1,1) -- (-1,1) -- cycle;
\draw[very thick,dotted,rounded corners=10] (-1.4,-1.4) -- (1.4,-1.4) -- (1.4,1.4) -- (-1.4,1.4) -- cycle;
\begin{scope}[yshift=-85]
\fill[white!] (1,0) circle (0.01pt)
node[right=4pt]{\color{black}\small $r_0^-$};
\draw[thick] (0.85,0.05)--(1.0,-0.1)--(1.15,0.05);
\draw[very thick,dotted,rounded corners=10] (-2,-1) -- (1,-1) -- (1,1) -- (-2,1) -- cycle;
\end{scope}
\begin{scope}[xshift=110,yshift=8,scale=1.2]
\fill[white!] (0,-1.4) circle (0.01pt)
node[below=4pt]{\color{black}\small $r_0^+$};
\fill[white!] (0,-1) circle (0.01pt)
node[above=4pt]{\color{black}\small $r_1^-$};
\draw[thick] (-0.05,-1.55)--(0.1,-1.4)--(-0.05,-1.25);
\draw[thick] (0.05,-1.15)--(-0.1,-1.0)--(0.05,-0.85);
\draw[very thick,dotted,rounded corners=10] (-1,-1) -- (1,-1) -- (1,1) -- (-1,1) -- cycle;
\draw[very thick,dotted,rounded corners=10] (-1.4,-1.4) -- (1.4,-1.4) -- (1.4,1.4) -- (-1.4,1.4) -- cycle;
\end{scope}

\end{tikzpicture}
\caption{Orientation of rapidity lines labelled $r$, on the surface $\sigma$ in Figure \ref{Glattice2}.}
\label{rnotation}
\end{figure}
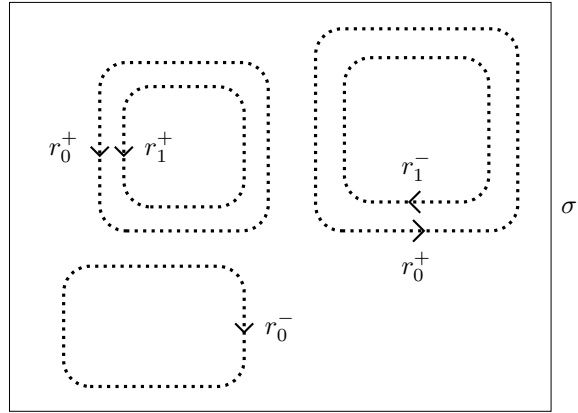

Let us now show the aforementioned property of Z-invariance.  Let $\sigma$ be a surface, containing the graph $\G$ of the edge interaction model, as defined in the previous subsection, and recall that $\sigma$ has the same boundary as some flat surface $\sigma_0$.  Now each of the elementary squares $\sigma_{ik},\sigma_{jk},\sigma_{ki},\sigma_{kj}\in F(\sigma)$, carries an instance of a rapidity line $r$, and this rapidity line $r$ forms a directed closed path that is either positively or negatively oriented.  In the following, two different orientations of $r$, will be distinguished by superscripts as $r^+$, and $r^-$, respectively.  Further to this, the notation $r_i$ will label a rapidity line $r$, that lies on the interior of instances of rapidity lines labelled $r_{i-1},r_{i-2},\ldots,r_0$, with $r_0$ lying on the interior of the boundary only.  For example, with this notation, the surface $\sigma$ in Figure \ref{Glattice2}, contains three rapidity lines labelled $r^-_0$, $r^+_0$, $r^+_0$, and two rapidity lines labelled $r^+_1$, $r^-_1$, as depicted in Figure \ref{rnotation}.  Different instances of the rapidity lines $r$ do not intersect each other, and the rapidity lines $r$ lie entirely within the boundary of $\sigma$.

Next, the deformations in Appendix \ref{app:str} are used, to add all rapidity lines $r$ that appear on $\sigma$.  First, for rapidity lines $r_0$, a deformation in Figure \ref{1cube} is used, to add four elementary squares $\sigma_{ik},\sigma_{jk},\sigma_{ki},\sigma_{kj}$ to $\sigma_0$, and a corresponding rapidity line $r$ on the latter squares.  Because of the condition \eqref{condition}, there can only be one pair of elementary squares $\sigma_{kj}(\n),\sigma_{ki}(\n+\jhat)\in F(\sigma)$, associated to a positively oriented line $r^+$, or one pair of elementary squares $\sigma_{kj}(\n+\ihat),\sigma_{ki}(\n)\in F(\sigma)$, associated to a negatively oriented line $r^-$.  This in fact fixes the choice of the initial deformation on $\sigma_0$.  For the positively oriented case, one of the top two deformations in Figure \ref{1cube} is applied to the square $\sigma_{ij}(\n)\in F(\sigma_0) $, while for the negatively oriented case, one of the bottom two deformations in Figure \ref{1cube} is applied to the square $\sigma_{ij}(\n)\in F(\sigma_0)$.

Once this is done, the remaining deformations in Figures \ref{2cube}, and \ref{3cube}, are then used to add additional elementary squares to $\sigma_0$, that stretch either a positively oriented rapidity line $r^+$ in the $+\ihat$ and $-\jhat$ directions, or a negatively oriented rapidity line $r^-$ in the $-\ihat$ and $+\jhat$ directions.  This is done until the same configuration of elementary squares $\sigma_{ik},\sigma_{jk},\sigma_{ki},\sigma_{kj}\in F(\sigma)$, with associated rapidity line $r_0$, appears on $\sigma_0$.

The above process is then repeated for rapidity lines $r_1,r_2,\ldots$, until all elementary squares $\sigma_{ik},\sigma_{jk},\sigma_{ki},\sigma_{kj}\in F(\sigma)$, with the associated rapidity lines $r_i$, also appear on $\sigma_0$.  This can always be done with the deformations in Appendix \ref{app:str}, and by following the above ordering of the deformations.  In this process, for positively oriented rapidity lines $r^+$, the deformations always involve adding to the surface, (at least) the triplets of elementary squares $\sigma_{ij}(\n+\khat)$, $\sigma_{ik}(\n)$, $\sigma_{jk}(\n+\ihat)$, while for negatively oriented rapidity lines $r^-$, the deformations always involve adding to the surface, (at least) the triplets of elementary squares $\sigma_{ij}(\n)$, $\sigma_{ik}(\n+\jhat)$, $\sigma_{jk}(\n)$.

Note also that the deformations that appear in Appendix \ref{app:str}, can never alter the boundary of $\sigma_0$.  After deforming the surface as described above, $\sigma_0$ will exactly coincide with $\sigma$, and consequently the edge interaction model on $L$, will coincide with the model defined on $\G$.  Since this has been done using only the deformations appearing in Appendix \ref{app:str}, that are a consequence of the star-triangle relation \eqref{STR}, and inversion relations \eqref{invrels}, the edge interaction models on $\G$, and $L$, respectively, have the same partition function up to simple factors coming from \eqref{STR}, and \eqref{invrels}. This is what was to be shown.

\subsection{Additional remarks}

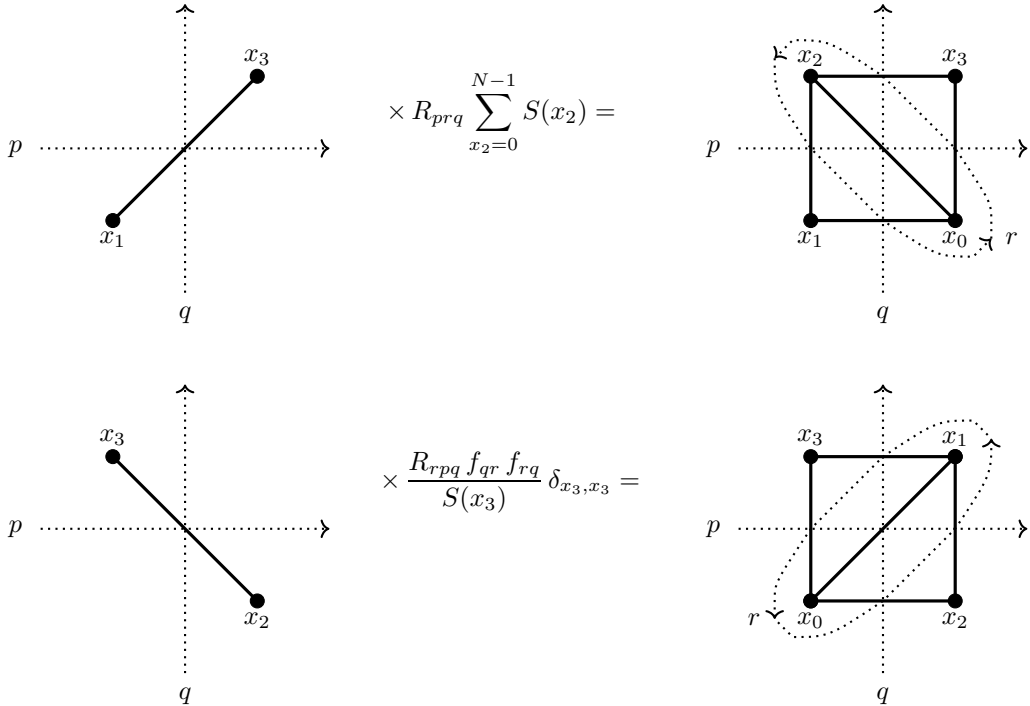
\begin{figure}[htb]
\centering
\begin{tikzpicture}[scale=2.4]

\draw[-,very thick] (-0.4,1.6)--(0.4,2.4);
\draw[->,thick,dotted,black] (0,1.2)--(0,2.8);
\draw[->,thick,dotted,black] (-0.8,2)--(0.8,2);
\fill[white!] (0,1.2) circle (0.01pt)
node[below=1.5pt]{\color{black}\small $q$};
\fill[white!] (-0.8,2) circle (0.01pt)
node[left=3.1pt]{\color{black}\small $p$};
\filldraw[fill=black,draw=black] (-0.4,1.6) circle (1.1pt)
node[below=0.5pt]{\color{black}\small $x_1$};
\filldraw[fill=black,draw=black] (0.4,2.4) circle (1.1pt)
node[above=0.5pt]{\color{black}\small $x_3$};

\draw[white] (1.75,2.5) circle (0.01pt)
node[below=1pt]{\color{black}\small $\ds\times\, R_{prq}\sum_{x_2=0}^{N-1} S(x_2)=$};

\begin{scope}[xshift=110]

\draw[->,thick,dotted,black] (0,1.2)--(0,2.8);
\draw[->,thick,dotted,black] (-0.8,2)--(0.8,2);
\fill[white!] (0.6,1.5) circle (0.01pt)
node[right=1.5pt]{\color{black}\small $r$};
\fill[white!] (0,1.2) circle (0.01pt)
node[below=1.5pt]{\color{black}\small $q$};
\fill[white!] (-0.8,2) circle (0.01pt)
node[left=3.1pt]{\color{black}\small $p$};
\filldraw[fill=black,draw=black] (-0.4,1.6) circle (1.1pt)
node[below=0.5pt]{\color{black}\small $x_1$};
\filldraw[fill=black,draw=black] (0.4,2.4) circle (1.1pt)
node[above=0.5pt]{\color{black}\small $x_3$};
\filldraw[fill=black,draw=black] (0.4,1.6) circle (1.1pt)
node[below=0.5pt]{\color{black}\small $x_0$};
\filldraw[fill=black,draw=black] (-0.4,2.4) circle (1.1pt)
node[above=0.0pt]{\color{black}\small $x_2$};
\draw[-,very thick] (-0.4,2.4)--(-0.4,1.6)--(0.4,1.6)--(-0.4,2.4)--(0.4,2.4)--(0.4,1.6);
\draw[->,thick,dotted,black] (0.6,1.5)--(0.6,1.6) .. controls (0.6,1.7) .. (0.4,2)--(0,2.4) .. controls (-0.3,2.6) .. (-0.5,2.6)-- (-0.6,2.5);
\draw[->,thick,dotted,black] (-0.6,2.5) .. controls (-0.6,2.3) .. (-0.4,2)--(0,1.6) .. controls (0.3,1.4) .. (0.5,1.4)--(0.6,1.5);

\end{scope}

\begin{scope}[yshift=-60]
\draw[-,very thick] (-0.4,2.4)--(0.4,1.6);
\draw[->,thick,dotted,black] (0,1.2)--(0,2.8);
\draw[->,thick,dotted,black] (-0.8,2)--(0.8,2);
\fill[white!] (0,1.2) circle (0.01pt)
node[below=1.5pt]{\color{black}\small $q$};
\fill[white!] (-0.8,2) circle (0.01pt)
node[left=3.1pt]{\color{black}\small $p$};
\filldraw[fill=black,draw=black] (-0.4,2.4) circle (1.1pt)
node[above=0.5pt]{\color{black}\small $x_3$};
\filldraw[fill=black,draw=black] (0.4,1.6) circle (1.1pt)
node[below=0.5pt]{\color{black}\small $x_2$};

\draw[white] (1.8,2.5) circle (0.01pt)
node[below=1pt]{\color{black}\small $\ds\times\, \frac{R_{rpq}\,f_{qr}\,f_{rq}}{S(x_3)}\,\delta_{x_3,x_3}=$};

\begin{scope}[xshift=110]

\draw[->,thick,dotted,black] (0,1.2)--(0,2.8);
\draw[->,thick,dotted,black] (-0.8,2)--(0.8,2);
\fill[white!] (-0.6,1.5) circle (0.01pt)
node[left=1.5pt]{\color{black}\small $r$};
\fill[white!] (0,1.2) circle (0.01pt)
node[below=1.5pt]{\color{black}\small $q$};
\fill[white!] (-0.8,2) circle (0.01pt)
node[left=3.1pt]{\color{black}\small $p$};
\filldraw[fill=black,draw=black] (-0.4,1.6) circle (1.1pt)
node[below=0.5pt]{\color{black}\small $x_0$};
\filldraw[fill=black,draw=black] (0.4,2.4) circle (1.1pt)
node[above=0.0pt]{\color{black}\small $x_1$};
\filldraw[fill=black,draw=black] (0.4,1.6) circle (1.1pt)
node[below=0.5pt]{\color{black}\small $x_2$};
\filldraw[fill=black,draw=black] (-0.4,2.4) circle (1.1pt)
node[above=0.5pt]{\color{black}\small $x_3$};
\filldraw[fill=black,draw=black] (0.4,2.4) circle (1.1pt);
\filldraw[fill=black,draw=black] (-0.4,1.6) circle (1.1pt);
\draw[-,very thick] (0.4,2.4)--(-0.4,2.4)--(-0.4,1.6)--(0.4,1.6)--(0.4,2.4)--(-0.4,1.6);
\draw[<-,thick,dotted,black] (-0.6,1.5)--(-0.6,1.6) .. controls (-0.6,1.7) .. (-0.4,2)--(0,2.4) .. controls (0.3,2.6) .. (0.5,2.6)-- (0.6,2.5);
\draw[<-,thick,dotted,black] (0.6,2.5) .. controls (0.6,2.3) .. (0.4,2)--(0,1.6) .. controls (-0.3,1.4) .. (-0.5,1.4)--(-0.6,1.5);

\end{scope}
\end{scope}

\end{tikzpicture}
\caption{A pair of deformations of edges of the square lattice, according to Figure \ref{1cube}.  The deformations are also valid when the orientation of the rapidity lines $r$ is reversed.  These deformations are central to the formulation of Z-invariance given in this paper.}
\label{newd}
\end{figure}

The deformations in Appendix \ref{app:str}, reduce to deformations included in the usual formulation of Z-invariance, except for the deformations in Figure \ref{1cube}, which add a new rapidity line $r$, that forms directed closed paths (both with itself, and in combination with $p$ and $q$ rapidity lines) in the rapidity graph.  When flattened onto the plane, the latter deformations have the form shown in Figure \ref{newd}.  In this figure, an edge connecting two vertices, is replaced with a ``square'' configuration of edges that connect four vertices, along with a new edge on a diagonal.  The deformations are also valid when $r$ has the opposite orientation.  This extension of Z-invariance, and corresponding reformulation in terms of ``cubic'' deformations appearing in Appendix \ref{app:str}, are the main results of this paper.

The Z-invariance described in this section, is valid for general integrable edge interaction models on a square lattice, that are defined by Boltzmann weights satisfying the star-triangle relation \eqref{STR}, and inversion relations \eqref{invrels}.  This includes, for example, the N-state chiral Potts, Kashiwara-Miwa, and Fateev-Zamolodchikov models, and also continuous spin models such as the Bazahnov-Sergeev master solution, and the Zamolodchikov fishing-net model.  In the latter continuous spin cases, all sums over discrete spin states in \eqref{zdef}, \eqref{STR}, and \eqref{invrels}, should be replaced with integrals over continuous spin states.

There are however some additional considerations that are needed, in order to deal with infinities that appear for the continuous spin models, which don't arise for discrete spin models.  For example, some of the continuous spin models, {\it e.g.} Zamolodchikov fishing-net model, and Faddeev-Volkov model, have spin variables taking values over the entire real line.  This means that the factor $\sum_{x_0}S(x_0)$, in \eqref{bc1eq}, \eqref{bc3eq}, becomes $\int_\mathbb{R}dx_0 S(x_0)$, which is generally a divergent quantity.  Also for all continuous spin models, the $\delta_{x,x}$ factors in \eqref{invrel2}, become $\delta(0)$ factors, which is again a divergent quantity.

Thus the Z-invariance in the continuous spin cases, appears to require some regularisation in order to avoid the infinite terms.  This may not pose too much of a problem depending on the desired application, as most quantities of interest are given in terms of derivatives of $\log(Z)$, and the divergent quantities have no dependence on any variables in $Z$.  It is also possible to avoid the divergence associated with the factor $\delta(0)$, by not using the inversion relation \eqref{invrel2}.  Then the deformations corresponding to Equations \eqref{bc2eq}, and \eqref{bc4eq}, cannot be used, which would require an additional restriction on the configuration of elementary squares on $\sigma$, similar to \eqref{condition}.  In any case, it is clear that additional care should be taken for the continuous spin case, compared to the discrete spin case.  It should again be emphasised, that there are no infinities at all for discrete spin models, whose spin states are generally restricted to values $\mathbb{Z}\mbox{ mod }N$, and for which the delta function in \eqref{invrel2} is simply a constant $\delta_{x,x}=1$.
~\\

Next recall that the condition \eqref{condition} was required, as there is no star-triangle relation where rapidity lines form a directed closed path around the interior vertex in Figure \ref{fig3}.  One possible way to avoid this condition, is to take the lattice model out of the physical regime in the following way.

First note that in all known cases, there exists a mapping $F(p)$, of a rapidity variable, such that Boltzmann weights are related by $\oW_{pq}(x_i,x_j)=W_{qF(p)}(x_i,x_j)$, and $W_{pq}(x_i,x_j)=\oW_{qF(p)}(x_j,x_i)$ \cite{Bax2,Bax02rip}.  Now consider the particular case when Boltzmann weights are reflection symmetric, such that $W_{pq}(x_i,x_j)=W_{pq}(x_j,x_i)$, $\oW_{pq}(x_i,x_j)=\oW_{pq}(x_j,x_i)$, and when the above mapping is given by $F(p)=p$.  For example, for models that satisfy crossing symmetry, this case corresponds to a non-physical regime obtained by taking the limit of the crossing parameter $\eta\to0^+$ (if this converges), while the rapidity variables $p$, $q$, are shifted off the real axis into the complex plane.  The Boltzmann weights are then related by $W_{pq}(x_i,x_j)=\oW_{qp}(x_i,x_j)$, and the second expression in the star-triangle relation \eqref{STR}, may be written as
\beq
\label{strnp}
\sum_{x_0}S(x_0)\,W_{rq}(x_0,x_1)\,W_{pr}(x_0,x_2)\,W_{qp}(x_0,x_3)=R_{pqr}\,\oW_{rq}(x_2,x_3)\,\oW_{pr}(x_3,x_1)\,\oW_{qp}(x_1,x_2)\,.
\eeq
This expression amounts to reversing the orientation of the rapidity line $q$ in Figure \ref{fig3}, in which case a directed closed path is formed around the central spin $x_0$.  Thus in this non-physical example, the condition \eqref{condition} would not be required.
~\\

Finally, there are also some straightforward generalisations that may be considered.  One is to allow different instances of rapidity lines $p,q,r$ to take different values, for example, these values may be distinguished with the labelling $p_1,p_2,\ldots$, $q_1,q_2,\ldots$, and $r_1,r_2,\ldots$.  This modification can be implemented without much difficulty.  Another generalisation is to consider a surface $\sigma$, which also includes oriented elementary squares of the type $\sigma_{ji}$.  The latter would require introducing additional $p$ and $q$ rapidity lines to the definition of the model, that form directed closed paths with themselves.  The additional deformations required for Z-invariance, would be identical to those already given in Appendix \ref{app:str}, up to a relabelling of spin variables and rapidity lines.  Consequently, this level of generality was not considered here in any detail, as it is not conceptually too different from the formulation of Z-invariance that was already described in this section.

\section{Quasi-classical limit and classical discrete Laplace equations}\label{sec:qcl}

This section considers the quasi-classical expansion of the edge interaction model, which is known to lead to classical discrete integrable equations \cite{BMS07a,Bazhanov:2010kz,BKS2}.  The quasi-classical limit is a low-temperature limit, in which the lattice model approaches a ground state configuration, that is determined by maximising a real valued action functional under fixed boundary conditions.  The equation of motion that comes from the quasi-classical expansion of the partition function \eqref{zdef}, corresponds to what is known as a classical discrete Laplace system of equations \cite{MR2467378,BG11}.  For a square lattice model these are five-point equations centered at each interior vertex of the lattice (see {\it e.g.} Figure \ref{figLaplace}).

For models with Boltzmann weights that satisfy the star-triangle relation \eqref{STR}, the quasi-classical limit leads to a {\it classical star-triangle relation} for two-point Lagrangians.  The latter Lagrangians arise in the leading order quasi-classical expansion of Boltzmann weights.  The classical star-triangle relation is satisfied on solutions of an equation of motion, that in the majority of known cases \cite{BKS2} is equivalent to a three-leg equation, that is associated to a discrete integrable equation in the Adler, Bobenko, Suris (ABS) classification \cite{AdlerBobenkoSuris}.

Table \ref{qcltable} provides an overview of most of the important properties of integrable models of statistical mechanics, and the related properties of classical discrete integrable equations, that arise through the quasi-classical limit.  The majority of items in this table, have appeared before in the study of various solutions of the star-triangle relation \cite{BMS07a,Bazhanov:2010kz,BKS2}.

The main result of this section corresponds to the last line in Table \ref{qcltable}.  First, it will be shown how the classical star-triangle relation, implies the classical Z-invariance property for a system of classical discrete Laplace equations.  Here the property of classical Z-invariance, is that an action functional for the classical discrete Laplace equations, is invariant under the deformations of a surface given in Appendix \ref{app:str}.  The classical Z-invariance is directly connected to a closure relation, that was first introduced by Lobb and Nijhoff \cite{LN09}, for particular classical discrete ABS systems involving three-point Lagrangians on elementary squares.  It is shown here how the classical star-triangle relation, may be interpreted as the corresponding closure relation for classical discrete Laplace systems, which involve two-point Lagrangians on diagonals of elementary squares.  This section uses the notations that were introduced in Sections \ref{sec:sig}, and \ref{sec:sigp}.

\begin{savenotes}
\begin{table}[htb!]
\centering
\begin{tabular}{c c}
Statistical Mechanics & Classical Discrete Laplace Equations \\[0.2cm]
\hline \\[-0.2cm]
Spins at vertices & Fields at vertices \\[0.3cm]
Rapidity variables & Parameters on edges \\[0.3cm]
Boltzmann weight (BW) & 2-point Lagrangian \\[0.3cm]
Inversion relations & Anti-symmetry of Lagrangians \\[0.3cm]
Star-triangle relation (STR) & Tetrahedron relation\footnote{``Tetrahedron relation'' refers to a characteristic property for ABS equations and associated Lagrangian functions \cite{AdlerBobenkoSuris,BS09}, and it is not related to the analogue of the Yang-Baxter equation for three-dimensional lattice models known as ``tetrahedron equation''.} \\[0.3cm]
STR saddle point equation & Three-leg equation \\[0.3cm]
Partition function (PF) &  Action functional \\[0.3cm]
PF saddle point equation & Discrete Laplace equations \\[0.3cm]
Z-invariance & Closure of action functional\\[0.3cm]
\hline 
\end{tabular}
\caption{Dictionary between integrable models of statistical mechanics, and classical discrete Laplace equations that appear in the quasi-classical limit.  The last line is the main focus of this section.}
\label{qcltable}
\end{table}

\subsection{Quasi-classical expansion of the star-triangle relation} \label{sec:qcl2}

The majority of this subsection is based on the results previously obtained in \cite{BKS2}.

Recall the Boltzmann weights which are denoted by $ W_{pq}(x_i,x_j)$, $\oW_{pq}(x_i,x_j)$.  These Boltzmann weights are assumed to implicitly depend on an an additional parameter, which will be denoted as $\hbar$.  This is a temperature-like parameter for the spin model (or Planck constant in the language of quantum field theory), where taking $\hbar\rightarrow0$ corresponds to the system approaching a classical ground state configuration.  The latter ground state configuration, will be seen to be determined by the solution of a classical discrete integrable equation. 

Let $f_\hbar(x)$ denote a scaling and translation of a variable $x$, that has a dependency on the parameter $\hbar$.  It is assumed that as $\hbar\rightarrow0$, the quasi-classical expansion of Boltzmann weights $W_{pq}(x_i,x_j)$, $\oW_{pq}(x_i,x_j)$, can be written as
\beq
\label{weight-exp}
\begin{array}{rcl}
\ds\log W_{f_\hbar(p)f_\hbar(q)}(f_\hbar(x_i),f_\hbar(x_j))&\!\!\!\!=\!\!\!\!&\ds  -\hbar^{-1}\, \lag_{pq}(x_i,x_j)+O(\hbar^0)\,,\\[0.3cm]
\ds\log \overline{ W}_{f_\hbar(p)f_\hbar(q)}(f_\hbar(x_i),f_\hbar(x_j))&\!\!\!\!=\!\!\!\!&\ds -\hbar^{-1}\,\ol_{pq}(x_i,x_j) -\frac{1}{2}\log\hbar+O(\hbar^0)\,.
\end{array}
\eeq

\end{savenotes}

This is the case for the majority of edge interaction models, under a suitable transformation $f_\hbar$, that is model dependent \cite{BKS2}.  Note that the asymptotics of $S(x)$ is included in the second equation of \eqref{weight-exp}, which is always possible through a redefinition of $\oW_{pq}(x_i,x_j)$.  Equation \eqref{weight-exp} is valid for models where the spin $x_i$ takes either continuous or discrete values, while in the quasi-classical limit, all spins $x_i$ become continuously valued.  Since for now we assume the edge interaction model to have real valued Boltzmann weights, the resulting Lagrangians $\Lambda_{pq}(x_i,x_j)$, $\olam_{pq}(x_i,x_j)$, are real valued functions of the spins $x_i,x_j\in\mathbb{R}$.

The inversion relations \eqref{invrels} satisfied by the Boltzmann weights, imply that the following anti-symmetry relations are satisfied by the Lagrangian functions
\beq
\label{clasinversion}
\ds\lag_{pq}(x_i ,x_j)+\lag_{qp}(x_j,x_i)=0\,,\quad
\ds \ol_{pq}(x_i,x_j)+\ol_{qp}(x_j,x_i)=0\,,
\eeq
for all $x_i,x_j$.

Using \eqref{weight-exp}, the quasi-classical expansion of the star-star relation \eqref{STR}, results in the following {\it classical star-triangle relation} at leading order
\beq
\label{cstar}
\begin{array}{l}
\ol_{qr}(x_1,x_0)+\lag_{pr}(x_2,x_0)+\ol_{pq}(x_0,x_3)=\lag_{qr}(x_2,x_3)+\ol_{pr}(x_1,x_3)+\lag_{pq}(x_2,x_1)\,,\\[0.3cm]
\ol_{qr}(x_0,x_1)+\lag_{pr}(x_0,x_2)+\ol_{pq}(x_3,x_0)=\lag_{qr}(x_3,x_2)+\ol_{pr}(x_3,x_1)+\lag_{pq}(x_1,x_2)\,,
\end{array}
\eeq
where the respective expressions in \eqref{cstar}, are satisfied on solutions of the three leg equations
\beq
\label{ceqmo}
\begin{array}{l}
\ds\frac{\partial}{\partial x_0}\left(\ol_{qr}(x_1,x_0)+\lag_{pr}(x_2,x_0)+\ol_{pq}(x_0,x_3)\right)=0\,,\\[0.5cm]
\ds\frac{\partial}{\partial x_0}\left(\ol_{qr}(x_0,x_1)+\lag_{pr}(x_0,x_2)+\ol_{pq}(x_3,x_0)\right)=0\,,
\end{array}
\eeq
for fixed values of $x_1,x_2,x_3,p,q,r$.  In \eqref{cstar}, the asymptotics of the factor $R_{\alpha\beta}$ are absorbed into the definitions of $\lag_{pq}(x_i,x_j)$, and $\ol_{pq}(x_i,x_j)$, which is always possible due to the factorisation of $R_{\alpha\beta}$ mentioned in Section \ref{sec:modeldef}.  It has previously been shown \cite{BKS2} that the three-leg forms \eqref{ceqmo}, satisfy ``3D-consistency'' \cite{nijhoffwalker,BoSur,AdlerBobenkoSuris}, as a consequence of the classical star-triangle relation \eqref{cstar}, and anti-symmetry relations \eqref{clasinversion}, and are explicitly identified as equations in the ABS classification \cite{AdlerBobenkoSuris}.

Note that it is usually the case that the Boltzmann weights will satisfy additional relations that can be used to simplify the above equations, however to be as general as possible, no additional relations or symmetries on the Boltzmann weights will be assumed for the remainder of this section.

\subsection{Classical discrete Laplace equations}

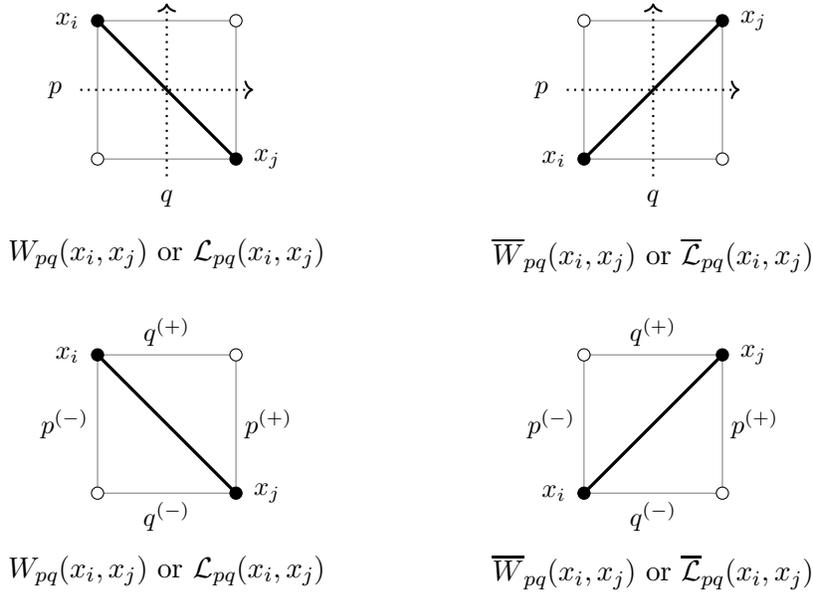
\begin{figure}[htb]
\centering
\begin{tikzpicture}[scale=2.3]
\draw[-,gray] (-0.4,1.6)--(0.4,1.6)--(0.4,2.4)--(-0.4,2.4)--(-0.4,1.6);
\draw[-,very thick] (0.4,1.6)--(-0.4,2.4);
\draw[->,thick,dotted,black] (0,1.5)--(0,2.5);
\draw[->,thick,dotted,black] (-0.5,2)--(0.5,2);
\fill[white!] (0,1.5) circle (0.01pt)
node[below=1.5pt]{\color{black}\small $q$};
\fill[white!] (-0.5,2) circle (0.01pt)
node[left=3.1pt]{\color{black}\small $p$};
\filldraw[fill=black,draw=black] (0.4,1.6) circle (1pt)
node[right=3pt]{\color{black}\small $x_j$};
\filldraw[fill=black,draw=black] (-0.4,2.4) circle (1pt)
node[left=3pt]{\color{black}\small $x_i$};
\filldraw[fill=white,draw=black] (-0.4,1.6) circle (1pt);
\filldraw[fill=white,draw=black] (0.4,2.4) circle (1pt);

\fill (0,1.2) circle(0.01pt)
node[below=0.05pt]{\color{black} $ W_{pq}(x_i,x_j)$ or $\lag_{pq}(x_i,x_j)$};

\begin{scope}[xshift=80pt]
\draw[-,gray] (-0.4,1.6)--(0.4,1.6)--(0.4,2.4)--(-0.4,2.4)--(-0.4,1.6);
\draw[-,very thick] (-0.4,1.6)--(0.4,2.4);
\draw[->,thick,dotted,black] (0,1.5)--(0,2.5);
\draw[->,thick,dotted,black] (-0.5,2)--(0.5,2);
\fill[white!] (0,1.5) circle (0.01pt)
node[below=1.5pt]{\color{black}\small $q$};
\fill[white!] (-0.5,2) circle (0.01pt)
node[left=3.1pt]{\color{black}\small $p$};
\filldraw[fill=black,draw=black] (-0.4,1.6) circle (1pt)
node[left=3pt]{\color{black}\small $x_i$};
\filldraw[fill=black,draw=black] (0.4,2.4) circle (1pt)
node[right=3pt]{\color{black}\small $x_j$};
\filldraw[fill=white,draw=black] (0.4,1.6) circle (1pt);
\filldraw[fill=white,draw=black] (-0.4,2.4) circle (1pt);

\fill (0,1.2) circle(0.01pt)
node[below=0.05pt]{\color{black} $\oW_{pq}(x_i,x_j)$ or $\ol_{pq}(x_i,x_j)$};

\end{scope}

\begin{scope}[yshift=-55pt]

\draw[-,gray] (-0.4,1.6)--(0.4,1.6)--(0.4,2.4)--(-0.4,2.4)--(-0.4,1.6);
\draw[-,very thick] (-0.4,2.4)--(0.4,1.6);
\fill[white!] (0,1.6) circle (0.01pt)
node[below=-0.5pt]{\color{black}\small $q^{(-)}$};
\fill[white!] (-0.4,2) circle (0.01pt)
node[left=-0.5pt]{\color{black}\small $p^{(-)}$};
\fill[white!] (0,2.4) circle (0.01pt)
node[above=-0.5pt]{\color{black}\small $q^{(+)}$};
\fill[white!] (0.4,2) circle (0.01pt)
node[right=-0.5pt]{\color{black}\small $p^{(+)}$};
\filldraw[fill=black,draw=black] (0.4,1.6) circle (1pt)
node[right=3pt]{\color{black}\small $x_j$};
\filldraw[fill=black,draw=black] (-0.4,2.4) circle (1pt)
node[left=3pt]{\color{black}\small $x_i$};
\filldraw[fill=white,draw=black] (-0.4,1.6) circle (1pt);
\filldraw[fill=white,draw=black] (0.4,2.4) circle (1pt);

\fill (0,1.3) circle(0.01pt)
node[below=0.05pt]{\color{black} $ W_{pq}(x_i,x_j)$ or $\lag_{pq}(x_i,x_j)$};

\begin{scope}[xshift=80pt]
\draw[-,gray] (-0.4,1.6)--(0.4,1.6)--(0.4,2.4)--(-0.4,2.4)--(-0.4,1.6);
\draw[-,very thick] (-0.4,1.6)--(0.4,2.4);
\fill[white!] (0,1.6) circle (0.01pt)
node[below=-0.5pt]{\color{black}\small $q^{(-)}$};
\fill[white!] (-0.4,2) circle (0.01pt)
node[left=-0.5pt]{\color{black}\small $p^{(-)}$};
\fill[white!] (0,2.4) circle (0.01pt)
node[above=-0.5pt]{\color{black}\small $q^{(+)}$};
\fill[white!] (0.4,2) circle (0.01pt)
node[right=-0.5pt]{\color{black}\small $p^{(+)}$};
\filldraw[fill=black,draw=black] (-0.4,1.6) circle (1pt)
node[left=3pt]{\color{black}\small $x_i$};
\filldraw[fill=black,draw=black] (0.4,2.4) circle (1pt)
node[right=3pt]{\color{black}\small $x_j$};
\filldraw[fill=white,draw=black] (0.4,1.6) circle (1pt);
\filldraw[fill=white,draw=black] (-0.4,2.4) circle (1pt);

\fill (0,1.3) circle(0.01pt)
node[below=0.05pt]{\color{black} $\oW_{pq}(x_i,x_j)$ or $\ol_{pq}(x_i,x_j)$};
\end{scope}
\end{scope}

\end{tikzpicture}
\caption{Configurations of directed rapidity lines on elementary squares (above), compared to a signed labelling of edges of elementary squares (below).}
\label{orientsquares}
\end{figure}

Consider now the the square lattice model on $\sigma_0$, defined in Section \ref{sec:sig}, with Boltzmann weights $W_{pq}(x_i,x_j)$, and $\oW_{pq}(x_i,x_j)$.  In the quasi-classical limit \eqref{weight-exp}, the Boltzmann weights on elementary squares, are replaced with Lagrangians $\lag_{pq}(x_i,x_j)$, $\ol_{pq}(x_i,x_j)$, on elementary squares, as depicted in the top half of Figure \ref{orientsquares}.

\begin{figure}[htb]
\centering
\begin{tikzpicture}[scale=1.8]

\draw[-,thick] (-1,-1)--(1,1);\draw[-,thick] (-1,1)--(1,-1);
\draw (-1.25,0.5) circle (0.1pt)
node[left=5pt]{\small $p$};
\draw[-{>[scale=1.5]},dotted] (-1.25,0.5)--(1.25,0.5);
\draw (-1.25,-0.5) circle (0.1pt)
node[left=5pt]{\small $p$};
\draw[-{>[scale=1.5]},dotted] (-1.25,-0.5)--(1.25,-0.5);
\draw (-0.5,-1.25) circle (0.1pt)
node[below=5pt]{\small $q$};
\draw[-{>[scale=1.5]},dotted] (-0.5,-1.25)--(-0.5,1.25);
\draw (0.5,-1.25) circle (0.1pt)
node[below=5pt]{\small $q$};
\draw[-{>[scale=1.5]},dotted] (0.5,-1.25)--(0.5,1.25);
\fill (1,1) circle (2pt)
node[right=5pt]{\small $x_{j_2}$};
\fill (-1,1) circle (2pt)
node[left=5pt]{\small $x_{j_1}$};
\fill (1,-1) circle (2pt)
node[right=5pt]{\small $x_{j_3}$};
\fill (-1,-1) circle (2pt)
node[left=5pt]{\small $x_{j_4}$};
\fill (0,0) circle (2pt)
node[below=5pt]{\small $x_i$};

\end{tikzpicture}
\caption{Edge configuration associated to the classical discrete Laplace equation \eqref{Euler-a}.}
\label{figLaplace}
\end{figure}
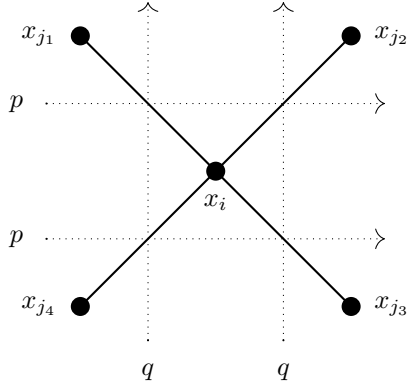

Using the quasi-classical expansion \eqref{weight-exp}, the expression for the partition function \eqref{zdef} may be evaluated with a saddle point method, leading to
\beq
\label{zqcl}
\log Z_0 = -\hbar^{-1}\, \mathcal{A}_0(\x_0^{(cl)})+{O}(\hbar^0)\,,
\eeq
where the classical action functional $\mathcal{A}_0(\x)$, is given by
\beq
\label{caction1}
\mathcal{A}_0(\x)=\sum_{(ij)\in E^{(1)}(L)} \lag_{pq}(x_i,x_j)+\sum_{(ij)\in E^{(2)}(L)} \ol_{pq}(x_i,x_j)\,.
\eeq
The $\x_0^{(cl)}$ solves the following classical equation of motion
\beq
\label{Euler-a}
\frac{\partial}{\partial x_i}\left(\lag_{pq}(x_{j_1},x_i)+\ol_{pq}(x_i,x_{j_2})+\lag_{pq}(x_i,x_{j_3})+\ol_{pq}(x_{j_4},x_i)\right)=0\,,\qquad i\in V_{int}(L)\,,
\eeq
where the $j_1,j_2,j_3,j_4$ are the four neighbouring vertices of $i$, such that $(ij_1),(ij_3)\in E^{(1)}(L)$, and $(ij_2),(ij_4)\in E^{(2)}(L)$, as depicted in Figure \ref{figLaplace}.  Equations \eqref{Euler-a}, comprise a set of constraints on each interior vertex $i\in V_{int}(L)$, under fixed boundary conditions, that may be interpreted as a general form of the classical discrete Laplace equations \cite{MR2467378,BG11}.

Next consider the edge interaction model on $\G$, defined in Section \ref{sec:sigp}.  The quasi-classical expansion of the partition function \eqref{zsig} of this model, is given by
\beq
\log Z=-\hbar^{-1}\mathcal{A}(\x^{(cl)})+O(\hbar^0)\,,	
\eeq
where the classical action functional $\mathcal{A}(\x)$, is
\beq
\label{caction2}
\ds\mathcal{A}(\x)=\sum_{(ij)\in E^{(1)}(\G)} \lag_{p_ip_j}(x_i,x_j)+\sum_{ (ij)\in E^{(2)}(\G)} \ol_{p_ip_j}(x_i,x_j)\,,
\eeq
and the $p_i,p_j\in\{p,q,r\}$, are rapidity variables associated to an edge $(ij)$, as depicted in Figures \ref{strsquare}, and \ref{4BW}.

The $\x^{(cl)}$ is the saddle point of the partition function, given by the solution of the following equation of motion
\beq
\label{Euler-c}
\frac{\partial}{\partial x_i}\mathcal{A}(\x)=0\,,\qquad i\in V_{int}(\G)\,.
\eeq

Equation \eqref{Euler-c} provides a constraint on each interior spin of   $\G$, which should be interpreted as a generalisation of the classical discrete Laplace equations \eqref{Euler-a} on $L$, to the graph $\G$.  Equation \eqref{Euler-c} will generally involve four/five/six/seven-point equations, centered at the different vertices $i\in V_{int}(\G)$.

As seen in Figure \ref{orientsquares}, the directed rapidity graph that was used for the statistical mechanical model in the previous section, is now naturally associated to the classical discrete Laplace equations, defined by \eqref{caction1}, \eqref{Euler-a}, or \eqref{caction2}, \eqref{Euler-c}.  As usual, the directed rapidity lines distinguish two types of edges of a graph, to which two different Lagrangians $\lag_{pq}(x_i,x_j)$, and $\ol_{pq}(x_i,x_j)$, are respectively associated.  This distinguishes the two types of elementary squares shown in the top half of Figure \ref{orientsquares}.  All elementary squares for the classical discrete Laplace system on $\sigma$, are obtained by relabelling/rotating either of these two elementary squares, as is seen for the lattice model in Figure \ref{4BW}.

The rapidity line configuration, may be equivalently replaced with oriented elementary squares having labelled edges, of the type shown on the bottom half of Figure \ref{orientsquares}.  The labelling of edges is signed, such that an edge that is common to two different elementary squares, is labelled $p^{(+)}$ with respect to one square, and $p^{(-)}$ with respect to the other square.  This is required to uniquely determine the Lagrangian as a function of an elementary square, because there are two types of Lagrangians that need to be distinguished, and also the ordering of spins needs to be taken into account ({\it i.e.},  the symmetry $\lag_{pq}(x_i,x_j)=\lag_{pq}(x_j,x_i)$, was not assumed here).

If an oriented elementary square, say $\sigma_{ij}$, has an associated Lagrangian $\lag_{pq}$ (or $\ol_{pq}$), the oppositely oriented elementary square $\sigma_{ji}$, has an associated Lagrangian $-\lag_{pq}$ (or $-\ol_{pq}$).  This is due to the anti-symmetry relations \eqref{clasinversion}, and is the usual case for such classical Lagrangian systems on elementary squares \cite{LN09}.  This is more simpler than the statistical mechanics situation, where the inversion relations \eqref{invrels} are multiplicative rather than additive, which means that a similar correspondence between Boltzmann weights doesn't hold once the orientation of a square is reversed.

Finally, observe that the classical discrete Laplace systems defined by Equations \eqref{caction1}, \eqref{Euler-a}, and \eqref{caction2}, \eqref{Euler-c}, involve the two different Lagrangians functions $\lag_{pq}$, and $\ol_{pq}$.  Usually the classical discrete Laplace equations for ABS systems \cite{MR2467378,BG11}, are defined in terms of only a single Lagrangian $\lag_{pq}$, which is equivalent to the case here when $\ol_{pq}=\lag_{pq}$.  A reason for this is that the building blocks of the lattice model are the two Boltzmann weights $W_{pq}$, and $\oW_{pq}$, and the quasi-classical limit directly results in Laplace equations given in terms of $\lag_{pq}$, and $\ol_{pq}$.  On the other hand, in the classical case the Lagrangians $\lag_{pq}$, and $\ol_{pq}$, are not used as building blocks of an ABS system on a quad graph, but rather their combination in a three-leg form is used.  Then a subsequent restriction of the latter system to a black coloured subgraph, results in classical discrete Laplace equations associated to a single ``long-leg'' Lagrangian function, $\lag_{pq}$.  This is usually also the case for most of the $Q$ equations (except for $Q_2$, and $Q1\delta=1$) that arise from the quasi-classical limit \cite{BKS2}, where the Lagrangians $\lag_{pq}$, and $\ol_{pq}$, are related to each other by the ``crossing symmetry'' property, in which case there is also just one independent Lagrangian function.

Classical discrete Laplace equations corresponding to $H$ equations, which involve two different Lagrangian functions, may also be obtained through a quasi-classical limit, similarly to the example of $H3\,\delta=0$ that was derived in the appendix of \cite{Bazhanov:2015gra}.  These $H$ cases are slightly more complicated, due to a different form of the star-triangle relation \eqref{STR}, and also because the Boltzmann weights will generally have a non-zero imaginary part, which leads to a complex valued saddle point method in the quasi-classical limit.  These cases are currently under investigation.

\subsection{Classical Z-invariance and closure relation}\label{sec:z-invar3}

The statement of classical Z-invariance, is that the action functionals $\mathcal{A}_0$, and $\mathcal{A}$, defined on graphs on $\sigma_0$, and $\sigma$ respectively, are equivalent, as a consequence of the classical star-triangle relation \eqref{cstar}, and inversion relations \eqref{clasinversion}.  In this case the equivalence is exact, due to the forms of the relations \eqref{cstar}, and \eqref{clasinversion}. 

The classical Z-invariance may be shown identically to the statistical mechanics case of the previous section, by the following simple consideration.  Recall that the classical star-triangle relation \eqref{cstar}, holds on solutions of the three-leg equations of motion \eqref{ceqmo}.  By Equation \eqref{Euler-c}, these equations of motion are always satisfied at a vertex that is common to three elementary squares, around which a combination of $p,q,r$ rapidity lines do not form a directed closed path.  This means that all of the deformations in Appendix \ref{app:str} also hold classically, as a consequence of \eqref{cstar}, and \eqref{clasinversion}.  This is enough to show Z-invariance for the system of classical discrete Laplace equations, analogously to that described in Section \ref{sec:z-invar}, for the edge interaction model of statistical mechanics.  

\begin{figure}[h]
\centering
\begin{tikzpicture}

\begin{scope}[scale=1.2]
\draw[-,gray] (-1.9,-1.5)--(-1.5,-1.5)--(-1.7,-1.7);\draw[-,gray] (0.3,-1.7)--(0.5,-1.5)--(0.9,-1.5);\draw[-,gray] (1.2,-0.8)--(1,-1)--(1.4,-1);\draw[-,gray] (1.2,1.2)--(1,1);\draw[-,gray] (-1.4,1)--(-1,1)--(-0.8,1.2);\draw[-,gray] (-1.5,0.5)--(-1.9,0.5);
\draw[-,gray] (-1,-1)--(1,-1)--(1,1)--(-1,1)--(-1,-1)--(-1,1)--(-1.5,0.5)--(-1.5,-1.5)--(-1,-1);
\draw[-,gray] (-1.5,-1.5)--(0.5,-1.5)--(1,-1);
\filldraw[fill=white,draw=black] (1,-1) circle (2.0pt);
\draw[white] (1.2,-1) circle (0.01pt)
node[above=1pt]{\color{black}\small $x_{2}$};
\filldraw[fill=white,draw=black] (-1,1) circle (2.0pt)
node[above=1pt]{\color{black}\small $x_{3}$};
\filldraw[fill=white,draw=black] (-1.5,-1.5) circle (2.0pt)
node[below=1pt]{\color{black}\small $x_{1}$};
\filldraw[fill=black,draw=black] (1,1) circle (2.0pt)
node[right=1.5pt]{\small $x_{23}$};
\filldraw[fill=black,draw=black] (-1,-1) circle (2.0pt)
node[left=-1pt]{\small $x$};
\filldraw[fill=black,draw=black] (-1.5,0.5) circle (2.0pt);
\draw[white] (-1.75,0.5) circle (0.01pt)
node[below=1.5pt]{\color{black}\small $x_{13}$};
\filldraw[fill=black,draw=black] (0.5,-1.5) circle (2.0pt)
node[below=2pt]{\small $x_{12}$};
\draw[-,very thick] (1,1)--(-1,-1)--(-1.5,0.5);
\draw[-,very thick] (-1,-1)--(0.5,-1.5);
\draw[->,thick,dotted] (-1.5,0.75)--(-1.25,0.75)--(-1.25,-1.25)--(1,-1.25);
\draw[<-,thick,dotted] (-1,-2)--(0,-1)--(0,1)--(0.25,1.25);
\draw[->,thick,dotted] (-1.75,-0.5)--(-1.5,-0.5)--(-1,0)--(1,0)--(1.25,0.25);
\draw[white] (-1.5,0.75) circle (0.01pt)
node[left=1.5pt]{\color{black}\small $q$};
\draw[white] (0.25,1.25) circle (0.01pt)
node[right=-1.5pt]{\color{black}\small $p$};
\draw[white] (-1.75,-0.5) circle (0.01pt)
node[left=1.5pt]{\color{black}\small $r$};
\end{scope}

\draw[white] (3.3,0) circle (0.01pt)
node[below=1pt]{\color{black}$=$};

\begin{scope}[scale=1.2,xshift=160,yshift=-15]
\draw[-,gray] (-1.4,-1)--(-1,-1)--(-1.2,-1.2);\draw[-,gray] (0.8,-1.2)--(1,-1)--(1.4,-1);\draw[-,gray] (1.7,-0.3)--(1.5,-0.5)--(1.9,-0.5);\draw[-,gray] (1.7,1.7)--(1.5,1.5);\draw[-,gray] (-0.9,1.5)--(-0.5,1.5)--(-0.3,1.7);\draw[-,gray] (-1,1)--(-1.4,1);
\draw[-,gray] (1.5,1.5)--(1,1)--(-1,1)--(-1,-1)--(1,-1)--(1,1);
\draw[-,gray] (1,-1)--(1.5,-0.5)--(1.5,1.5)--(-0.5,1.5)--(-1,1);
\draw[-,gray,thick,dotted] (-0.5,1.5)--(-0.5,-0.5)--(1.5,-0.5);
\draw[-,gray,thick,dotted] (-1,-1)--(-0.5,-0.5);
\filldraw[fill=white,draw=black] (1,1) circle (2.0pt);
\draw[white] (0.75,1) circle (0.01pt)
node[below=1pt]{\color{black}\small $x_{123}$};
\draw[white] (1.7,-0.5) circle (0.01pt)
node[above=1pt]{\color{black}\small $x_{2}$};
\filldraw[fill=white,draw=black] (-0.5,1.5) circle (2.0pt)
node[above=1pt]{\color{black}\small $x_{3}$};
\filldraw[fill=white,draw=black] (-1,-1) circle (2.0pt)
node[below=1pt]{\color{black}\small $x_{1}$};
\filldraw[fill=black,draw=black] (-1,1) circle (2.0pt);
\draw[white] (-1.0,0.8) circle (0.01pt)
node[left=-2pt]{\color{black}\small $x_{13}$};
\filldraw[fill=black,draw=black] (1,-1) circle (2.0pt)
node[below=1.5pt]{\small $x_{12}$};
\filldraw[fill=black,draw=black] (1.5,1.5) circle (2.0pt)
node[right=1.5pt]{\small $x_{23}$};
\draw[-,very thick] (1,-1)--(-1,1);
\draw[-,very thick] (1,-1)--(1.5,1.5);
\draw[-,very thick] (-1,1)--(1.5,1.5);
\draw[->,thick,dotted] (-1,1.25)--(1.25,1.25)--(1.25,-0.75)--(2.0,-0.75);
\draw[<-,thick,dotted] (-0.5,-1.5)--(0,-1)--(0,1)--(0.75,1.75);
\draw[->,thick,dotted] (-1.25,0)--(1,0)--(1.75,0.75);
\draw[white] (-1,1.25) circle (0.01pt)
node[left=1.5pt]{\color{black}\small $q$};
\draw[white] (0.75,1.75) circle (0.01pt)
node[right=-1.5pt]{\color{black}\small $p$};
\draw[white] (-1.25,0) circle (0.01pt)
node[left=1.5pt]{\color{black}\small $r$};
\end{scope}

\end{tikzpicture}
\caption{Classical star-triangle relation \eqref{cstar} as a ``closure relation'' \eqref{cybe2}.}
\label{cube}
\end{figure}
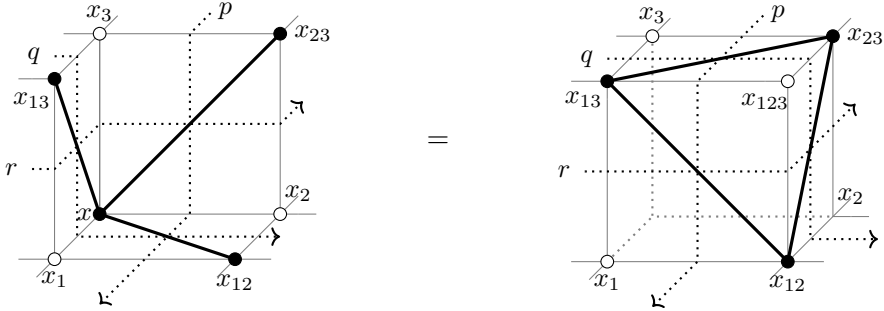

Next it will be shown how the classical Z-invariance, as a consequence of the classical star-triangle relation \eqref{cstar}, may be equivalently interpreted as a closure property \cite{LN09} of the action functionals \eqref{Euler-a}, \eqref{Euler-c}.  This ends up being rather straightforward.  Consider first the configuration of elementary squares shown in Figure \ref{cube} (related to the deformation in Figure \ref{3cube}).  This is a graphical representation of the first expression for the classical star-triangle relation in \eqref{cstar}, which is easily seen after the following change of variables (in \eqref{cstar})
\beq
\ds x_0\to x\,,\quad x_1\to x_{13}\,,\quad x_2\to x_{23}\,,\quad x_3\to x_{12}\,,\quad p\to r\,,\quad r\to p\,.
\eeq

Then note that the first expression for the classical star-triangle relation in Equation \eqref{cstar}, may be written in the equivalent form
\beq
\label{cybe2}
\Delta_i\lag_{pr}(x,x_2,x_{23},x_3)+\Delta_j\lag_{qr}(x,x_3,x_{13},x_1)+\Delta_k\lag_{pq}(x,x_1,x_{12},x_2)=0\,,
\eeq
where
\beq
\begin{array}{l}
\ds\Delta_i\lag_{pr}(x,x_2,x_{23},x_3):=\ol_{pr}(x_{13},x_{12})-\lag_{pr}(x_{23},x)\,,\\[0.3cm]
\ds\Delta_j\lag_{qr}(x,x_3,x_{13},x_1):=\lag_{qr}(x_{23},x_{12})-\ol_{qr}(x_{13},x)\,,\\[0.3cm]
\ds\Delta_k\lag_{pq}(x,x_1,x_{12},x_2):=\lag_{pq}(x_{23},x_{13})-\ol_{pq}(x,x_{12})\,,
\end{array}
\eeq
and where $x$ is required to satisfy the saddle point equation \eqref{ceqmo}.  

In the form \eqref{cybe2}, the classical star-triangle relation \eqref{cstar}, as pictured in Figure \ref{cube}, clearly represents a local closure relation\footnote{Compare the closure relation \eqref{cybe2} (this paper) for two-point Lagrangians, with the closure relation Equation (3.1) (reference \cite{LN09}) for three-point Lagrangians.}, for the classical discrete Laplace system with two-point Lagrangians on diagonals of elementary squares.  Thus the classical Z-invariance for  the action functionals \eqref{caction1}, \eqref{caction2}, may be equivalently interpreted in terms of either of the relations \eqref{cstar}, or \eqref{cybe2}.  Note also that the classical star-triangle relation, was previously used \cite{BS09} in deriving a closure relation, for particular ABS systems that involve three-point Lagrangians on elementary squares.   In comparison, for the classical discrete Laplace systems considered here involving two-point Lagrangians on elementary squares, it is the classical star-triangle relation \eqref{cstar} itself that is naturally identified as the corresponding closure relation \eqref{cybe2}.

\subsection{An example: $Q1\delta=0$}

The results of this section will be illustrated with the example of the cross-ratio equation.  This is known in the ABS classification as $Q1\delta=0$, and is one of the simplest examples of a discrete integrable equation that is obtained in the quasi-classical limit.  The $Q1\delta=0$ equation is known to arise through the quasi-classical limit of the $D=1$ Zamolodchikov fishing-net model, the latter coming from the study of fishing-net diagrams of quantum field theories \cite{Zam-fish}.  The fishing-net model has also been shown to arise in a certain limit of the chiral Potts model \cite{AuYang:1999hd}.  The quasi-classical limit of the fishing-net models for general $D\geq1$ was previously considered in \cite{BKS2}.

In this case the Boltzmann weights depend only on the difference of the two rapidities $p$, and $q$, and so are written in terms of the spectral variable $\alpha=p-q$, as
\beq
W_\alpha(x_i,x_j):=W_{pq}(x_i,x_j)\,\quad \overline{W}_{\alpha}(x_i,x_j):=\overline{W}_{pq}(x_i,x_j)\,.
\eeq
The two Boltzmann weights $W_\alpha(x_i,x_j)$, and $\overline{W}_\alpha(x_i,x_j)$, are also related by the crossing symmetry
\beq
\overline{W}_\alpha(x_i,x_j)=W_{\eta-\alpha}(x_i,x_j)\,,
\eeq
where $\eta=\pi$ is the crossing parameter.  Thus the lattice model may be described in terms of the single Boltzmann weight $W_\alpha(x_i,x_j)$.  The explicit Boltzmann weights are given by
\beq
S(x)=1\,,\qquad W_\alpha(x_i,x_j)=\left|x_i-x_j\right|^{-\alpha/\pi},
\eeq
and satisfy the star-triangle relation
\beq
\label{Q1qstr}
\int_\mathbb{R}dx_0\,W_{\eta-\alpha}(x_1,x_0)W_{\alpha+\beta}(x_2,x_0)W_{\eta-\beta}(x_3,x_0)=R_{\alpha\beta}\,W_\alpha(x_2,x_3)W_{\eta-\alpha-\beta}(x_1,x_3)W_\beta(x_1,x_2)\,,
\eeq
for all $x_1,x_2,x_3\in\mathbb{R}$, and $0<\re(\alpha),\re(\beta)<\eta$.  The factor $R_{\alpha\beta}$ in \eqref{Q1qstr} is given by
\beq
R_{\alpha\beta}=\sqrt{\pi}\,\frac{\Gamma\!\left(\frac{\alpha}{2\pi}\right)\,\Gamma\!\left(\frac{\pi-(\alpha+\beta)}{2\pi}\right)\,\Gamma\!\left(\frac{\beta}{2\pi}\right)}{\Gamma\!\left(\frac{\pi-\alpha}{2\pi}\right)\,\Gamma\!\left(\frac{\alpha+\beta}{2\pi}\right)\,\Gamma\!\left(\frac{\pi-\beta}{2\pi}\right)}\,.
\eeq

The quasi-classical limit involves rescaling the imaginary part of the spectral variables as $\im(\alpha)\to\hbar^{-1}\im(\alpha)$, $\im(\beta)\to\hbar^{-1}\im(\beta)$, and then considering \eqref{Q1qstr} as $\hbar\to0$.  This results in the following Lagrangian at leading order
\beq
\label{q1d0lag}
\lag_\alpha(x,y)=\alpha\log\left|x-y\right|-\frac{\alpha}{2}\log\left|\alpha\right|\,,
\eeq
according to \eqref{weight-exp}.  Note that the asympotics of the factor $R_{\alpha\beta}$, are included in the definition of $\lag_\alpha(x,y)$.  Since in the quasi-classical limit only the imaginary part of $\alpha$ and $\beta$ is scaled by $\hbar^{-1}$, the classical parameter $\alpha$ in \eqref{q1d0lag} should be a real number.

The Lagrangian function \eqref{q1d0lag} satisfies the following classical star-triangle relation
\beq
\label{q1d0cstr}
\lag_{\alpha}(x_1,x^{(cl)}_0)-\lag_{\alpha+\beta}(x_2,x^{(cl)}_0)+\lag_{\beta}(x_3,x^{(cl)}_0)=-\lag_{\alpha}(x_2,x_3)+\lag_{\alpha+\beta}(x_1,x_3)-\lag_{\beta}(x_1,x_2)\,,
\eeq
where $x^{(cl)}_0$ is the solution of the classical equation of motion (3-leg equation centered at $x_0$)
\beq
\label{3legq1d01}
\left.\frac{\partial}{\partial x_0}\left(\lag_{\alpha}(x_1,x_0)-\lag_{\alpha+\beta}(x_2,x_0)+\lag_{\beta}(x_3,x_0)\right)\right|_{x_0=x_0^{(cl)}}\hspace{-0.1cm}=\frac{\alpha}{x_1-x^{(cl)}_0}-\frac{\alpha+\beta}{x_2-x^{(cl)}_0}+\frac{\beta}{x_3-x^{(cl)}_0}=0\,,
\eeq
for fixed values of $x_1,x_2,x_3$, and $\alpha,\beta$.  In this case, after taking the quasi-classical limit, the crossing parameter $\eta\to0$.

The quasi-classical asymptotics of the partition function \eqref{zdef}, are given by \eqref{zqcl}, where
\beq
\label{q1d0action}
\mathcal{A}_0(\x)=\sum_{(ij)\in E^{(1)}(L)}\lag_{\alpha}(x_i,x_j)+\sum_{(ij)\in E^{(2)}(L)}\lag_{-\alpha}(x_i,x_j)\,,
\eeq
and $\x_0^{(cl)}$ solves the system of classical discrete Laplace equations, explicitly given by
\beq
\label{q1d0laplace}
\frac{\partial A_0(\x)}{\partial x_i}=\frac{\alpha}{x_i-x_{j_1}}-\frac{\alpha}{x_i-x_{j_2}}+\frac{\alpha}{x_i-x_{j_3}}-\frac{\alpha}{x_i-x_{j_4}}=0\,,\quad\forall i\in V_{int}(L)\,,
\eeq
with the configuration of edges for \eqref{q1d0laplace} as shown in Figure \ref{figLaplace}, with $p-q=\alpha$.

Equations \eqref{q1d0cstr}, \eqref{3legq1d01}, \eqref{q1d0action}, \eqref{q1d0laplace}, are the main equations for the system of classical discrete Laplace equations corresponding to $Q1\delta=0$.  Following the results of this section, the classical star-triangle relation \eqref{q1d0cstr} implies a closure relation of the form \eqref{cybe2}, along with classical Z-invariance of the action \eqref{q1d0action} on the square lattice, under the deformations given in Appendix \ref{app:str}.  Some other examples of lattice models which lead to Lagrangian functions for $Q$ equations in the quasi-classical limit, were previously considered in \cite{BKS2}.

\section{Conclusion}

In this paper a new formulation of the Z-invariance property \cite{Bax1,Baxter:1986prs} is given, for exactly solved two-dimensional lattice models of statistical mechanics that satisfy the star-triangle relation.  Specifically, in Section \ref{sec:z3model}, it was shown how the integrable square lattice model can be extended onto general planar graphs, with edges connecting a subset of next nearest neighbour vertices of $\mathbb{Z}^3$, whereupon the partition function remains invariant up to simple factors that come from the star-triangle relation \eqref{STR}, and the inversion relations \eqref{invrels}.  This also extends Baxter's original formulation of Z-invariance \cite{Bax1,Baxter:1986prs,Bax2}, by allowing for instances of oriented rapidity lines which form directed closed paths in the rapidity graph, as is depicted in Figure \ref{newd}, and in Figure \ref{1cube}.

The quasi-classical limit was also considered, where an analogous Z-invariance property was shown to be satisfied by the resulting system of classical discrete Laplace equations.  Specifically, it was shown in Section \ref{sec:qcl}, how the quasi-classical expansion of the partition function results in an action functional for the classical discrete Laplace system.  This action functional was shown to be invariant under cubic deformations of the underlying surface of elementary squares, identically to the case of the lattice model of statistical mechanics under Z-invariance.

In Section \ref{sec:qcl}, it was also shown that a classical star-triangle relation for the Laplace system implies a local closure relation \cite{LN09} for the action.  For ABS systems, the closure relation has only previously been studied for particular systems of three-point Lagrangians \cite{LN09,BS09}, which are quite different from the classical discrete Laplace equations considered in this paper, that involve two-point Lagrangian functions.  It would be interesting if such three-point Lagrangian systems, also arise as the quasi-classical limit of a lattice model satisfying the Yang-Baxter equation, perhaps on a triangular lattice.

Finally, note that that the focus of this paper has been exclusively on models that satisfy the star-triangle relation form of the Yang-Baxter equation.  It is expected that the formulation of Z-invariance given here, is also applicable to exactly solved models that satisfy other forms of the Yang-Baxter equation, such as those for the vertex and interaction-round-a-face (IRF) models \cite{Baxterbook,ABF84}.  This may also lead to new forms of classical discrete Laplace systems, with associated classical Yang-Baxter equation, Z-invariance, closure relation {\it etc.}, arising in the quasi-classical limit.  The appearance of Z-invariance for the case of other forms of the Yang-Baxter equation is currently under investigation.

\section*{Acknowledgements}

The author is an overseas researcher under Postdoctoral Fellowship of Japan Society for the Promotion of Science (JSPS).  This work was initiated during a stay at the Technische Univerit\"{a}t Berlin, where the author was supported by DFG Collaborative Research Center TRR 109, ``Discretization in Geometry and Dynamics''.  The author thanks Alexander Bobenko, Raphael Boll, Nikolay Dimitrov, and Yuri Suris, for fruitful discussions during the early stages of the project.

\begin{appendices}
\numberwithin{equation}{section}

\section{Deformations of a square lattice model} \label{app:str}

This appendix uses the notations that were introduced in Section \ref{sec:z3model}.  The deformations given in this appendix, are derived with the use of only the star-triangle relation \eqref{STR}, and the inversion relations \eqref{invrels} (or \eqref{invrel2}).  Note that the deformations in Figure \ref{1cube} (also Figure \ref{newd}) were not previously associated with Z-invariance, and involve the addition of instances of oriented rapidity lines that form closed directed paths.  The deformations in this appendix are used in Section \ref{sec:z-invar}, to show Z-invariance of the partition function for edge interaction models on a general planar graph, with edges connecting a subset of next nearest neighbour vertices of $\mathbb{Z}^3$.  

The deformations here, may be considered to exchange $m$ elementary squares, with $n$ different elementary squares, where $m=1,2,3,4,5$, $n=6-m$, and the union of the $n$ and $m$ squares forms a cube.  Consequently, such deformations will sometimes be referred to as ``cubic flips''.  Note that a deformation that adds (or ``stretches'') a positively oriented rapidity line $r$, always involves translating an elementary square $\sigma_{ij}(\n)$ to $\sigma_{ij}(\n+\khat)$, while a deformation that adds (or stretches) a negatively oriented rapidity line $r$, always involves translating an elementary square $\sigma_{ij}(\n)$ to $\sigma_{ij}(\n-\khat)$.  The orientation of the rapidity line $r$ is chosen here as convention, and the deformations are also valid for the case when all $r$ lines have the opposite orientation.

\subsection{$m=1$ square $\leftrightarrow$ $n=5$ squares}

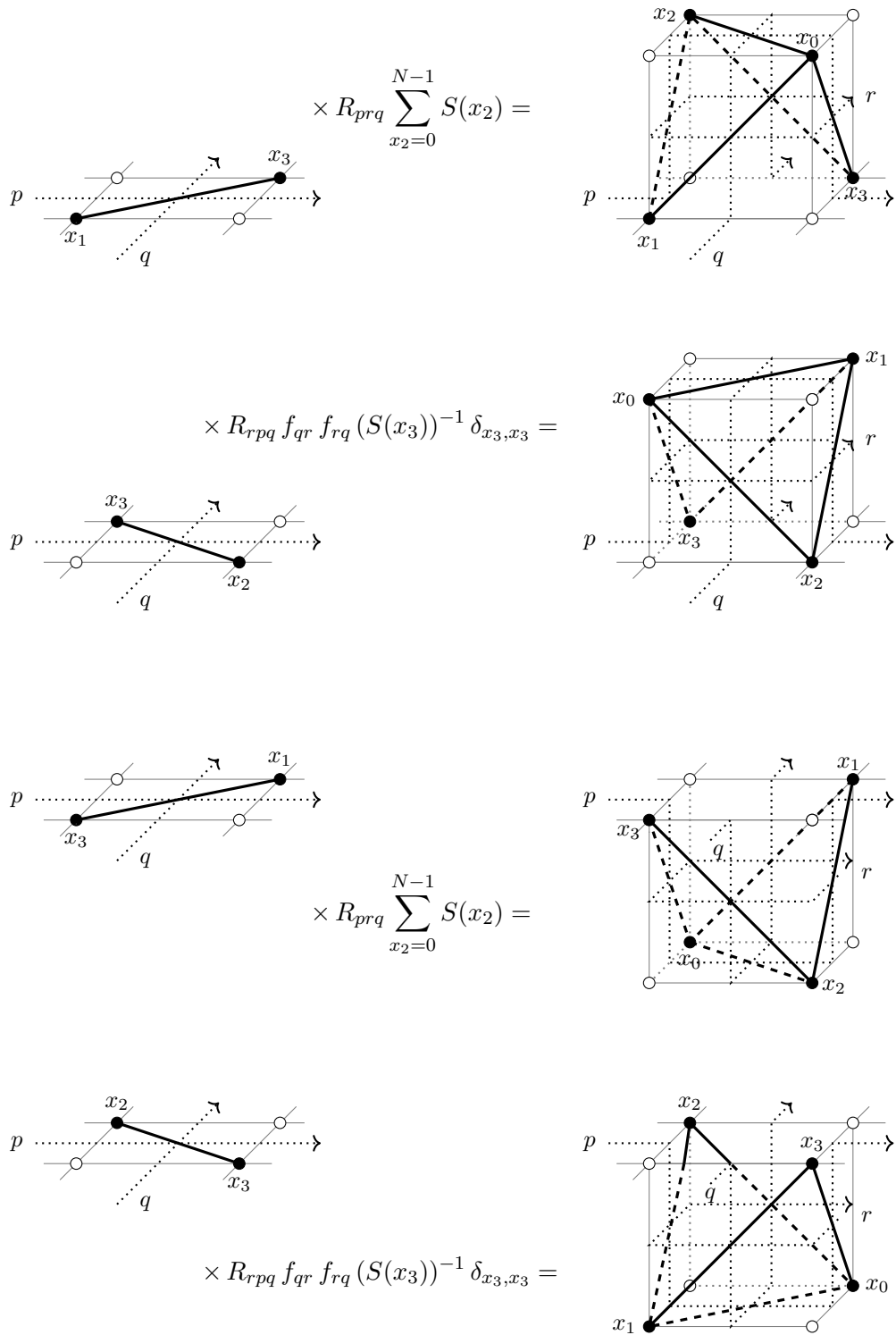
\begin{figure}[htb!]
\centering

\begin{tikzpicture}

\begin{scope}[scale=1.213]

\draw[-,gray] (-1.4,-1)--(1.4,-1);\draw[-,gray] (0.8,-1.2)--(1.7,-0.3);\draw[-,gray,thick,dotted] (1.5,-0.5)--(-0.9,-0.5);\draw[-,gray,thick,dotted] (-0.3,-0.3)--(-1,-1);\draw[-,gray] (-1,-1)--(-1.2,-1.2);
\draw[-,gray] (1.5,1.5)--(1,1)--(-1,1)--(-1,-1)--(1,-1)--(1,1);
\draw[-,gray] (1.9,-0.5)--(1.5,-0.5)--(1.5,1.5)--(-0.5,1.5)--(-1,1);
\draw[-,gray,thick,dotted] (-0.5,1.5)--(-0.5,-0.5);
\filldraw[fill=black,draw=black] (-1,-1) circle (2.0pt);
\draw[circle] (-1,-1.05) circle (0.01pt)
node[below=-2.5pt]{\color{black}\small $x_1$};
\filldraw[fill=black,draw=black] (1,1) circle (2.0pt);
\draw[white] (0.95,1.05) circle (0.01pt)
node[above=-2pt]{\color{black}\small $x_0$};
\filldraw[fill=black,draw=black] (1.5,-0.5) circle (2.0pt);
\draw[white] (1.55,-0.55) circle (0.01pt)
node[below=-1.5pt]{\color{black}\small $x_3$};
\filldraw[fill=black,draw=black] (-0.5,1.5) circle (2.0pt)
node[left=1.5pt]{\small $x_2$};
\filldraw[fill=white,draw=black] (-1,1) circle (2.0pt);
\filldraw[fill=white,draw=black] (1,-1) circle (2.0pt);
\filldraw[fill=white,draw=black] (-0.5,-0.5) circle (2.0pt);
\filldraw[fill=white,draw=black] (1.5,1.5) circle (2.0pt);
\draw[-,very thick] (-1,-1)--(1,1)--(1.5,-0.5);\draw[-,very thick] (-0.5,1.5)--(1,1);
\draw[-,dashed,very thick] (-1,-1)--(-0.5,1.5)--(1.5,-0.5);
\draw[->,thick,dotted] (-1.5,-0.75)--(-0.75,-0.75)--(-0.75,1.25)--(1.25,1.25)--(1.25,-0.75)--(2.0,-0.75);
\draw[->,thick,dotted] (-0.5,-1.5)--(0,-1)--(0,1)--(0.5,1.5)--(0.5,-0.5)--(0.75,-0.25);
\draw[->,thick,dotted] (1.2,0.5)--(-0.5,0.5)--(-1,0)--(1,0)--(1.5,0.5);
\draw[white] (-1.5,-0.75) circle (0.01pt)
node[left=1.5pt]{\color{black}\small $p$};
\draw[white] (-0.15,-1.25) circle (0.01pt)
node[below=1.5pt]{\color{black}\small $q$};
\draw[white] (1.5,0.5) circle (0.01pt)
node[right=1.5pt]{\color{black}\small $r$};

\draw[white] (-3.8,1) circle (0.01pt)
node[below=1pt]{\color{black}$\ds\times\, R_{prq}\sum_{x_2=0}^{N-1} S(x_2)=$};

\begin{scope}[xshift=-200]
\draw[-,gray] (-1.4,-1)--(1.4,-1);\draw[-,gray] (0.8,-1.2)--(1.7,-0.3);\draw[-,gray] (1.8,-0.5)--(-0.9,-0.5);\draw[-,gray] (-0.3,-0.3)--(-1.2,-1.2);
\filldraw[fill=black,draw=black] (-1,-1) circle (2.0pt);
\draw[white] (-1,-1.05) circle (0.01pt)
node[below=0pt]{\color{black}\small $x_1$};
\filldraw[fill=black,draw=black] (1.5,-0.5) circle (2.0pt);
\draw[white] (1.5,-0.45)
node[above=0pt]{\color{black}\small $x_3$};
\filldraw[fill=white,draw=black] (1,-1) circle (2.0pt);
\filldraw[fill=white,draw=black] (-0.5,-0.5) circle (2.0pt);
\draw[-,very thick] (-1,-1)--(1.5,-0.5);
\draw[->,thick,dotted] (-1.5,-0.75)--(2.0,-0.75);
\draw[->,thick,dotted] (-0.5,-1.5)--(0.75,-0.25);
\draw[white] (-1.5,-0.75) circle (0.01pt)
node[left=1.5pt]{\color{black}\small $p$};
\draw[white] (-0.15,-1.25) circle (0.01pt)
node[below=1.5pt]{\color{black}\small $q$};
\end{scope}
\end{scope}

\begin{scope}[scale=1.213,yshift=-120pt]
\draw[-,gray] (-1.4,-1)--(1.4,-1);\draw[-,gray] (0.8,-1.2)--(1.7,-0.3);\draw[-,gray,thick,dotted] (1.5,-0.5)--(-0.9,-0.5);\draw[-,gray,thick,dotted] (-0.3,-0.3)--(-1,-1);\draw[-,gray] (-1,-1)--(-1.2,-1.2);
\draw[-,gray] (1.5,1.5)--(1,1)--(-1,1)--(-1,-1)--(1,-1)--(1,1);
\draw[-,gray] (1.9,-0.5)--(1.5,-0.5)--(1.5,1.5)--(-0.5,1.5)--(-1,1);
\draw[-,gray,thick,dotted] (-0.5,1.5)--(-0.5,-0.5);
\filldraw[fill=white,draw=black] (-1,-1) circle (2.0pt);
\filldraw[fill=white,draw=black] (1,1) circle (2.0pt);
\filldraw[fill=white,draw=black] (1.5,-0.5) circle (2.0pt);
\filldraw[fill=white,draw=black] (-0.5,1.5) circle (2.0pt);
\filldraw[fill=black,draw=black] (-1,1) circle (2.0pt)
node[left=1.5pt]{\small $x_0$};
\filldraw[fill=black,draw=black] (1,-1) circle (2.0pt)
node[below=1.5pt]{\small $x_2$};
\filldraw[fill=black,draw=black] (-0.5,-0.5) circle (2.0pt)
node[below=1.5pt]{\small $x_3$};
\filldraw[fill=black,draw=black] (1.5,1.5) circle (2.0pt)
node[right=1.5pt]{\small $x_1$};
\draw[-,very thick] (-1,1)--(1.5,1.5)--(1,-1)--(-1,1);
\draw[-,dashed, very thick] (-1,1)--(-0.5,-0.5)--(1.5,1.5);
\draw[->,thick,dotted] (-1.5,-0.75)--(-0.75,-0.75)--(-0.75,1.25)--(1.25,1.25)--(1.25,-0.75)--(2.0,-0.75);
\draw[->,thick,dotted] (-0.5,-1.5)--(0,-1)--(0,1)--(0.5,1.5)--(0.5,-0.5)--(0.75,-0.25);
\draw[->,thick,dotted] (1.2,0.5)--(-0.5,0.5)--(-1,0)--(1,0)--(1.5,0.5);
\draw[white] (-1.5,-0.75) circle (0.01pt)
node[left=1.5pt]{\color{black}\small $p$};
\draw[white] (-0.15,-1.25) circle (0.01pt)
node[below=1.5pt]{\color{black}\small $q$};
\draw[white] (1.5,0.5) circle (0.01pt)
node[right=1.5pt]{\color{black}\small $r$};
\filldraw[fill=white,draw=black] (1,1) circle (2.0pt);

\draw[white] (-4.3,1) circle (0.01pt)
node[below=1pt]{\color{black}$\ds\times\, R_{rpq}\,f_{qr}\,f_{rq}\,(S(x_3))^{-1}\,\delta_{x_3,x_3}=$};

\begin{scope}[xshift=-200]
\draw[-,gray] (-1.4,-1)--(1.4,-1);\draw[-,gray] (0.8,-1.2)--(1.7,-0.3);\draw[-,gray] (1.8,-0.5)--(-0.9,-0.5);\draw[-,gray] (-0.3,-0.3)--(-1.2,-1.2);
\filldraw[fill=white,draw=black] (-1,-1) circle (2.0pt);
\filldraw[fill=white,draw=black] (1.5,-0.5) circle (2.0pt);
\filldraw[fill=black,draw=black] (1,-1) circle (2.0pt)
node[below=2pt]{\small $x_2$};
\filldraw[fill=black,draw=black] (-0.5,-0.5) circle (2.0pt)
node[above=1.5pt]{\small $x_3$};
\draw[-,very thick] (1,-1)--(-0.5,-0.5);
\draw[->,thick,dotted] (-1.5,-0.75)--(2.0,-0.75);
\draw[->,thick,dotted] (-0.5,-1.5)--(0.75,-0.25);
\draw[white] (-1.5,-0.75) circle (0.01pt)
node[left=1.5pt]{\color{black}\small $p$};
\draw[white] (-0.15,-1.25) circle (0.01pt)
node[below=1.5pt]{\color{black}\small $q$};
\end{scope}
\end{scope}

\begin{scope}[scale=1.213,yshift=-210pt]
\draw[-,gray] (-1.4,-1)--(1.4,-1);\draw[-,gray] (0.8,-1.2)--(1.7,-0.3);\draw[-,gray] (1.5,-0.5)--(-0.9,-0.5);\draw[-,gray] (-0.3,-0.3)--(-1,-1);\draw[-,gray] (-1,-1)--(-1.2,-1.2);\draw[-,gray] (1.5,-2.5)--(1.5,-0.5)--(1.9,-0.5);
\draw[-,gray] (1.5,-2.5)--(1,-3)--(-1,-3)--(-1,-1)--(1,-1);
\draw[-,gray,thick,dotted](1.5,-2.5)--(-0.5,-2.5)--(-0.5,-0.5);
\draw[-,gray,thick,dotted] (-1,-3)--(-0.5,-2.5);
\filldraw[fill=black,draw=black] (-1,-1) circle (2.0pt);
\draw[circle] (-1,-1.15) circle (0.01pt)
node[left=-2.5pt]{\color{black}\small $x_3$};
\filldraw[fill=black,draw=black] (1,-3) circle (2.0pt);
\draw[white] (0.95,-3.05) circle (0.01pt)
node[right=2pt]{\color{black}\small $x_2$};
\filldraw[fill=black,draw=black] (1.5,-0.5) circle (2.0pt);
\draw[white] (1.45,-0.55) circle (0.01pt)
node[above=1.5pt]{\color{black}\small $x_1$};
\filldraw[fill=black,draw=black] (-0.5,-2.5) circle (2.0pt)
node[below=1.5pt]{\small $x_0$};
\filldraw[fill=white,draw=black] (-1,-3) circle (2.0pt);
\filldraw[fill=white,draw=black] (1,-1) circle (2.0pt);
\filldraw[fill=white,draw=black] (-0.5,-0.5) circle (2.0pt);
\filldraw[fill=white,draw=black] (1.5,-2.5) circle (2.0pt);
\draw[-,very thick] (-1,-1)--(1,-3)--(1.5,-0.5);
\draw[-,dashed,very thick] (1.5,-0.5)--(-0.5,-2.5)--(-1,-1);\draw[-,dashed,very thick] (-0.5,-2.5)--(1,-3);
\draw[->,thick,dotted] (-1.5,-0.75)--(-0.75,-0.75)--(-0.75,-2.75)--(1.25,-2.75)--(1.25,-0.75)--(2.0,-0.75);
\draw[->,thick,dotted] (-0.25,-1.25)--(0,-1)--(0,-3)--(0.5,-2.5)--(0.5,-0.5)--(0.75,-0.25);
\draw[<-,thick,dotted] (1.5,-1.5)--(-0.5,-1.5)--(-1,-2)--(1,-2)--(1.35,-1.65);
\draw[white] (-1.5,-0.75) circle (0.01pt)
node[left=1.5pt]{\color{black}\small $p$};
\draw[white] (-0.15,-1.15) circle (0.01pt)
node[below=1.5pt]{\color{black}\small $q$};
\draw[white] (1.55,-1.65) circle (0.01pt)
node[right=-2pt]{\color{black}\small $r$};
\filldraw[fill=white,draw=black] (1,-1) circle (2.0pt);

\draw[white] (-3.8,-1.5) circle (0.01pt)
node[below=1pt]{\color{black}$\ds\times\, R_{prq}\sum_{x_2=0}^{N-1} S(x_2)=$};

\begin{scope}[xshift=-200]
\draw[-,gray] (-1.4,-1)--(1.4,-1);\draw[-,gray] (0.8,-1.2)--(1.7,-0.3);\draw[-,gray] (1.8,-0.5)--(-0.9,-0.5);\draw[-,gray] (-0.3,-0.3)--(-1.2,-1.2);
\filldraw[fill=black,draw=black] (-1,-1) circle (2.0pt);
\draw[white] (-1,-1.05) circle (0.01pt)
node[below=0pt]{\color{black}\small $x_3$};
\filldraw[fill=black,draw=black] (1.5,-0.5) circle (2.0pt);
\draw[white] (1.5,-0.45)
node[above=0pt]{\color{black}\small $x_1$};
\filldraw[fill=white,draw=black] (1,-1) circle (2.0pt);
\filldraw[fill=white,draw=black] (-0.5,-0.5) circle (2.0pt);
\draw[-,very thick] (-1,-1)--(1.5,-0.5);
\draw[->,thick,dotted] (-1.5,-0.75)--(2.0,-0.75);
\draw[->,thick,dotted] (-0.5,-1.5)--(0.75,-0.25);
\draw[white] (-1.5,-0.75) circle (0.01pt)
node[left=1.5pt]{\color{black}\small $p$};
\draw[white] (-0.15,-1.25) circle (0.01pt)
node[below=1.5pt]{\color{black}\small $q$};
\end{scope}
\end{scope}

\begin{scope}[scale=1.213,yshift=-330pt]
\draw[white] (0,-3.1) circle (0.1pt);

\draw[-,gray] (-1.4,-1)--(1.4,-1);\draw[-,gray] (0.8,-1.2)--(1.7,-0.3);\draw[-,gray] (1.5,-0.5)--(-0.9,-0.5);\draw[-,gray] (-0.3,-0.3)--(-1,-1);\draw[-,gray] (-1,-1)--(-1.2,-1.2);\draw[-,gray] (1.5,-2.5)--(1.5,-0.5)--(1.9,-0.5);
\draw[-,gray] (1.5,-2.5)--(1,-3)--(-1,-3)--(-1,-1)--(1,-1);
\draw[-,gray,thick,dotted](1.5,-2.5)--(-0.5,-2.5)--(-0.5,-0.5);
\draw[-,gray,thick,dotted] (-1,-3)--(-0.5,-2.5);
\filldraw[fill=white,draw=black] (-1,-1) circle (2.0pt);
\filldraw[fill=white,draw=black] (1,-3) circle (2.0pt);
\filldraw[fill=white,draw=black] (1.5,-0.5) circle (2.0pt);
\filldraw[fill=white,draw=black] (-0.5,-2.5) circle (2.0pt);
\filldraw[fill=black,draw=black] (-1,-3) circle (2.0pt)
node[left=1.5pt]{\small $x_1$};
\filldraw[fill=black,draw=black] (1,-1) circle (2.0pt)
node[above=1.5pt]{\small $x_3$};
\filldraw[fill=black,draw=black] (-0.5,-0.5) circle (2.0pt)
node[above=1.5pt]{\small $x_2$};
\filldraw[fill=black,draw=black] (1.5,-2.5) circle (2.0pt)
node[right=1.5pt]{\small $x_0$};
\draw[-,very thick] (-1,-3)--(1,-1)--(1.5,-2.5);\draw[-,very thick] (0,-1)--(-0.5,-0.5)--(-0.572,-1);
\draw[-,dashed,very thick] (-0.572,-1)--(-1,-3)--(1.5,-2.5)--(0,-1);
\draw[->,thick,dotted] (-1.5,-0.75)--(-0.75,-0.75)--(-0.75,-2.75)--(1.25,-2.75)--(1.25,-0.75)--(2.0,-0.75);
\draw[->,thick,dotted] (-0.25,-1.25)--(0,-1)--(0,-3)--(0.5,-2.5)--(0.5,-0.5)--(0.75,-0.25);
\draw[<-,thick,dotted] (1.5,-1.5)--(-0.5,-1.5)--(-1,-2)--(1,-2)--(1.35,-1.65);
\draw[white] (-1.5,-0.75) circle (0.01pt)
node[left=1.5pt]{\color{black}\small $p$};
\draw[white] (-0.25,-1.15) circle (0.01pt)
node[below=1.5pt]{\color{black}\small $q$};
\draw[white] (1.55,-1.65) circle (0.01pt)
node[right=-2pt]{\color{black}\small $r$};

\draw[white] (-4.3,-2) circle (0.01pt)
node[below=1pt]{\color{black}$\ds\times\, R_{rpq}\,f_{qr}\,f_{rq}\,(S(x_3))^{-1}\,\delta_{x_3,x_3}=$};

\begin{scope}[xshift=-200]
\draw[-,gray] (-1.4,-1)--(1.4,-1);\draw[-,gray] (0.8,-1.2)--(1.7,-0.3);\draw[-,gray] (1.8,-0.5)--(-0.9,-0.5);\draw[-,gray] (-0.3,-0.3)--(-1.2,-1.2);
\filldraw[fill=white,draw=black] (-1,-1) circle (2.0pt);
\filldraw[fill=white,draw=black] (1.5,-0.5) circle (2.0pt);
\filldraw[fill=black,draw=black] (1,-1) circle (2.0pt)
node[below=2pt]{\small $x_3$};
\filldraw[fill=black,draw=black] (-0.5,-0.5) circle (2.0pt)
node[above=1.5pt]{\small $x_2$};
\draw[-,very thick] (1,-1)--(-0.5,-0.5);
\draw[->,thick,dotted] (-1.5,-0.75)--(2.0,-0.75);
\draw[->,thick,dotted] (-0.5,-1.5)--(0.75,-0.25);
\draw[white] (-1.5,-0.75) circle (0.01pt)
node[left=1.5pt]{\color{black}\small $p$};
\draw[white] (-0.15,-1.25) circle (0.01pt)
node[below=1.5pt]{\color{black}\small $q$};
\end{scope}
\end{scope}

\end{tikzpicture}
\caption{Exchanging one elementary square, with five elementary squares, corresponding respectively to Equations \eqref{bc1eq}, \eqref{bc2eq}, \eqref{bc3eq}, \eqref{bc4eq}, from top to bottom.}
\label{1cube}
\end{figure}

Four different ways to replace one elementary square, with five elementary squares, are shown in Figure \ref{1cube}.  When flattened onto the plane, these deformations have the form given in Figure \ref{newd}.  These deformations introduce a closed directed rapidity line labelled $r$, on the right hand side of each diagram in Figure \ref{1cube}.  For the top two deformations in Figure \ref{1cube}, this line positively oriented, while for the bottom two deformations in Figure \ref{1cube}, this line is negatively oriented.

For the first deformation in Figure \ref{1cube}, the contribution to the partition function coming from the right hand side is 
\beq
\begin{array}{l}
\ds\sum_{x_0=0}^{N-1}\sum_{x_2=0}^{N-1} S(x_0)\, S(x_2)\, W_{qr}(x_2,x_3)\, W_{rp}(x_2,x_1)\,\oW_{rq}(x_1,x_0)\, W_{pq}(x_2,x_0)\,\oW_{pr}(x_0,x_3) \\[0.4cm]
\ds\phantom{\sum\ldots}=\oW_{pq}(x_1,x_3)\, R_{prq}\sum_{x_2=0}^{N-1} S(x_2)\, W_{qr}(x_2,x_3)\, W_{rp}(x_2,x_1)\, W_{rq}(x_2,x_3)\, W_{pr}(x_2,x_1) \\[0.6cm]
\ds\phantom{\sum\ldots}=\oW_{pq}(x_1,x_3)\left( R_{prq}\sum_{x_2=0}^{N-1} S(x_2)\right).
\end{array}
\label{bc1eq}
\eeq

The first star-triangle relation in \eqref{STR}, and first inversion relation in \eqref{invrels}, have been used in \eqref{bc1eq}, to show that the contribution to the partition function of the left and right hand sides of the first deformation in Figure \ref{1cube}, are equal up to a factor $ R_{prq}\sum_{x_2=0}^{N-1} S(x_2)$.

For the second deformation in Figure \ref{1cube}, the contribution to the partition function coming from the right hand side is
\beq
\begin{array}{l}
\ds\sum_{x_0=0}^{N-1}\sum_{x_1=0}^{N-1} S(x_0)\, S(x_1)\,\oW_{qr}(x_1,x_3)\, W_{pr}(x_1,x_2)\,\oW_{pq}(x_0,x_1)\, W_{rq}(x_0,x_2)\,\oW_{rp}(x_3,x_0) \\[0.4cm]
\ds\phantom{\sum\ldots}= W_{pq}(x_3,x_2)\, R_{rpq}\sum_{x_1=0}^{N-1} S(x_1)\,\oW_{qr}(x_1,x_3)\, W_{pr}(x_1,x_2)\,\oW_{rq}(x_3,x_1)\, W_{rp}(x_1,x_2) \\[0.6cm]
\ds\phantom{\sum\ldots}= W_{pq}(x_3,x_2)\left( R_{rpq}\,f_{qr}\,f_{rq}\,( S(x_3))^{-1}\,\delta_{x_3,x_3}\right).
\end{array}
\label{bc2eq}
\eeq

The second star-triangle relation in \eqref{STR}, and both inversion relations in \eqref{invrels}, have been used in \eqref{bc2eq}, to show that the contribution to the partition function of the left and right hand sides of the second deformation in Figure \ref{1cube}, are equal up to a factor $ R_{rpq}\,f_{qr}\,f_{rq}\,(S(x_3))^{-1}\,\delta_{x_3,x_3}$.

For the third deformation in Figure \ref{1cube}, the contribution to the partition function coming from the right hand side is
\beq
\begin{array}{l}
\ds\sum_{x_0=0}^{N-1}\sum_{x_2=0}^{N-1} S(x_0)\, S(x_2)\, W_{qr}(x_3,x_2)\, W_{rp}(x_1,x_2)\,\oW_{rq}(x_0,x_1)\, W_{pq}(x_0,x_2)\,\oW_{pr}(x_3,x_0) \\[0.4cm]
\ds\phantom{\sum\ldots}=\oW_{pq}(x_3,x_1)\, R_{prq}\sum_{x_2=0}^{N-1} S(x_2)\, W_{qr}(x_3,x_2)\, W_{rp}(x_1,x_2)\, W_{rq}(x_3,x_2)\, W_{pr}(x_1,x_2) \\[0.6cm]
\ds\phantom{\sum\ldots}=\oW_{pq}(x_3,x_1)\left( R_{prq}\sum_{x_2=0}^{N-1} S(x_2)\right).
\end{array}
\label{bc3eq}
\eeq

The second star-triangle relation in \eqref{STR}, and first inversion relation in \eqref{invrels}, have been used in \eqref{bc3eq}, to show that the contribution to the partition function of the left and right hand sides of the third deformation in Figure \ref{1cube}, are equal up to a factor $ R_{prq}\sum_{x_2=0}^{N-1} S(x_2)$.  Note that \eqref{bc3eq}, is equivalent to \eqref{bc1eq}, after exchanging the spins appearing in each Boltzmann weight.

For the fourth deformation in Figure \ref{1cube}, the contribution to the partition function coming from the right hand side is
\beq
\begin{array}{l}
\ds\sum_{x_0=0}^{N-1}\sum_{x_1=0}^{N-1} S(x_0)\, S(x_1)\,\oW_{qr}(x_3,x_1)\, W_{pr}(x_2,x_1)\,\oW_{pq}(x_1,x_0)\, W_{rq}(x_2,x_0)\,\oW_{rp}(x_0,x_3) \\[0.4cm]
\ds\phantom{\sum\ldots}= W_{pq}(x_2,x_3)\, R_{rpq}\sum_{x_1=0}^{N-1} S(x_1)\, \oW_{qr}(x_3,x_1)\, W_{pr}(x_2,x_1)\, \oW_{rq}(x_1,x_3)\, W_{rp}(x_2,x_1) \\[0.6cm]
\ds\phantom{\sum\ldots}= W_{pq}(x_2,x_3)\left( R_{rpq}\,f_{qr}\,f_{rq}\,(S(x_3))^{-1}\,\delta_{x_3,x_3}\right).
\end{array}
\label{bc4eq}
\eeq

The first star-triangle relation in \eqref{STR}, and both inversion relations in \eqref{invrels}, have been used in \eqref{bc4eq}, to show that the contribution to the partition function of the left and right hand sides of the fourth deformation in Figure \ref{1cube}, are equal up to a factor $ R_{rpq}\,f_{qr}\,f_{rq}\,(S(x_3))^{-1}\,\delta_{x_3,x_3}$.  Note that \eqref{bc4eq}, is equivalent to \eqref{bc2eq}, after exchanging the spins appearing in each Boltzmann weight.

Recall that the Z-invariance property, is usually only valid for lattices in which the rapidity graph contains no closed directed paths.  Notice that for the example of the first and third deformations of Figure \ref{1cube}, there exists directed closed paths around vertices with the spin $x_2$, that is formed by the three rapidity lines $p,q,r$.  This means that a deformation involving the vertex with the spin $x_2$ cannot be made directly, since a corresponding star-triangle relation is not satisfied.  Fortunately in these cases the vertex may be added, since it is accompanied by the addition of a vertex with the spin $x_0$, to which the star-triangle relation can be applied.  A consequence of this is the appearance of the constant $\sum_{x_2}{S}(x_2)$.

\subsection{$m=2$ squares $\leftrightarrow$ $n=4$ squares}

The deformations in Figure \ref{1cube}, {\it add} rapidity lines $r$ to the surface.  The remaining deformations given in this and the following sub-appendix {\it stretch} rapidity lines $r$, and are obtained with simpler calculations of the form appearing in \eqref{bc1eq}-\eqref{bc4eq}, and hence wont be written explicitly.  When $r$ is positively oriented, the following deformations involve adding (removing) three elementary squares $\sigma_{ij}(\n+\khat)$, $\sigma_{ik}(\n)$, $\sigma_{jk}(\n+\ihat)$ to $\sigma$, and when $r$ is negatively oriented, the following deformations involve adding (removing) three elementary squares $\sigma_{ij}(\n)$, $\sigma_{ik}(\n+\jhat)$, $\sigma_{jk}(\n)$, to (from) $\sigma$.  This has the effect of stretching (contracting) a rapidity line $r$, that appears on the associated elementary squares.

\begin{figure}[htb!]
\centering
\begin{tikzpicture}

\begin{scope}[scale=1.2]
\draw[-,gray] (-1.2,-1.2)--(-1,-1)--(1,-1)--(1,1)--(-1,1)--(-1,-1)--(-1.4,-1);
\draw[-,gray] (1.5,-0.5)--(1.5,1.5)--(-0.5,1.5)--(-1,1);
\draw[-,gray,thick,dotted] (-0.5,1.5)--(-0.5,-0.5)--(1.5,-0.5);
\draw[-,gray,thick,dotted] (-1,-1)--(-0.5,-0.5);
\draw[-,gray] (1,-1)--(1.5,-0.5);\draw[-,gray] (1,1)--(1.5,1.5);
\draw[-,gray] (0.8,-1.2)--(1,-1)--(3,-1)--(3.5,-0.5)--(1.5,-0.5)--(1.7,-0.3);
\draw[-,gray] (2.8,-1.2)--(3,-1)--(3.4,-1);\draw[-,gray] (3.7,-0.3)--(3.5,-0.5)--(3.9,-0.5);
\filldraw[fill=black,draw=black] (-1,-1) circle (2.0pt);
\filldraw[fill=black,draw=black] (1,1) circle (2.0pt)
node[right=2pt]{\small $x_3$};
\filldraw[fill=black,draw=black] (1.5,-0.5) circle (2.0pt);
\draw[white] (1.55,-0.4) circle (0.1pt)
node[right=2pt]{\color{black}\small $x_1$};
\filldraw[fill=black,draw=black] (-0.5,1.5) circle (2.0pt);
\filldraw[fill=white,draw=black] (-1,1) circle (2.0pt);
\filldraw[fill=white,draw=black] (1,-1) circle (2.0pt);
\filldraw[fill=white,draw=black] (-0.5,-0.5) circle (2.0pt);
\filldraw[fill=white,draw=black] (1.5,1.5) circle (2.0pt);
\draw[-,very thick] (-1,-1)--(1,1)--(-0.5,1.5);\draw[-,very thick] (1,1)--(1.5,-0.5)--(3,-1);
\draw[-,dashed,very thick] (1.5,-0.5)--(-0.5,1.5)--(-1,-1);
\draw[->,thick,dotted] (-1.5,-0.75)--(-0.75,-0.75)--(-0.75,1.25)--(1.25,1.25)--(1.25,-0.75)--(3.5,-0.75);
\draw[->,thick,dotted] (-0.5,-1.5)--(0,-1)--(0,1)--(0.5,1.5)--(0.5,-0.5)--(0.75,-0.25);
\draw[->,thick,dotted] (1.2,0.5)--(-0.5,0.5)--(-1,0)--(1,0)--(1.5,0.5);
\draw[->,thick,dotted] (1.5,-1.5)--(2.75,-0.25);

\filldraw[fill=black,draw=black] (3,-1) circle (2.0pt);
\filldraw[fill=white,draw=black] (3.5,-0.5) circle (2.0pt);
\filldraw[fill=black,draw=black] (3,-1) circle (2.0pt)
node[below=1.5pt]{\color{black}\small $x_2$};
\draw[white] (-1.5,-0.75) circle (0.01pt)
node[left=1.5pt]{\color{black}\small $p$};
\draw[white] (1.5,-1.5) circle (0.01pt)
node[below=1.5pt]{\color{black}\small $q$};
\draw[white] (1.5,0.5) circle (0.01pt)
node[right=1.5pt]{\color{black}\small $r$};
\end{scope}

\draw[white] (5.0,1) circle (0.01pt)
node[below=1pt]{\color{black}$\times\, R_{pqr}=$};

\begin{scope}[scale=1.2,xshift=190]
\draw[-,gray] (-1,1)--(-1,-1)--(1,-1)--(1,1)--(-1,1)--(-0.5,1.5)--(1.5,1.5);
\draw[-,gray,thick,dotted] (-0.5,1.5)--(-0.5,-0.5)--(1.5,-0.5)--(1.5,1.5);
\draw[-,gray,thick,dotted] (-1,-1)--(-0.5,-0.5);
\draw[-,gray,thick,dotted] (1,-1)--(1.5,-0.5)--(3.5,-0.5);
\draw[-,gray] (0.8,-1.2)--(1,-1)--(3,-1)--(3,1)--(1,1)--(1.5,1.5)--(3.5,1.5)--(3,1);
\draw[-,gray] (3,-1)--(3.5,-0.5)--(3.5,1.5);
\draw[-,gray] (-1.4,-1)--(-1,-1)--(-1.2,-1.2);\draw[-,gray] (2.8,-1.2)--(3,-1)--(3.4,-1);\draw[-,gray] (3.7,-0.3)--(3.5,-0.5)--(3.9,-0.5);
\filldraw[fill=black,draw=black] (-1,-1) circle (2.0pt);
\filldraw[fill=black,draw=black] (1,1) circle (2.0pt)
node[above=2pt]{\small $x_3$};
\filldraw[fill=black,draw=black] (1.5,-0.5) circle (2.0pt)
node[below=1.5pt]{\small $x_1$};
\filldraw[fill=black,draw=black] (-0.5,1.5) circle (2.0pt);
\filldraw[fill=white,draw=black] (-1,1) circle (2.0pt);
\filldraw[fill=white,draw=black] (1,-1) circle (2.0pt);
\filldraw[fill=white,draw=black] (-0.5,-0.5) circle (2.0pt);
\filldraw[fill=white,draw=black] (1.5,1.5) circle (2.0pt);
\draw[-,very thick] (-1,-1)--(1,1)--(3,-1)--(3.5,1.5)--(1,1)--(-0.5,1.5);
\draw[-,dashed,very thick] (-1,-1)--(-0.5,1.5)--(1.5,-0.5)--(3.5,1.5);
\draw[->,thick,dotted] (-1.5,-0.75)--(-0.75,-0.75)--(-0.75,1.25)--(3.25,1.25)--(3.25,-0.75)--(3.75,-0.75);
\draw[->,thick,dotted] (-0.5,-1.5)--(0,-1)--(0,1)--(0.5,1.5)--(0.5,-0.5)--(0.75,-0.25);
\draw[->,thick,dotted] (3.2,0.5)--(-0.5,0.5)--(-1,0)--(3,0)--(3.5,0.5);
\draw[->,thick,dotted] (1.5,-1.5)--(2,-1)--(2,1)--(2.5,1.5)--(2.5,-0.5)--(2.75,-0.25);
\filldraw[fill=black,draw=black] (3.5,1.5) circle (2.0pt)
node[right=1.5pt]{\color{black}\small $x_0$};
\filldraw[fill=black,draw=black] (3,-1) circle (2.0pt)
node[below=1.5pt]{\color{black}\small $x_2$};
\filldraw[fill=white,draw=black] (3.5,-0.5) circle (2.0pt);
\filldraw[fill=white,draw=black] (3,1) circle (2.0pt);
\draw[white] (-1.5,-0.75) circle (0.01pt)
node[left=1.5pt]{\color{black}\small $p$};
\draw[white] (1.5,-1.5) circle (0.01pt)
node[below=1.5pt]{\color{black}\small $q$};
\draw[white] (3.5,0.5) circle (0.01pt)
node[right=1.5pt]{\color{black}\small $r$};
\end{scope}

\begin{scope}[yshift=-140pt]
\begin{scope}[scale=1.2]
\draw[-,gray] (-1.2,-1.2)--(-1,-1)--(1,-1)--(1,1)--(-1,1)--(-1,-1)--(-1.4,-1);
\draw[-,gray] (1.5,-0.5)--(1.5,1.5)--(-0.5,1.5)--(-1,1);
\draw[-,gray,thick,dotted] (-0.5,1.5)--(-0.5,-0.5)--(1.5,-0.5);
\draw[-,gray,thick,dotted] (-1,-1)--(-0.5,-0.5);
\draw[-,gray] (1,-1)--(1.5,-0.5);\draw[-,gray] (1,1)--(1.5,1.5);
\draw[-,gray] (0.8,-1.2)--(1,-1)--(3,-1)--(3.5,-0.5)--(1.5,-0.5)--(1.7,-0.3);
\draw[-,gray] (2.8,-1.2)--(3,-1)--(3.4,-1);\draw[-,gray] (3.7,-0.3)--(3.5,-0.5)--(3.9,-0.5);
\filldraw[fill=white,draw=black] (-1,-1) circle (2.0pt);
\filldraw[fill=white,draw=black] (1.5,-0.5) circle (2.0pt);
\filldraw[fill=white,draw=black] (-0.5,1.5) circle (2.0pt);
\filldraw[fill=black,draw=black] (-1,1) circle (2.0pt);
\filldraw[fill=black,draw=black] (1,-1) circle (2.0pt)
node[below=1.5pt]{\small $x_1$};
\filldraw[fill=black,draw=black] (-0.5,-0.5) circle (2.0pt);
\filldraw[fill=black,draw=black] (1.5,1.5) circle (2.0pt)
node[right=1.5pt]{\small $x_2$};
\draw[-,very thick] (1,-1)--(1.5,1.5)--(-1,1)--(1,-1)--(3.5,-0.5);
\draw[-,dashed, very thick] (1.5,1.5)--(-0.5,-0.5)--(-1,1);
\filldraw[fill=white,draw=black] (1,1) circle (2.0pt);
\draw[->,thick,dotted] (-1.5,-0.75)--(-0.75,-0.75)--(-0.75,1.25)--(1.25,1.25)--(1.25,-0.75)--(3.75,-0.75);
\draw[->,thick,dotted] (-0.5,-1.5)--(0,-1)--(0,1)--(0.5,1.5)--(0.5,-0.5)--(0.75,-0.25);
\draw[->,thick,dotted] (1.2,0.5)--(-0.5,0.5)--(-1,0)--(1,0)--(1.5,0.5);
\draw[->,thick,dotted] (1.5,-1.5)--(2.75,-0.25);
\draw[-,gray] (1,-1)--(3,-1)--(3.5,-0.5)--(1.5,-0.5)--(1,-1);
\filldraw[fill=white,draw=black] (3,-1) circle (2.0pt);
\filldraw[fill=black,draw=black] (3.5,-0.5) circle (2.0pt);
\draw[white] (3.6,-0.4) circle (0.1pt)
node[right=1.0pt]{\color{black}\small $x_3$};
\draw[white] (-1.5,-0.75) circle (0.01pt)
node[left=1.5pt]{\color{black}\small $p$};
\draw[white] (1.5,-1.5) circle (0.01pt)
node[below=1.5pt]{\color{black}\small $q$};
\draw[white] (1.5,0.5) circle (0.01pt)
node[right=1.5pt]{\color{black}\small $r$};
\end{scope}

\draw[white] (5.0,1) circle (0.01pt)
node[below=1pt]{\color{black}$\times\, R_{prq}=$};

\begin{scope}[scale=1.2,xshift=190]
\draw[-,gray] (-1,1)--(-1,-1)--(1,-1)--(1,1)--(-1,1)--(-0.5,1.5)--(1.5,1.5);
\draw[-,gray,thick,dotted] (-0.5,1.5)--(-0.5,-0.5)--(1.5,-0.5)--(1.5,1.5);
\draw[-,gray,thick,dotted] (-1,-1)--(-0.5,-0.5);
\draw[-,gray,thick,dotted] (1,-1)--(1.5,-0.5)--(3.5,-0.5);
\draw[-,gray] (0.8,-1.2)--(1,-1)--(3,-1)--(3,1)--(1,1)--(1.5,1.5)--(3.5,1.5)--(3,1);
\draw[-,gray] (3,-1)--(3.5,-0.5)--(3.5,1.5);
\draw[-,gray] (-1.4,-1)--(-1,-1)--(-1.2,-1.2);\draw[-,gray] (2.8,-1.2)--(3,-1)--(3.4,-1);\draw[-,gray] (3.7,-0.3)--(3.5,-0.5)--(3.9,-0.5);
\filldraw[fill=white,draw=black] (-1,-1) circle (2.0pt);
\filldraw[fill=white,draw=black] (1.5,-0.5) circle (2.0pt);
\filldraw[fill=white,draw=black] (-0.5,1.5) circle (2.0pt);
\filldraw[fill=black,draw=black] (-1,1) circle (2.0pt);
\filldraw[fill=black,draw=black] (1,-1) circle (2.0pt)
node[below=1.5pt]{\small $x_1$};
\filldraw[fill=black,draw=black] (-0.5,-0.5) circle (2.0pt);
\filldraw[fill=black,draw=black] (1.5,1.5) circle (2.0pt)
node[above=1.5pt]{\small $x_2$};
\draw[-,very thick] (-1,1)--(1,-1)--(3,1)--(1.5,1.5)--(-1,1);\draw[-,very thick] (3,1)--(3.5,-0.5);
\draw[-,dashed,very thick] (-1,1)--(-0.5,-0.5)--(1.5,1.5)--(3.5,-0.5);
\draw[->,thick,dotted] (-1.5,-0.75)--(-0.75,-0.75)--(-0.75,1.25)--(3.25,1.25)--(3.25,-0.75)--(3.75,-0.75);
\draw[->,thick,dotted] (-0.5,-1.5)--(0,-1)--(0,1)--(0.5,1.5)--(0.5,-0.5)--(0.75,-0.25);
\draw[->,thick,dotted] (3.2,0.5)--(-0.5,0.5)--(-1,0)--(3,0)--(3.5,0.5);
\draw[->,thick,dotted] (1.5,-1.5)--(2,-1)--(2,1)--(2.5,1.5)--(2.5,-0.5)--(2.75,-0.25);
\filldraw[fill=white,draw=black] (3.5,1.5) circle (2.0pt);
\filldraw[fill=white,draw=black] (3,-1) circle (2.0pt);
\filldraw[fill=black,draw=black] (3.5,-0.5) circle (2.0pt);
\draw[white] (3.6,-0.4) circle (0.1pt)
node[right=1.0pt]{\color{black}\small $x_3$};
\filldraw[fill=black,draw=black] (3,1) circle (2.0pt)
node[right=2.0pt]{\small $x_0$};
\filldraw[fill=white,draw=black] (1,1) circle (2.0pt);
\draw[white] (-1.5,-0.75) circle (0.01pt)
node[left=1.5pt]{\color{black}\small $p$};
\draw[white] (1.5,-1.5) circle (0.01pt)
node[below=1.5pt]{\color{black}\small $q$};
\draw[white] (3.5,0.5) circle (0.01pt)
node[right=1.5pt]{\color{black}\small $r$};
\end{scope}
\end{scope}

\end{tikzpicture}
\caption{Two types of deformations, that exchange two elementary squares with four elementary squares.}
\label{2cube}
\end{figure}
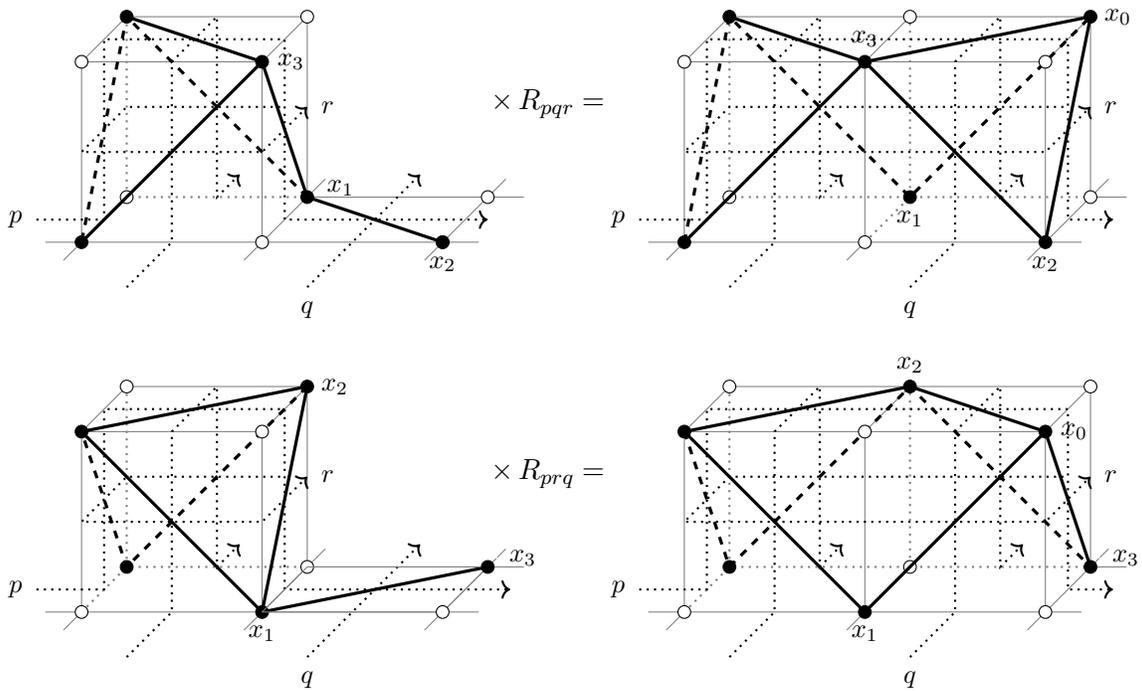

From the right hand sides of the deformations in Figure \ref{1cube}, there are eight different ways to deform the surface in the manner depicted in Figure \ref{2cube}, through the use of the star-triangle relation \eqref{STR}, and inversion relations \eqref{invrels}.

The deformations in Figure \ref{2cube}, are for positively oriented rapidity lines $r$.  In addition to Figure \ref{2cube}, two more deformations for positively oriented rapidity lines $r$, may be obtained by reversing the orientation of the rapidity line $p$, and then exchanging $p\leftrightarrow q$.

The corresponding four deformations for when $r$ is negatively oriented, are obtained from the four deformations that involve a positively oriented rapidity line $r$, by exchanging the spins appearing in each Boltzmann weight.  This gives the eight ``$2\leftrightarrow4$'' deformations that are used in this paper.

\subsection{$m=3$ squares $\leftrightarrow$ $n=3$ squares}

\begin{figure}[htb!]
\centering
\begin{tikzpicture}

\begin{scope}[scale=1.2]
\draw[-,gray] (-1.9,-1.5)--(-1.5,-1.5)--(-1.7,-1.7);\draw[-,gray] (0.3,-1.7)--(0.5,-1.5)--(0.9,-1.5);\draw[-,gray] (1.2,-0.8)--(1,-1)--(1.4,-1);\draw[-,gray] (1.2,1.2)--(1,1);\draw[-,gray] (-1.4,1)--(-1,1)--(-0.8,1.2);\draw[-,gray] (-1.5,0.5)--(-1.9,0.5);
\draw[-,gray] (-1,-1)--(1,-1)--(1,1)--(-1,1)--(-1,-1)--(-1,1)--(-1.5,0.5)--(-1.5,-1.5)--(-1,-1);
\draw[-,gray] (-1.5,-1.5)--(0.5,-1.5)--(1,-1);
\filldraw[fill=white,draw=black] (1,-1) circle (2.0pt);
\filldraw[fill=white,draw=black] (-1,1) circle (2.0pt);
\filldraw[fill=white,draw=black] (-1.5,-1.5) circle (2.0pt);
\filldraw[fill=black,draw=black] (1,1) circle (2.0pt)
node[above=1.5pt]{\small $x_1$};
\filldraw[fill=black,draw=black] (-1,-1) circle (2.0pt)
node[left=1pt]{\small $x_0$};
\filldraw[fill=black,draw=black] (-1.5,0.5) circle (2.0pt);
\draw[white] (-1.65,0.45) circle (0.01pt)
node[below=1.5pt]{\color{black}\small $x_3$};
\filldraw[fill=black,draw=black] (0.5,-1.5) circle (2.0pt)
node[below=2pt]{\small $x_2$};
\draw[-,very thick] (1,1)--(-1,-1)--(-1.5,0.5);
\draw[-,very thick] (-1,-1)--(0.5,-1.5);
\draw[->,thick,dotted] (-1.5,0.75)--(-1.25,0.75)--(-1.25,-1.25)--(1,-1.25);
\draw[->,thick,dotted] (-1,-2)--(0,-1)--(0,1)--(0.25,1.25);
\draw[->,thick,dotted] (-1.75,-0.5)--(-1.5,-0.5)--(-1,0)--(1,0)--(1.25,0.25);
\draw[white] (-1.5,0.75) circle (0.01pt)
node[left=1.5pt]{\color{black}\small $p$};
\draw[white] (-1,-2) circle (0.01pt)
node[below=1.5pt]{\color{black}\small $q$};
\draw[white] (-1.75,-0.5) circle (0.01pt)
node[left=1.5pt]{\color{black}\small $r$};
\end{scope}

\draw[white] (3.3,0) circle (0.01pt)
node[below=1pt]{\color{black}$= R_{prq}\,\times$};

\begin{scope}[scale=1.2,xshift=160,yshift=-15]
\draw[-,gray] (-1.4,-1)--(-1,-1)--(-1.2,-1.2);\draw[-,gray] (0.8,-1.2)--(1,-1)--(1.4,-1);\draw[-,gray] (1.7,-0.3)--(1.5,-0.5)--(1.9,-0.5);\draw[-,gray] (1.7,1.7)--(1.5,1.5);\draw[-,gray] (-0.9,1.5)--(-0.5,1.5)--(-0.3,1.7);\draw[-,gray] (-1,1)--(-1.4,1);
\draw[-,gray] (1.5,1.5)--(1,1)--(-1,1)--(-1,-1)--(1,-1)--(1,1);
\draw[-,gray] (1,-1)--(1.5,-0.5)--(1.5,1.5)--(-0.5,1.5)--(-1,1);
\draw[-,gray,thick,dotted] (-0.5,1.5)--(-0.5,-0.5)--(1.5,-0.5);
\draw[-,gray,thick,dotted] (-1,-1)--(-0.5,-0.5);
\filldraw[fill=white,draw=black] (-1,-1) circle (2.0pt);
\filldraw[fill=white,draw=black] (1,1) circle (2.0pt);
\filldraw[fill=white,draw=black] (1.5,-0.5) circle (2.0pt);
\filldraw[fill=white,draw=black] (-0.5,1.5) circle (2.0pt);
\filldraw[fill=black,draw=black] (-1,1) circle (2.0pt);
\draw[white] (-1.05,0.8) circle (0.01pt)
node[left=-2pt]{\color{black}\small $x_3$};
\filldraw[fill=black,draw=black] (1,-1) circle (2.0pt)
node[below=1.5pt]{\small $x_2$};
\filldraw[fill=black,draw=black] (1.5,1.5) circle (2.0pt)
node[right=1.5pt]{\small $x_1$};
\draw[-,very thick] (1,-1)--(-1,1);
\draw[-,very thick] (1,-1)--(1.5,1.5);
\draw[-,very thick] (-1,1)--(1.5,1.5);
\draw[->,thick,dotted] (-1,1.25)--(1.25,1.25)--(1.25,-0.75)--(2.0,-0.75);
\draw[->,thick,dotted] (-0.5,-1.5)--(0,-1)--(0,1)--(0.75,1.75);
\draw[->,thick,dotted] (-1.25,0)--(1,0)--(1.75,0.75);
\draw[white] (-1,1.25) circle (0.01pt)
node[left=1.5pt]{\color{black}\small $p$};
\draw[white] (-0.5,-1.5) circle (0.01pt)
node[below=1.5pt]{\color{black}\small $q$};
\draw[white] (-1.25,0) circle (0.01pt)
node[left=1.5pt]{\color{black}\small $r$};
\end{scope}

\begin{scope}[yshift=140pt]
\begin{scope}[scale=1.2]
\draw[-,gray] (-1.9,-1.5)--(-1.5,-1.5)--(-1.7,-1.7);\draw[-,gray] (0.3,-1.7)--(0.5,-1.5)--(0.9,-1.5);\draw[-,gray] (1.2,-0.8)--(1,-1)--(1.4,-1);\draw[-,gray] (1.2,1.2)--(1,1);\draw[-,gray] (-1.4,1)--(-1,1)--(-0.8,1.2);\draw[-,gray] (-1.5,0.5)--(-1.9,0.5);
\draw[-,gray] (-1,-1)--(1,-1)--(1,1)--(-1,1)--(-1,-1)--(-1,1)--(-1.5,0.5)--(-1.5,-1.5)--(-1,-1);
\draw[-,gray] (-1.5,-1.5)--(0.5,-1.5)--(1,-1);
\filldraw[fill=black,draw=black] (1,-1) circle (2.0pt);
\draw[white] (1.1,-0.9) circle (0.1pt)
node[right=1pt]{\color{black}\small $x_3$};
\filldraw[fill=black,draw=black] (-1,1) circle (2.0pt)
node[above=1.5pt]{\small $x_2$};
\filldraw[fill=black,draw=black] (-1.5,-1.5) circle (2.0pt)
node[below=1.5pt]{\small $x_1$};
\filldraw[fill=white,draw=black] (1,1) circle (2.0pt);
\filldraw[fill=white,draw=black] (-1,-1) circle (2.0pt);
\filldraw[fill=white,draw=black] (-1.5,0.5) circle (2.0pt);
\filldraw[fill=white,draw=black] (0.5,-1.5) circle (2.0pt);
\draw[-,very thick] (-1.5,-1.5)--(-1,1);
\draw[-,very thick] (-1,1)--(1,-1);
\draw[-,very thick] (1,-1)--(-1.5,-1.5);
\draw[->,thick,dotted] (-1.5,0.75)--(-1.25,0.75)--(-1.25,-1.25)--(1,-1.25);
\draw[->,thick,dotted] (-1,-2)--(0,-1)--(0,1)--(0.25,1.25);
\draw[->,thick,dotted] (-1.75,-0.5)--(-1.5,-0.5)--(-1,0)--(1,0)--(1.25,0.25);
\draw[white] (-1.5,0.75) circle (0.01pt)
node[left=1.5pt]{\color{black}\small $p$};
\draw[white] (-1,-2) circle (0.01pt)
node[below=1.5pt]{\color{black}\small $q$};
\draw[white] (-1.75,-0.5) circle (0.01pt)
node[left=1.5pt]{\color{black}\small $r$};
\end{scope}

\draw[white] (3.3,0) circle (0.01pt)
node[below=1pt]{\color{black}$\times\, R_{prq}=$};

\begin{scope}[scale=1.2,xshift=160,yshift=-10]
\draw[-,gray] (-1.4,-1)--(-1,-1)--(-1.2,-1.2);\draw[-,gray] (0.8,-1.2)--(1,-1)--(1.4,-1);\draw[-,gray] (1.7,-0.3)--(1.5,-0.5)--(1.9,-0.5);\draw[-,gray] (1.7,1.7)--(1.5,1.5);\draw[-,gray] (-0.9,1.5)--(-0.5,1.5)--(-0.3,1.7);\draw[-,gray] (-1,1)--(-1.4,1);
\draw[-,gray] (1.5,1.5)--(1,1)--(-1,1)--(-1,-1)--(1,-1)--(1,1);
\draw[-,gray] (1,-1)--(1.5,-0.5)--(1.5,1.5)--(-0.5,1.5)--(-1,1);
\draw[-,gray,thick,dotted] (-0.5,1.5)--(-0.5,-0.5)--(1.5,-0.5);
\draw[-,gray,thick,dotted] (-1,-1)--(-0.5,-0.5);
\filldraw[fill=black,draw=black] (-1,-1) circle (2.0pt)
node[below=1.5pt]{\small $x_1$};
\filldraw[fill=black,draw=black] (1,1) circle (2.0pt)
node[right=0pt]{\small $x_0$};
\filldraw[fill=black,draw=black] (1.5,-0.5) circle (2.0pt);
\draw[white] (1.6,-0.4) circle (0.1pt)
node[right=1pt]{\color{black}\small $x_3$};
\filldraw[fill=black,draw=black] (-0.5,1.5) circle (2.0pt)
node[above=1.5pt]{\small $x_2$};
\filldraw[fill=white,draw=black] (-1,1) circle (2.0pt);
\filldraw[fill=white,draw=black] (1,-1) circle (2.0pt);
\filldraw[fill=white,draw=black] (1.5,1.5) circle (2.0pt);
\draw[-,very thick] (1,1)--(-1,-1);
\draw[-,very thick] (1,1)--(1.5,-0.5);
\draw[-,very thick] (1,1)--(-0.5,1.5);
\draw[->,thick,dotted] (-1,1.25)--(1.25,1.25)--(1.25,-0.75)--(2.0,-0.75);
\draw[->,thick,dotted] (-0.5,-1.5)--(0,-1)--(0,1)--(0.75,1.75);
\draw[->,thick,dotted] (-1.25,0)--(1,0)--(1.75,0.75);
\draw[white] (-1,1.25) circle (0.01pt)
node[left=1.5pt]{\color{black}\small $p$};
\draw[white] (-0.5,-1.5) circle (0.01pt)
node[below=1.5pt]{\color{black}\small $q$};
\draw[white] (-1.25,0) circle (0.01pt)
node[left=1.5pt]{\color{black}\small $r$};
\end{scope}
\end{scope}

\end{tikzpicture}
\caption{Exchanging three elementary squares, with three elementary squares, using the star-triangle relation \eqref{STR}.}
\label{3cube}
\end{figure}
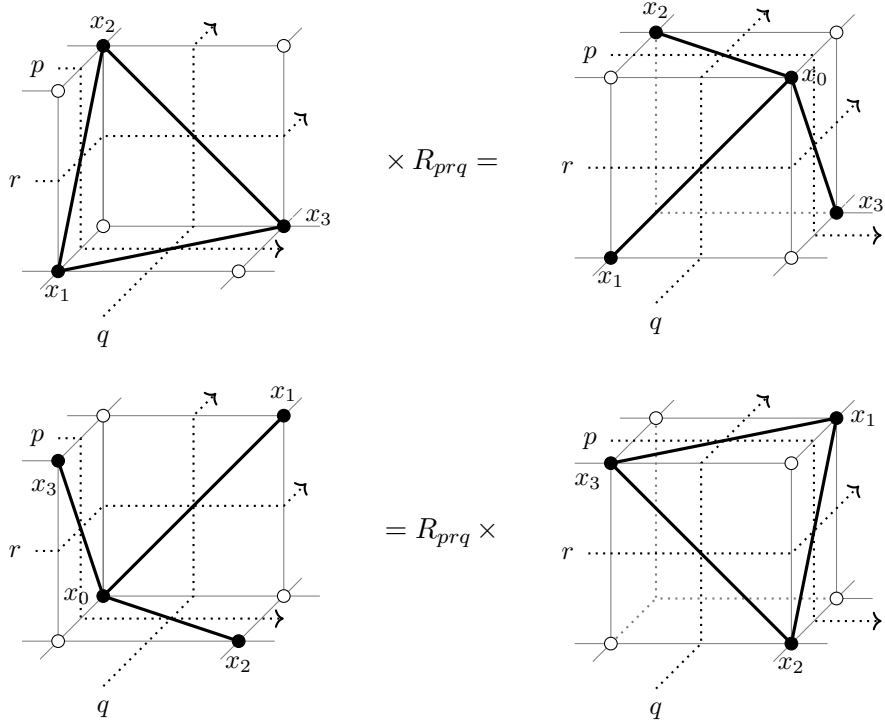

Figure \ref{3cube} depicts two types of deformations, that exchange 3 elementary squares with 3 other elementary squares.  These deformations involve a direct application of the star-triangle relation \eqref{STR}, and can be applied when $r$ is either positively or negatively oriented.  This gives the four ``$3\leftrightarrow3$'' deformations that are used in this paper.

\subsection{A star-triangle type relation arising from deformations}

By equating different configurations of Boltzmann weights that arise from the deformations, it is possible to obtain summation formulas.  Sometimes the resulting formulas are not obvious, particularly when equating two similar configurations that cannot be deformed into each another in a simple way, due to the non-commutativity of the deformations.

\begin{figure}[htb!]
\centering
\begin{tikzpicture}

\begin{scope}[scale=1.8]

\draw[white!] (1,2.1) circle (0.01pt)
node[above=0.1pt,rotate=45]{\color{black}\huge $\Leftarrow$};

\draw[-,gray] (-1.5,-0.5)--(-1.5,0.5)--(-0.5,0.5)--(-0.5,-0.5);\draw[-,gray] (-0.5,0.5)--(0,1)--(0,0);\draw[-,gray] (0,1)--(1,1)--(1,0);\draw[-,gray] (1,1)--(1.5,1.5)--(1.5,0.5)--(1,0);\draw[-,gray](1.5,1.5)--(0.5,1.5)--(0,1)--(-1,1)--(-1.5,0.5);\draw[-,gray] (-1.5,-0.5)--(0.5,-0.5)--(1,0)--(0,0)--(-0.5,-0.5);
\draw[-,very thick] (-1.5,-0.5)--(-0.5,0.5)--(-1,1);\draw[-,very thick] (-0.5,0.5)--(0,0)--(0.5,-0.5);\draw[-,very thick] (0,0)--(1,1)--(0.5,1.5);\draw[-,very thick] (1,1)--(1.5,0.5);
\draw[<-,thick,dotted] (1.5,1)--(1,0.5)--(0,0.5)--(-0.5,0)--(-1.5,0)
node[left=2pt]{\color{black}\small $r$};
\draw[->,thick,dotted] (-0.5,1)--(-1,0.5)--(-1,-0.5)--(-1.25,-0.75);
\draw[->,thick,dotted] (1,1.5)--(0.5,1)--(0.5,0)--(-0.25,-0.75);
\draw[->,thick,dotted] (1.6,0.25)--(1.25,0.25)--(1.25,1.25)--(0.25,1.25);
\draw[->,thick,dotted] (1.1,-0.25)--(-0.25,-0.25)--(-0.25,0.75)--(-1.25,0.75);
\draw[white!] (-0.5,1) circle (0.01pt)
node[above=0.1pt]{\color{black}\small $q$};
\draw[white!] (1,1.5) circle (0.01pt)
node[above=0.1pt]{\color{black}\small $q$};
\draw[white!] (1.1,-0.25) circle (0.01pt)
node[right=0.1pt]{\color{black}\small $p$};
\draw[white!] (1.6,0.25) circle (0.01pt)
node[right=0.1pt]{\color{black}\small $p$};
\filldraw[fill=black,draw=black] (0,0) circle (1.4pt)
node[below=2pt]{\color{black}\small $x_2$};
\filldraw[fill=black,draw=black] (-1.5,-0.5) circle (1.4pt)
node[below=2pt]{\color{black}\small $x_4$};
\filldraw[fill=black,draw=black] (-0.5,0.5) circle (1.4pt)
node[above=1.5pt]{\color{black}\small $x'$};
\filldraw[fill=black,draw=black] (-1,1) circle (1.4pt)
node[left=2pt]{\color{black}\small $x'''$}; 
\filldraw[fill=black,draw=black] (1.5,0.5) circle (1.4pt)
node[right=2pt]{\color{black}\small $x_1$};
\filldraw[fill=black,draw=black] (1,1) circle (1.4pt)
node[above=2pt]{\color{black}\small $x$};
\filldraw[fill=black,draw=black] (0.5,1.5) circle (1.4pt)
node[left=2pt]{\color{black}\small $x''$};
\filldraw[fill=black,draw=black] (0.5,-0.5) circle (1.4pt)
node[below=2pt]{\color{black}\small $x_3$};
\end{scope}

\begin{scope}[scale=1.8,xshift=145]

\draw[white!] (-0.3,2.1) circle (0.01pt)
node[above=0.1pt,rotate=-45]{\color{black}\huge $\Rightarrow$};

\draw[-,gray] (-1.5,-0.5)--(-1.5,0.5)--(-0.5,0.5)--(-0.5,-0.5);\draw[-,gray] (-0.5,0.5)--(0.5,0.5)--(0.5,-0.5);\draw[-,gray] (0.5,0.5)--(1,1)--(1,0);\draw[-,gray] (1,1)--(1.5,1.5)--(1.5,0.5);\draw[-,gray](1.5,1.5)--(0.5,1.5)--(0,1)--(-1,1)--(-1.5,0.5);\draw[-,gray](-1.5,-0.5)--(0.5,-0.5)--(1.5,0.5);\draw[-,gray] (-0.5,0.5)--(0,1)--(1,1);
\draw[-,very thick] (-1.5,-0.5)--(-0.5,0.5)--(-1,1);\draw[-,very thick] (-0.5,0.5)--(0.5,-0.5)--(1,1)--(-0.5,0.5);\draw[-,very thick] (0.5,1.5)--(1,1)--(1.5,0.5);
\draw[<-,thick,dotted] (1.5,1)--(0.5,0)--(-1.5,0)
node[left=2pt]{\color{black}\small $r$};
\draw[->,thick,dotted] (-0.5,1)--(-1,0.5)--(-1,-0.5)--(-1.25,-0.75);
\draw[->,thick,dotted] (1,1.5)--(0,0.5)--(0,-0.5)--(-0.25,-0.75);
\draw[->,thick,dotted] (1.6,0.25)--(1.25,0.25)--(1.25,1.25)--(0.25,1.25);
\draw[->,thick,dotted] (1.1,-0.25)--(0.75,-0.25)--(0.75,0.75)--(-1.25,0.75);
\draw[white!] (-0.5,1) circle (0.01pt)
node[above=0.1pt]{\color{black}\small $q$};
\draw[white!] (1,1.5) circle (0.01pt)
node[above=0.1pt]{\color{black}\small $q$};
\draw[white!] (1.1,-0.25) circle (0.01pt)
node[right=0.1pt]{\color{black}\small $p$};
\draw[white!] (1.6,0.25) circle (0.01pt)
node[right=0.1pt]{\color{black}\small $p$};
\filldraw[fill=black,draw=black] (-1.5,-0.5) circle (1.4pt)
node[below=2pt]{\color{black}\small $x_4$};
\filldraw[fill=black,draw=black] (-0.5,0.5) circle (1.4pt)
node[above=1.5pt]{\color{black}\small $x'$};
\filldraw[fill=black,draw=black] (-1,1) circle (1.4pt)
node[left=2pt]{\color{black}\small $x'''$};
\filldraw[fill=black,draw=black] (1.5,0.5) circle (1.4pt)
node[right=2pt]{\color{black}\small $x_1$};
\filldraw[fill=black,draw=black] (1,1) circle (1.4pt)
node[above=2pt]{\color{black}\small $x$};
\filldraw[fill=black,draw=black] (0.5,1.5) circle (1.4pt)
node[left=2pt]{\color{black}\small $x''$};
\filldraw[fill=black,draw=black] (0.5,-0.5) circle (1.4pt)
node[below=2pt]{\color{black}\small $x_3$};
\end{scope}

\begin{scope}[scale=1.8,xshift=95,yshift=90]
\draw[-,gray] (-1.5,-0.5)--(-0.5,0.5)--(1.5,0.5)--(0.5,-0.5)--(-1.5,-0.5);
\draw[-,gray] (-0.5,-0.5)--(0.5,0.5);\draw[-,gray] (-1,0)--(1,0);
\draw[-,very thick] (-1.5,-0.5)--(1.5,0.5);\draw[-,very thick] (-0.5,0.5)--(0.5,-0.5);
\draw[<-,thick,dotted] (-1.25,-0.75)--(0.25,0.75);\draw[<-,thick,dotted] (-0.25,-0.75)--(1.25,0.75);
\draw[<-,thick,dotted] (-1.6,-0.25)--(1.1,-0.25);\draw[<-,thick,dotted] (-1.1,0.25)--(1.6,0.25);
\draw[white!] (0.25,0.75) circle (0.01pt)
node[above=0.1pt]{\color{black}\small $q$};
\draw[white!] (1.25,0.75) circle (0.01pt)
node[above=0.1pt]{\color{black}\small $q$};
\draw[white!] (1.1,-0.25) circle (0.01pt)
node[right=0.1pt]{\color{black}\small $p$};
\draw[white!] (1.6,0.25) circle (0.01pt)
node[right=0.1pt]{\color{black}\small $p$};
\filldraw[fill=black,draw=black] (0,0) circle (1.4pt)
node[below=2pt]{\color{black}\small $x_2$};
\filldraw[fill=black,draw=black] (-1.5,-0.5) circle (1.4pt)
node[below=2pt]{\color{black}\small $x_4$};
\filldraw[fill=black,draw=black] (-0.5,0.5) circle (1.4pt);
\filldraw[fill=black,draw=black] (1.5,0.5) circle (1.4pt)
node[right=2pt]{\color{black}\small $x_1$};
\filldraw[fill=black,draw=black] (0.5,-0.5) circle (1.4pt)
node[below=2pt]{\color{black}\small $x_3$};
\end{scope}

\end{tikzpicture}
\caption{Two different deformations of three elementary squares (not all edges that connect black vertices are shown).  The double arrows mean that the top diagram is proportional to both of the bottom two diagrams.  Equating the proportionality terms gives the star-triangle type identity \eqref{str0}.}
\label{3square}
\end{figure}
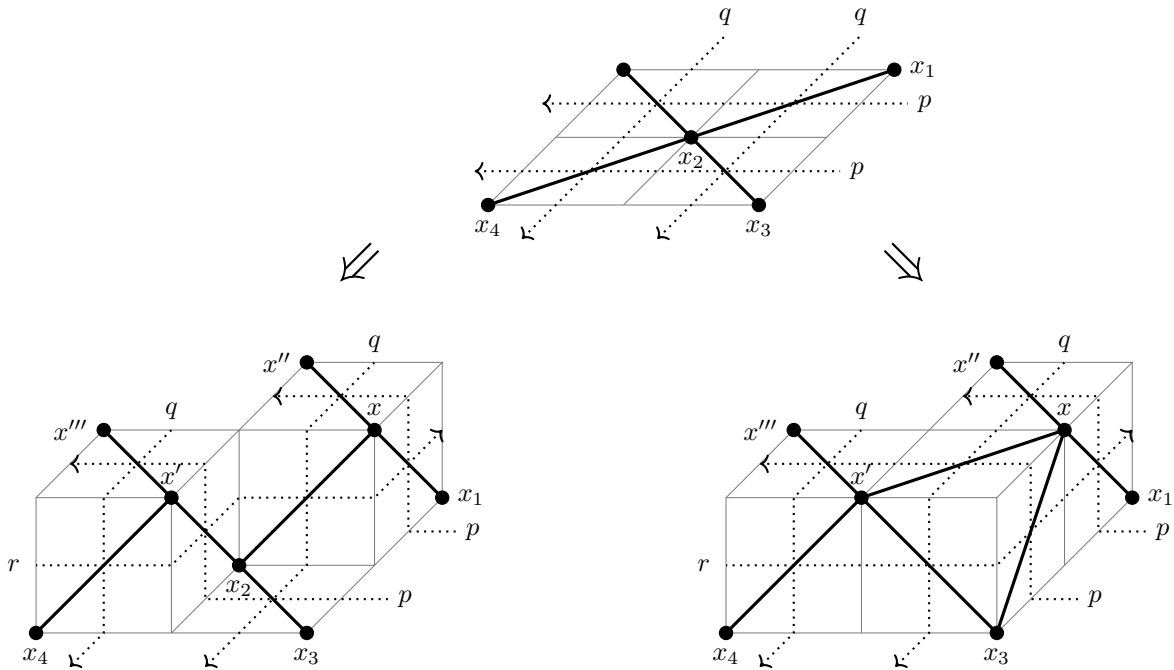

As an example, consider two different sequences of deformations on three elementary squares $\sigma_{ij}(\n),\sigma_{ij}(\n+\jhat),\sigma_{ij}(\n+\ihat+\jhat)$, shown in Figure \ref{3square}.

The first sequence involves the first deformation in Figure \ref{1cube}, applied to both of the elementary squares $\sigma_{ij}(\n),\sigma_{ij}(\n+\ihat+\jhat)$.\footnote{Note that this deformation does not lead to a valid surface $\sigma$ as defined in Section \ref{sec:sigp}, since one edge is now common to four different elementary squares.}  The second sequence involves first applying the second deformation of Figure \ref{1cube}, to $\sigma_{ij}(\n+\jhat)$, followed by applying the second deformation in Figure \ref{2cube}, to both $\sigma_{ij}(\n)$, and $\sigma_{ij}(\n+\ihat+\jhat)$.

Comparing the above two sequences of deformations, leads to the following ``star-triangle type'' relation, centered at $x_2$
\beq
\label{str0}
\begin{array}{l}
\ds\sum_{x,x',x'',x'''}f(x,x',x'',x''',x_1,x_2,x_4)\,\oW_{pq}(x,x')\,\oW_{rp}(x_3,x)\,\oW_{qr}(x',x_3) \\[0.7cm]
\ds\phantom{xxxx}=C\!\sum_{x,x',x'',x'''}\!f(x,x',x'',x''',x_1,x_2,x_4)\, W_{rp}(x',x_2)\, W_{qr}(x,x_2)\, W_{pq}(x_3,x_2)\,,
\end{array}
\eeq
where $C= R_{rpq}\,f_{qr}\,f_{rq}\,(S(x_3))^{-1}\,\delta_{x_3,x_3}\,(\sum_{x_2}S(x_2))^{-1}$, and
\beq
\begin{array}{l}
\ds f(x,x',x'',x''',x_1,x_2,x_4)= S(x)\, S(x')\, S(x'')\, S(x''') \\[0.3cm]
\ds\phantom{f(x,x',x'',x''',x_1,x_2,x_4)=xx}\times W_{rp}(x,x_1)\,\oW_{rq}(x_1,x'')\, W_{pq}(x,x'')\,\oW_{pr}(x'',x_2) \\[0.3cm]
\ds\phantom{f(x,x',x'',x''',x_1,x_2,x_4)=xx}\times W_{qr}(x',x_4)\,\oW_{rq}(x_2,x''')\, W_{pq}(x',x''')\,\oW_{pr}(x''',x_4)\,.
\end{array}
\eeq

Equation \eqref{str0} is valid for any edge interaction model with Boltzmann weights satisfying the star-triangle relation \eqref{STR}, and inversion relations \eqref{invrels}.  Graphically \eqref{str0} resembles the star-triangle relation in Figure \ref{fig3} (with some additional edges and vertices), although it is not straightforwardly derived from the corresponding equation \eqref{STR}, or inversion relations \eqref{invrels}.

\end{appendices}

\bibliography{total32}
\bibliographystyle{vvb-bibstyle}

\end{document}